\RequirePackage{fix-cm}
\documentclass[smallextended]{svjour3}       
\smartqed  
\usepackage{latexsym,graphics,amssymb,amsmath}

\usepackage[pdfstartview=FitH,colorlinks=true,citecolor=blue,bookmarks=false]{hyperref}

\usepackage{subcaption}

\usepackage{supertabular}

\usepackage{xcolor}

\usepackage{overpic}

\usepackage{graphicx,float,wrapfig}
\definecolor{Saddle}{HTML}{8b4513}
\definecolor{NewGrey}{HTML}{85888D}

\newcommand{\unit}[1]{\ensuremath{\, \mathrm{#1}}}
\newcommand\Pow{\textit{Pow}}

\renewcommand{\epsilon}{\varepsilon}
\renewcommand{\leq}{\leqslant}
\renewcommand{\geq}{\geqslant}

\newcommand{\be}{\begin{equation}}
\newcommand{\ee}{\end{equation}}
\newcommand{\eq}[1]{\eqref{#1}}

\renewcommand{\phi}{\varphi}
\renewcommand{\epsilon}{\varepsilon}

\newcommand{\NP}{N_P}

\newcommand{\cO}{\mathcal{O}}

\newcommand{\Qhomeo}{Q^{*}}
\newcommand{\Nhomeo}{N^{*}}
\newcommand{\Gonehomeo}{G_1^*}
\newcommand{\Gtwohomeo}{G_2^*}
\newcommand{\Gprod}{G_{\!\textit{prod}}}
\newcommand{\VN}{V_{\!N_{\!M\!}}}
\newcommand{\aNM}{a_{N_{\!M}}}
\newcommand{\tauNM}{\tau_{N_{\!M\!}}}
\newcommand{\gammaNM}{\gamma_{N_{\!M}}}
\newcommand{\bNP}{b_{N_{\!P}}}
\newcommand{\tauNP}{\tau_{N_{\!P}}}
\newcommand{\etaNP}{\eta_{N_{\!P}}}
\newcommand{\etaNPhomeo}{\eta_{N_{\!P}}^*}
\newcommand{\etaNPinf}{\eta_{N_{\!P}}^\textit{inf}}
\newcommand{\tauQ}{\tau_Q}
\newcommand{\tauN}{\tau_N}

\newcommand{\kappaNhomeo}{\kappa^*}
\newcommand{\kappaN}{\kappa}
\newcommand{\NR}{N_{\!R}}
\newcommand{\NRhomeo}{N_R^{*}}
\newcommand{\Ntot}{N_{tot}}
\newcommand{\Ntott}{[\NR(t)+N(t)]}
\newcommand{\Ntothomeo}{[\NRhomeo+\Nhomeo]}
\newcommand{\gammaNR}{\gamma_{N_{\!R}}}
\newcommand{\tauNRhomeo}{\tau_{N_{\!R}}^*}
\newcommand{\tauNChomeo}{\tau_{N_C}^*}
\newcommand{\tauNChalf}{\tau_{1/2}}
\newcommand{\ftrans}{\phi_{N_{\!R}}}
\newcommand{\ftranshomeo}{\phi_{N_{\!R}}^*}
\newcommand{\Nkohomeo}{N_{ko}^*}
\newcommand{\NRkohomeo}{N_{Rko}^*}
\newcommand{\GonehomeoPow}{(G_1^{*})^\Pow}

\newcommand{\mug}{\unit{\mu g}}

\newcommand{\Cf}{C_{\!f}}

\journalname{Bulletin of Mathematical Biology}

\def\makeheadbox{}

\begin{document}

\title{A mathematical model of granulopoiesis incorporating the negative feedback dynamics and kinetics
of G-CSF/neutrophil binding and internalisation
}

\titlerunning{Modelling granulopoiesis with G-CSF/neutrophil feedback}        

\author{M.~Craig \and
            A.R.~Humphries \and
            M.C.~Mackey
}


\institute{M. Craig \at
              Facult{\'e} de Pharmacie, Universit{\'e} de Montr{\'e}al, Montr{\'e}al, QC, Canada H3T 1J4 \\
              \email{morgan.craig@umontreal.ca}          
           \and
           A.R. Humphries \at
              Department of Mathematics and Statistics, McGill University, Montr{\'e}al, QC, Canada, H3A 0B9\\
              \email{tony.humphries@mcgill.ca}
             \and
             M.C. Mackey \at
             Departments of Mathematics, Physics and Physiology, McGill University, Montr\'eal, QC, Canada, H3G 1Y6\\
              \email{michael.mackey@mcgill.ca}
}


\maketitle

\begin{abstract}

We develop a physiological model of granulopoiesis which includes
explicit modelling of the kinetics of the cytokine granulocyte colony-stimulating factor (G-CSF) incorporating both the freely circulating concentration
and the concentration of the cytokine bound to mature neutrophils. G-CSF concentrations are used to
directly regulate neutrophil production, with the rate of differentiation of stem cells to neutrophil precursors,
the effective proliferation rate in mitosis, the maturation time, and the release rate from the mature marrow reservoir into circulation all dependent on the level of G-CSF in the system. The dependence of the maturation time
on the cytokine concentration introduces a state-dependent delay into our differential equation model, and we show
how this is derived from an age-structured partial differential equation model of the mitosis and
maturation, and also detail the derivation of the rest of our model.
 The model and its estimated parameters are shown to successfully predict the neutrophil and G-CSF responses to a variety of treatment scenarios, including the combined administration of chemotherapy and exogenous G-CSF. This concomitant treatment was reproduced {\it without any additional fitting} to characterise drug-drug interactions.
\end{abstract}

\section{Introduction}
\label{sec:Intro}

We present a new model of granulopoiesis, in which the production of neutrophils
is governed by a negative feedback loop between the neutrophils and
granulocyte colony stimulating factor (G-CSF). G-CSF is the principal cytokine
known to regulate neutrophil production and in our model it is used to moderate
differentiation of stem cells, apoptosis of proliferating neutrophil precursors,
the speed at which neutrophils mature and the rate that mature neutrophils are released
from the marrow reservoir. To facilitate this, we derive not only new functions
for the pharmacodynamic effects of G-CSF, but also a new model of the
G-CSF kinetics which incorporates cytokine binding and internalisation by
the neutrophils. We dispense with the mass action law assumption made in some previous
models and directly model the concentration of both circulating G-CSF and G-CSF bound
to neutrophils. This improved kinetic model furnishes us with G-CSF concentrations
which are considerably more accurate than our previous models so we are able
to use them to directly drive the pharmacodynamic effects and finally form a fully closed
cytokine-neutrophil feedback loop.

At homeostasis the dominant removal
mechanism for G-CSF is internalisation by neutrophils after it binds to
receptors on these cells \cite{Layton2006}.
This gives rise to a negative feedback mechanism on the G-CSF pharmacokinetics (PKs)
whereby large concentrations of neutrophils result
in G-CSF being removed from circulation, in turn leading to low concentrations of circulating G-CSF.
On the other hand if neutrophil concentrations are reduced then G-CSF is not cleared from circulation
as quickly and circulating concentrations increase as a result. The feedback loop is
completed by the pharmacodynamic (PD) effects of the G-CSF:  elevated (depressed) G-CSF levels lead to increased (decreased) neutrophil production.
 Due to this feedback, using the simple paradigm
that neutrophil concentration
is a cipher for the cytokine concentration (with one low when the other is high), it is possible to derive
granulopoiesis models without explicitly modelling the cytokines. This is particularly useful because it is not universally agreed where or how the multitude of identified cytokines all act.

The mathematical modelling of granulopoiesis has a long and rich history
 \cite{Bernard2003,Brooks2012,Colijn:2,Craig2015,Foley:06,Foley:09,Hearn1998,kazarinoff79,kingsmith70,Schirm2014,schmitz88,schmitz90,scholz2005,shvitra83,wichlof88,Vainas2012,Vainstein2005,vonschulthess82} but one of the earliest and most complete treatments is that of  Rubinow~\cite{Rubinow1975} which  incorporates a number of features that we retain in our model, including active proliferation, maturation, a marrow reservoir and free exchange between the circulating and marginal blood neutrophil pools.  Rubinow's
model, however, predates the discovery and characterisation of G-CSF and so it uses neutrophil concentrations as a cipher for the cytokine and its effects. Subsequent physiological models
have also all incorporated at least some elements of this cytokine paradigm in their
modelling. Some authors have been principally interested in neutrophil pathologies, including cyclical neutropenia, chronic myeloid leukemia, and myelosuppression during chemotherapy, while others have primarily studied the effects of G-CSF mimetics. Many models of cyclic neutropenia, including \cite{Colijn:2,Foley:06,Hearn1998,Lei:10} and \cite{Schmitz1996} acknowledge the role of G-CSF in neutrophil production and pathologies but rely on the cytokine paradigm to drive the pharmacodynamic responses. A number of modelling approaches have been proposed, including compartmental ODE models \cite{Schirm2014,Friberg2003,Quartino2012,GonzalezSales2012,Krzyzanski2010,Wang2001}, delay differential equations (DDEs) incorporating statistical distributions to model delays \cite{Vainstein2005,Vainas2012}, and DDEs derived from age-structured partial differential equation (PDE) models, like the one developed in this work \cite{Brooks2012,Craig2015,Foley:09}.
%

In recent years, synthetic forms of G-CSF have been developed and are administered
to patients for a variety of reasons, including to treat cyclical neutropenia or as an adjuvant during chemotherapy \cite{Dale2015,Dale2011,Molineux2012}.
However, the administration of exogenous G-CSF breaks the cytokine paradigm and it is possible
for neutrophil and G-CSF concentrations to both be elevated at the same time. This breakdown of the natural feedback relationship can cause physiological models that use the paradigm
to mischaracterise the elimination dynamics of G-CSF. For example, both \cite{Krzyzanski2010} and \cite{Craig2015} overestimate the renal clearance of G-CSF so much as
to essentially eliminate the contribution of neutrophil-mediated internalisation, even though they each
include this nonlinear clearance in their models.
If elevated neutrophil concentrations are used to drive the system dynamics on the assumption
that corresponding G-CSF concentrations are reduced when they are in fact elevated, the modelled effects will act in the opposite sense to the physiology. As a consequence, the model will either
develop instabilities and/or give a poor fit to observed dynamics. 


The mischaracterisation of G-CSF elimination dynamics was the impetus for the current work.
Consequently, we will not use the neutrophil concentration as a cipher for the
G-CSF concentration, but will model both the G-CSF pharmacokinetics and pharmacodynamics (PK/PD) in detail.
For this, we develop a novel pharmacokinetic model of G-CSF which includes both unbound and bound blood concentrations.
The G-CSF concentrations given by this kinetic model are then used to drive the
pharmacodynamic effects of the cytokine, in a fully formed negative feedback loop.

We begin by summarising the granulopoiesis model in Section~\ref{sec:Model}. Its development is then extensively detailed in Section~\ref{sec:moddev}, beginning from the stem cells in Section~\ref{sec:HSCmodel}. The novel pharmacokinetic G-CSF model incorporating bound and unbound blood concentrations is motivated and developed in Section~\ref{sec:GCSFModel}. There we show how the hypothesis of an equilibrium between bound and unbound concentrations is not satisfied for G-CSF, necessitating the inclusion of more complex kinetics in its pharmacokinetic model. Next, the derivation of the DDE granulopoiesis model is given in Section~\ref{sec:PDEDerivation} and the  pharmacodynamic model of G-CSF is developed in Section~\ref{sec:PDUpdates}. Models of the exogenous drugs considered in our study are detailed in Section~\ref{sec:exogcsf}. Having laid the foundations of our model, the various methods of parameter estimation and fitting used for our analyses are subsequently explained in Section~\ref{sec:ParEstim}. These approaches include model-specific constraints, as seen in Sections~\ref{sec:NeutConstr} and \ref{sec:KO}, while fitting procedures from published data are described in Sections~\ref{sec:GCSFParameters}, \ref{sec:NeutFit}, and \ref{sec:ChemoEstimation}. The resulting parameters are then summarized in Section~\ref{sec:ParVals}. Finally in Section \ref{sec:ModelEvaluation} we put our model to the acid test of {\it predicting (not fitting)} the population neutrophil response in a group of patients undergoing simultaneous chemotherapy and G-CSF administration \cite{Pfreundschuh2004b,Pfreundschuh2004a} and obtain excellent agreement between the model predicted behavior and the clinical data.  
We conclude with a short discussion in Section~\ref{sec:Discussion}.

\section{Model Summary}
\label{sec:Model}

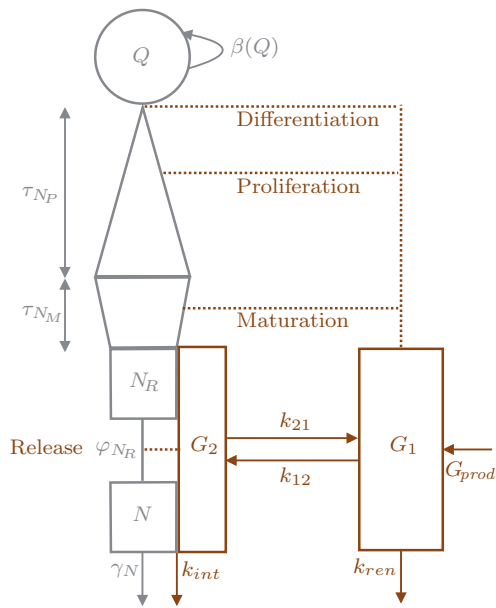
\begin{figure}[t]
\centering
\begin{overpic}[scale=0.95]{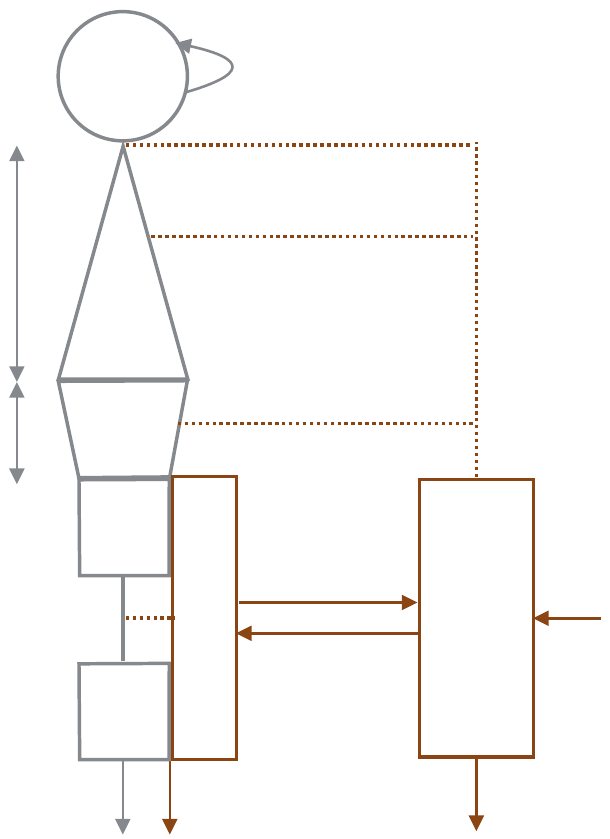}
\put(13,90){{\color{NewGrey}$Q$}}
\put(29,91.5){{\color{NewGrey}$\beta(Q)$}}
\put(-5,68){{\color{NewGrey}$\tauNP$}}
\put(-5,48.5){{\color{NewGrey}$\tauNM$}}
\put(7,26){{\color{NewGrey}$\ftrans$}}
\put(12.5,36.5){{\color{NewGrey}$\NR$}}
\put(13,14.5){{\color{NewGrey}$N$}}
\put(9.5,6){{\color{NewGrey}$\gamma_{N}$}}
\put(22.5,26){{\color{Saddle}$G_2$}}
\put(21,5.5){{\color{Saddle}$k_{int}$}}
\put(55,26){{\color{Saddle}$G_1$}}
\put(49,6){{\color{Saddle}$k_{ren}$}}
\put(64,22){{\color{Saddle}\small{$\Gprod$}}}
\put(30,79.5){{\color{Saddle}\small{Differentiation}}}
\put(30,68.5){{\color{Saddle}\small{Proliferation}}}
\put(30,46.5){{\color{Saddle}\small{Maturation}}}
\put(-7,25.5){{\color{Saddle}\small{Release}}}
\put(37,30){{\color{Saddle}$k_{21}$}}
\put(37,21){{\color{Saddle}$k_{12}$}}
\end{overpic}
\caption{Schematic representation of the production of circulating neutrophils in the bone marrow and the interaction of the system with G-CSF. Hematopoietic stem cells (HSCs-$Q$) enter the neutrophil lineage, the other blood lines, or are removed from the HSC pool. Differentiated HSCs then undergo successive divisions during the proliferative phase. Cells then mature 
before being stored in the marrow reservoir, or dying off during maturation. Neutrophils remain in the reservoir until they are removed randomly or enter the circulation, where they disappear rapidly from the blood.
Freely circulating G-CSF may bind to receptors on the neutrophils.
The concentration of bound G-CSF drives its pharmacodynamic effects. The concentration of
G-CSF bound to mature neutrophils, $G_2$, determines the rate of release from the marrow reservoir.
The concentration of G-CSF bound to neutrophil precursors, assumed proportional to $G_1$ the concentration
of freely circulating G-CSF, determines the rate of differentiation from the HSCs, the speed of maturation,
and the rate of proliferation. For all four effects, speed and rates increase with increasing G-CSF concentration.}
\label{fig:ModelSchematic}
\end{figure}

Here we define the variables and summarise the equations that define our granulopoiesis model. A detailed derivation is contained in Section~\ref{sec:moddev}.
Figure~\ref{fig:ModelSchematic} shows a schematic diagram describing the main elements of the hematopoietic system that we model.

The hematopoietic stem cell (HSC), neutrophil and G-CSF model is a set of five differential equations
including constant and state-dependent delays.
Let $Q(t)$ be the concentration of HSCs at time $t$, $\NR(t)$ be the concentration of mature neutrophils in the marrow reservoir, and $N(t)$ be the concentration of the total blood neutrophil pool (TBNP) at time $t$ (which includes both circulating and marginated neutrophils).
Further, let $G_1(t)$ be the concentration of unbound, circulating G-CSF and $G_2(t)$ be the concentration of G-CSF bound to receptors on mature neutrophils (in the reservoir or in the blood neutrophil pool).

The production of neutrophils from the HSCs is modelled by
\begin{eqnarray}
\tfrac{d}{dt}Q(t) &=& -\bigl(\kappaN(G_1(t)) +\kappa_\delta + \beta(Q(t))\bigr)Q(t) \notag \\
&&\mbox{}\qquad + A_Q(t)\beta\left(Q(t-\tauQ)\right)Q(t-\tauQ)   \label{eq:HSCs} \\
\tfrac{d}{dt}\NR(t) &=&  A_N(t) \kappaN(G_1(t-\tauN(t))) Q(t-\tauN(t))
 \frac{\VN(G_1(t))}{\VN(G_1(t-\tauNM(t)))} \notag \\
&&\mbox{}\qquad-\bigl (\gammaNR +\ftrans(G_1(t))\bigr )\NR(t) \label{eq:Reservoir} \\
\tfrac{d}{dt}N(t) &=&   \ftrans(G_{BF}(t))\NR(t)-\gamma_N N(t), \label{eq:Neutrophils}
\end{eqnarray}
with the concentrations of G-CSF (unbound and bound to neutrophil G-CSF receptors) given by
\begin{align} \notag
\tfrac{d}{dt}G_{1\hspace{-0.1em}}(t) & = I_G(t) 
+\Gprod-k_{ren}G_{1\hspace{-0.1em}}(t)\\ \label{eq:FreeGCSF}
& \mbox{}\qquad\quad-k_{12}(\Ntott V-G_{2\hspace{-0.05em}}(t))G_{1\hspace{-0.1em}}(t)^{\Pow}\hspace{-0.1em}
+k_{21}G_{2\hspace{-0.05em}}(t) \\ \label{eq:BoundGCSF}
\tfrac{d}{dt}G_{2\hspace{-0.05em}}(t) & = -k_{int}G_{2\hspace{-0.05em}}(t)+k_{12}\bigl(\hspace{-0.05em}
[N_{\!R}(t) \hspace{-0.1em}+\hspace{-0.05em}N\hspace{-0.1em}(t)] V\hspace{-0.4em}-G_{2\hspace{-0.05em}}(t)\hspace{-0.1em}\bigr)G_{1\hspace{-0.1em}}(t)^{\Pow}\hspace{-0.3em}-k_{21}G_{2\hspace{-0.05em}}(t),
\end{align}
where $I_G(t)$ indicates input of exogenous G-CSF, which we assume is filgrastim (the most common bio-similar
exogenous form of G-CSF). Filgrastim has very similar PK/PD properties to endogenous G-CSF,
so we will not distinguish between the two types of G-CSF in our model.

The derivation of these equations is given in Section~\ref{sec:moddev}.
In Section~\ref{sec:PDEDerivation}, particular attention is paid to the
derivation of the state-dependent delay terms in \eqref{eq:Reservoir} from an age-structured partial differential equation (PDE) model
of the mitosis and maturation with variable aging rate of the neutrophil precursors.
The G-CSF equations \eqref{eq:FreeGCSF},\eqref{eq:BoundGCSF} are explained in detail in Section~\ref{sec:GCSFModel}.

In the stem cell equation \eqref{eq:HSCs}, as explained in Section~\ref{sec:HSCmodel}, we have
\begin{gather}
\label{eq:betaQ}
\beta(Q) = f_Q\frac{\theta_2^{s_2}}{\theta_2^{s_2} + Q^{s_2}},\\
\label{eq:AQ}
A_Q(t) =  A_Q^* = 2e^{-\gamma_Q\tau_Q}.
\end{gather}
Only in the case of administration of chemotherapy is the stem cell amplification factor $A_Q(t)$ non-constant.
During chemotherapeutic treatment $A_Q(t)$ will be modified by replacing \eq{eq:AQ}
with \eqref{eq:HSCchemo} as discussed in Section~\ref{sec:exogcsf}.
Stem cells commit to differentiate to neutrophil precursors at a rate given by
\begin{equation} \label{eq:NewKappa}
\kappaN(G_1) = \kappaNhomeo + (\kappaNhomeo-\kappaN^\textit{min})\left[\frac{G_1^{s_1}-(\Gonehomeo)^{s_1}}{G_1^{s_1}+(\Gonehomeo)^{s_1}}\right].
\end{equation}
Here, and throughout, the superscript $^*$ denotes the homeostasis value of a quantity.
The rationale for using \eq{eq:NewKappa} to describe the pharmacodynamic effect of the G-CSF on the
differentiation of the HSCs, along with the other $G_1$-dependent functions
is explained in Section~\ref{sec:PDUpdates}.

After entering the neutrophil lineage, cells undergo mitosis at a variable rate ($\etaNP(G_1(t))$) given by
\be \label{eq:etaNP}
\etaNP(G_1(t)) =  \etaNPhomeo + (\etaNPhomeo-\etaNP^\textit{min})\frac{\bNP}{\Gonehomeo}
\left(\frac{G_1(t)-\Gonehomeo}{G_1(t)+\bNP}\right)
\ee
for a proliferation time $\tauNP$, considered to be constant. Cells subsequently mature
at a variable aging rate given by
\be \label{eq:Vn}
\VN(G_1(t)) = 1+(V_{max}-1)\frac{G_1(t)-\Gonehomeo}{G_1(t)-\Gonehomeo+b_V},
\ee
until they reach age $\aNM$ so the time $\tauNM(t)$ it  takes for a neutrophil maturing at time $t$ to mature satisfies the
integral relationship
\be \label{eq:tauNM1}
\int_{t-\tauNM(t)}^{t}\VN(G_1(s))ds   =   \aNM.
\ee
At homeostasis, $\VN(\Gonehomeo)=1$, and thus $\aNM$ is the homeostatic maturation time.
The total time it takes a neutrophil to be produced (from HSC differentiation
to release into the reservoir pool) is
\begin{equation}
\label{eq:tauN}
\tau_N (t)=\tauNP+\tauNM(t),
\end{equation}
and we can differentiate equation \eqref{eq:tauNM1} to obtain the following DDE for both $\tau_N$ and $\tauNM$
\begin{equation}
\label{eq:DerivTaus}
\tfrac{d}{dt}\tau_N(t)=\tfrac{d}{dt}\tauNM(t)  =   1- \frac{\VN(G_1(t))}{\VN(G_1(t-\tauNM(t)))}.
\end{equation}
Maturing neutrophils are assumed to die at a constant rate given by $\gammaNM$. The amplification factor $A_N(t)$ between
differentiation from HSCs to maturation that appears in \eqref{eq:Reservoir} is then given by
\begin{equation} \label{eq:AN}
A_N(t)
= \exp \left[\int_{t-\tau_N(t)}^{t-\tauNM(t)} \etaNP(G_1(s)) d s-\gammaNM\tauNM(t)\right]
\end{equation}
as derived in Section~\ref{sec:PDEDerivation}. 
Numerical implementation of the neutrophil amplification rate is obtained by differentiating the integral expressions in \eqref{eq:AN}
using Leibniz's Rule to obtain
\begin{align} \notag
\tfrac{d}{dt}A_{N\!}(t) = A_{N\!}(t)\Bigl[\bigl(1-\tfrac{d}{dt}\tauNM(t)\bigr)\bigl(\etaNP(G_1(t\!-\!\tauNM(t)))&-\etaNP(G_1(t\!-\!\tau_N(t)))\bigr)\\
& -\gammaNM\tfrac{d}{dt}\tauNM(t)\Bigr]. \label{eq:dANdt}
\end{align}
After maturation neutrophils are sequestered into the marrow neutrophil reservoir.
Mature neutrophils exit the reservoir either by dying with constant rate $\gammaNR$, or by being released into circulation with a rate $\ftrans$ depending on
the fraction $G_{BF}(t)$ of neutrophil receptors that are bound by G-CSF. We define
\be \label{eq:GBF}
G_{BF}(t)=\frac{G_2(t)}{V\Ntott}\in[0,1], \qquad G_{BF}^*=\frac{\Gtwohomeo}{V[\NRhomeo+\Nhomeo]},
\ee
and let
\begin{equation}
\label{eq:nu}
\ftrans(G_{BF}(t))=\ftranshomeo+(\ftrans^\textit{max}-\ftranshomeo)\frac{G_{BF}(t)-G_{BF}^*}{G_{BF}(t)-G_{BF}^*+b_G}.
\end{equation}
Neutrophils are removed from circulation with constant rate $\gamma_N$.

In equations \eq{eq:HSCs}--\eq{eq:BoundGCSF} we use units of $10^9$ cells per kilogram (of body mass) for
the reservoir and circulating neutrophils, and \unit{10^6cell/kg} for the stem cells. The scaling factors
ensure that computations are performed with numbers of similar magnitude which improves numerical stability.
Circulating and bound G-CSF concentrations are measured in standard units of nanograms per millilitre of blood.
The differing units for neutrophils and G-CSF
are only problematical in equations \eq{eq:FreeGCSF},\eq{eq:BoundGCSF} where quantities
in both units appear; see Section~\ref{sec:GCSFParameters} for the derivation of the conversion factor $V$.

Its also important to note that $N(t)$ measures the total blood neutrophil pool, including both the circulating and marginated neutrophils. To convert $N(t)$ to an absolute neutrophil count/circulating neutrophil numbers $N_C(t)$ (or \emph{vice versa}) there
is a conversion factor; see \eq{eq:ANC}.

\section{Model Development}
\label{sec:moddev}

Here we describe the development of our granulopoiesis model leading to the equations presented in Section~\ref{sec:Model}. The equation for the stem cells \eq{eq:HSCs} is described briefly
in Section~\ref{sec:HSCmodel}. The size of the mature neutrophil reservoir is described by \eq{eq:Reservoir}.
The first term on the right-hand side of this equation gives the rate that mature neutrophils enter the reservoir.
This term is derived from an age-structured PDE model described in Section~\ref{sec:PDEDerivation} below.
Neutrophils are assumed to leave the reservoir either by dying at rate $\gammaNR$ or by entering into
circulation at rate $\ftrans$, and are removed from circulation at a constant rate $\gamma_N$.
In Section~\ref{sec:GCSFModel} we describe our new G-CSF model \eq{eq:FreeGCSF},\eq{eq:BoundGCSF}
of the unbound freely circulating G-CSF ($G_1$), and the G-CSF bound to receptors
on the neutrophils ($G_2$). This model allows us to model the pharmacodynamic effects of the G-CSF
directly as detailed in Section~\ref{sec:PDUpdates}. Finally, Section~\ref{sec:exogcsf} outlines our models for the exogenous drugs we will consider in later sections.

\subsection{Stem Cells}
\label{sec:HSCmodel}

Equation \eq{eq:HSCs} for the stem cell dynamics was previously used in \cite{Mackey:01,Colijn:1,Colijn:2,PujoMenjouet2005,Foley:09,Lei:10,Brooks2012,Craig2015}. In particular, see \cite{Bernard2003} for a detailed derivation.
Here, we remove the dependence of $\gamma_Q$ upon G-CSF as the HSC population is relatively stable and infrequently dividing \cite{Riether2015,Durand2015} and, to our knowledge, no direct evidence of G-CSF's action upon HSC apoptosis currently exists.
Craig \cite{Craig2015} uses
\be \label{eq:AQCraig}
A_Q(t) =  2 \exp\left[-\int_{t-\tau_Q}^{t}\hspace{-1em}\gamma_Q(s) ds\right],
\ee
and in the absence of chemotherapy we take the apoptotic rate $\gamma_Q$ to be constant so this becomes \eqref{eq:AQ}.

\subsection{A physiologically constructed pharmacokinetic G-CSF model}
\label{sec:GCSFModel}

A new pharmacokinetic model of G-CSF, already stated in \eqref{eq:FreeGCSF},\eqref{eq:BoundGCSF}
is used to model the concentrations of both unbound and bound G-CSF.
We do not distinguish between endogenous and exogenous G-CSF in the model, which
constrains us to only consider biosimilar forms of exogenous G-CSF.
Accordingly, we focus on filgrastim, the most widely-available form of exogenous G-CSF. However, other less common forms of biosimilar exogenous G-CSF are available and include lenograstim and Nartograstim\textsuperscript{\textregistered} \cite{Molineux2012}.
The pegylated form of rhG-CSF has greatly reduced renal clearance
relative to endogenous G-CSF, which would require a different model,
so we will not consider it in this work.


In equations \eqref{eq:FreeGCSF},\eqref{eq:BoundGCSF}
$G_1$ is the concentration of freely circulating G-CSF and $G_2$ is the concentration
of G-CSF which is bound to receptors on the neutrophils. Since the bone marrow is well
perfused. G-CSF can bind to mature neutrophils in the marrow reservoir
as well as neutrophils in circulation. In the model $k_{ren}$ denotes the nonsaturable
removal rate of circulating G-CSF (mainly renal). $k_{int}$ denotes the removal rate
of bound-G-CSF, which we refer to as the effective internalisation rate. This term models
the removal of bound G-CSF both by internalisation after binding
and through the removal of the neutrophil itself from circulation (along with its bound
G-CSF molecules). $k_{12}$ is the rate of binding of free G-CSF to the neutrophils,
and $\Pow$ is the effective binding coefficient. The G-CSF receptor has a 2:2 stoichiometry in \textit{in vitro} studies \cite{Layton2006}, so a simple chemical reaction model would suggest $\Pow=2$. However, the number of ligands binding to a receptor only provides an upper bound on the corresponding Hill coefficient \cite{Santillan2008}. Accordingly, we use an effective binding coefficient $\Pow\in[1,2]$.


In this model the bound G-CSF
concentration is saturable, with $V \Ntott$ being the capacity of this
compartment. $G_2=V \Ntott$ would indicate that every receptor on every neutrophil
in the reservoir and circulation was bound to two G-CSF molecules. Thus the removal
rate of neutrophils by internalisation is saturable.
G-CSF also binds to immature neutrophils and precursors, which will be important for the
pharmacodynamics, but since these cells are fewer in number and/or have fewer receptors
than the mature neutrophils we neglect this effect on the pharmacokinetics.
Finally, $k_{21}$ is the rate of unbinding (transformation from bound G-CSF to circulating G-CSF),
and $I_G(t)$ denotes exogenous administration of G-CSF, discussed in Section~\ref{sec:exogcsf}.

If we were to assume that
there is no net transfer between the bound and circulating G-CSF then
letting $\tilde{N}(t)=\Ntott$, equations \eqref{eq:FreeGCSF},\eqref{eq:BoundGCSF} imply
\be \label{eq:gcsfbal}
k_{12}(V \tilde{N}(t) -G_2)G_1^{\Pow}-k_{21}G_2\approx0.
\ee
Rearranging \eq{eq:gcsfbal} we obtain
$$G_2(t)\approx\frac{[G_1(t)]^\Pow}{[G_1(t)]^\Pow+k_{21}/k_{12}}V\tilde{N}(t).$$
Now, adding \eq{eq:FreeGCSF} and \eq{eq:BoundGCSF}
$$\tfrac{d}{dt}(G_1+G_2)\approx I_G(t)+\Gprod-k_{ren}G_1-k_{int}G_2,$$
and assuming that $G_1\gg G_2$ and that $\tfrac{d}{dt}(G_1+G_2)\approx \tfrac{d}{dt}G_1$,
and finally replacing the $\approx$ by an equality
we have
\be \label{eq:gcsf1eq}
\tfrac{d}{dt}G_1 = I_G(t)+\Gprod-k_{ren}G_1-k_{int}V\tilde{N}(t)\frac{[G_1(t)]^\Pow}{[G_1(t)]^\Pow+k_{21}/k_{12}}.
\ee
Equations similar to \eq{eq:gcsf1eq} have been used to model G-CSF pharmacokinetics in many papers including \cite{Craig2015,Brooks2012,Foley:09,Krzyzanski2010,Krinner2013,Wang2001}, but usually
with $\tilde{N}(t)=N(t)$ the concentration of circulating neutrophils,
as opposed to $\tilde{N}(t)=\Ntott$ as \eqref{eq:FreeGCSF},\eqref{eq:BoundGCSF} suggest.

The usual derivation of \eq{eq:gcsf1eq} is from the law of mass action, but this is equivalent
to the assumption \eq{eq:gcsfbal} that the bound and circulating G-CSF are in quasi-equilibrium.
However, the equilibrium hypothesis \eq{eq:gcsfbal} cannot hold at homeostasis, since if
\eq{eq:gcsfbal} holds and $k_{int}>0$ then $\tfrac{d}{dt}G_2<0$ which is contradictory.
Clinical evidence \cite{Sarkar2003,Terashi1999}
suggests that at homeostasis, binding and internalisation is
the dominant removal mechanism for G-CSF, so not only does \eq{eq:gcsfbal} not hold but
the net transfer from unbound to bound G-CSF  should be more than $0.5\times\Gprod$.
Another important situation where \eq{eq:gcsfbal} will fail is during exogenous administration of G-CSF,
which will initially increase the concentration of unbound G-CSF (often by orders of magnitude).

\begin{figure}[t]
\begin{center}
\includegraphics[scale=1]{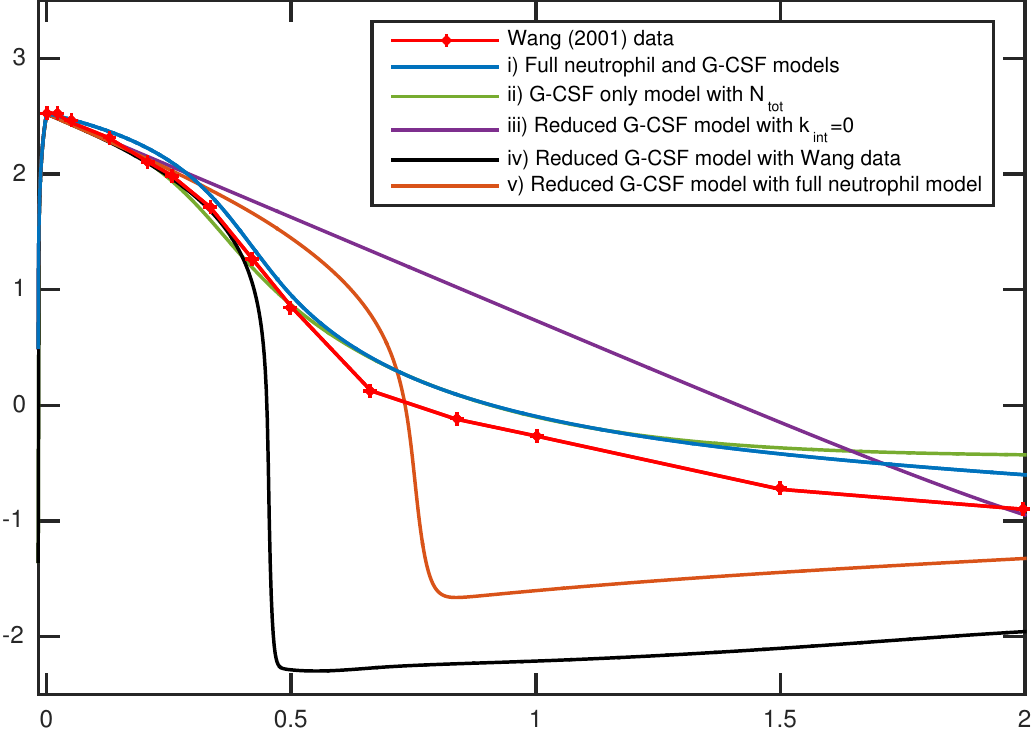}
\put(-305,191){\rotatebox[origin=c]{90}{log($G_1(t))$}}
\put(-20,-5){Days}
\end{center}
\caption{Data from Wang \cite{Wang2001} for G-CSF concentrations after a 750\unit{\mu g} 25 minute IV infusion and five different simulations: (i) the full neutrophil and G-CSF model \eq{eq:HSCs}--\eq{eq:BoundGCSF}
(ii) the G-CSF only model \eq{eq:UnboundGCSFsimp},\eq{eq:BoundGCSFsimp}, (iii) the reduced G-CSF model \eq{eq:gcsf1eq}
with $k_{int}=0$, (iv) the reduced G-CSF model \eq{eq:gcsf1eq} with $k_{int}=30$ and $\tilde{N}(t)=N(t)$ and
neutrophil concentrations taken from the Wang \cite{Wang2001} and (v) the full neutrophil model
\eq{eq:HSCs}--\eq{eq:Neutrophils} and the reduced G-CSF model \eq{eq:gcsf1eq} with  $k_{int}=25$ and $\tilde{N}(t)=\Ntott$.
In ii) $N_{tot}=4.1457$ and $G_2^*$ and $G_{prod}$ are determined by Equations~\eqref{eq:G2Calculation} and \eqref{eq:GprodCalculation}, respectively. In (ii), (iv) and (v) $k_{ren}=4.12$ and
 $G_{prod}$ is determined by \eqref{eq:gcsf1eq}.
 All other parameters take values specified in the third columns of Tables~\ref{tab:PKParams} and \ref{tab:NeutrophilValues}.}
\label{fig:gcsffails}
\end{figure}

Figure~\ref{fig:gcsffails} illustrates some of the issues involved in modelling the kinetics of G-CSF.
This figure shows data from a 750 $\mu$g intravenous (IV) infusion digitised from Figure~6 of Wang \cite{Wang2001},
along with a number of simulations of the protocol using different G-CSF kinetic models.
The data in Figure~\ref{fig:gcsffails} seems to have at least two different slopes, suggesting that the
G-CSF time course could be approximated by the sum of two exponentials. This naturally leads
to two compartment pharmacokinetic models \cite{DiPiro2010}.
Such a two-compartment G-CSF model was previously considered by Kuwabara \cite{Kuwabara1994} for Nartograstim\textsuperscript{\textregistered}.
Consistent with general two-compartment models in pharmacology, the two compartments corresponded to the blood
and the tissues, and  generic saturable and nonsaturable removal of the G-CSF both occurred from the blood compartment.
This differs from our model where elimination occurs from the two compartments (which instead represent unbound and bound G-CSF concentrations), both of which are subject to linear elimination. By contrast, in our model one compartment is saturable with nonsaturable elimination (the bound G-CSF), which corresponds to known G-CSF removal mechanisms. The assignment of elimination to the first or second compartments also has significant effects on the estimation of corresponding pharmacokinetic parameters so the mischaracterisation of these elimination dynamics could have significant effects on the model's predictions and behaviours \cite{Wu2015b}.

The circulating G-CSF concentration time course for a simulation of our full model \eq{eq:HSCs}--\eq{eq:BoundGCSF}
tracks the measured G-CSF data very closely in Figure~\ref{fig:gcsffails}.
It slightly overestimates the G-CSF, but it is important to note that
the data points are average values from a number of subjects and we will see in Section~\ref{sec:GCSFParameters}
that our G-CSF concentrations are well within the data range for several of administration protocols.

Also shown in Figure~\ref{fig:gcsffails} is a simulation of a simplified version of the G-CSF equations
\eq{eq:FreeGCSF},\eq{eq:BoundGCSF} where the time dependent neutrophil term $\Ntott$ is replaced by a constant
$\Ntot$, so the G-CSF kinetic equations become independent of the neutrophil dynamics. The
resulting equations are stated as \eq{eq:UnboundGCSFsimp},\eq{eq:BoundGCSFsimp} in Section~\ref{sec:GCSFParameters}
where they are used to determine the pharmacokinetic parameters that appear in \eq{eq:FreeGCSF},\eq{eq:BoundGCSF}.
The constant $\Ntot$ can be thought of as a time average of the term $\Ntott$.
As seen in Figure~\ref{fig:gcsffails}, this stand-alone
simplified G-CSF model gives G-CSF concentrations very close to those of the full model, which justifies
using it to determine the kinetic parameters.

Three different simulations of the single G-CSF equation \eq{eq:gcsf1eq} are also shown in Figure~\ref{fig:gcsffails}
to illustrate the difficulties in dealing with reduced models. One simulation has $k_{int}=0$ so that the elimination
of G-CSF is purely renal and it is clear that the nuances of the G-CSF kinetics are lost.


A simulation of \eq{eq:gcsf1eq} with $k_{int}>0$ and $\tilde{N}(t)=N(t)$ (with values for $N(t)$ taken from the
Wang data) gives even worse results than the purely renal elimination case. The problem with this model is that
for the first few hours while the neutrophil concentration is low, the elimination of the G-CSF is mainly renal
and the solution closely tracks the results from the purely renal elimination simulation. But as soon as the circulating
neutrophil concentrations get high enough the elimination of G-CSF by binding becomes dominant and quickly drives the
G-CSF concentration to very low levels. Similar results are seen if our full neutrophil model
\eq{eq:HSCs}--\eq{eq:Neutrophils} is coupled to \eq{eq:gcsf1eq} with $\tilde{N}(t)=\Ntott$.

The tendency of the internalisation term to quickly drive the G-CSF concentrations down, along with the
propensity for parameter fitting with linear scales resulted in several previous models using
versions of \eq{eq:gcsf1eq} to take kinetic parameters for which the elimination of G-CSF is
always renal dominated. This is seen both when the G-CSF kinetics is coupled to
physiological models as in \cite{Brooks2012,Craig2015} and when using
traditional empirical models as in \cite{Wang2001,Krzyzanski2010}, which consequently all
have elimination dynamics which are always renal dominated.

This is true in both the models of Craig~\cite{Craig2015}, which used \eq{eq:gcsf1eq}
with $\tilde{N}(t)=N(t)$, and Krzyzanski~\cite{Krzyzanski2010} which used an equation similar to
\eq{eq:gcsf1eq} but taking account of binding to all available receptors. In both,
elimination by internalisation is included in the mathematical models but occurs at an insignificant rate
compared to the renal elimination, contrary to the clinical understanding that elimination of G-CSF
by internalisation is the dominant removal mechanism at homeostasis.

From our numerical experiments it seems impossible to fit the single G-CSF equation \eq{eq:gcsf1eq} to data
when $\tilde{N}(t)$ is taken to be $N(t)$. The mature marrow neutrophil reservoir is an order of magnitude larger than the total blood neutrophil pool, and the receptors on the mature neutrophils need to be taken into account in the kinetics
as in \eqref{eq:FreeGCSF},\eqref{eq:BoundGCSF} to obtain a good fit to data. But taking account of all the receptors
is not sufficient to obtain a model that fits the physiology closely. This is evidenced by the very poor
fit obtained
in Figure~\ref{fig:gcsffails} when coupling our neutrophil model to the reduced G-CSF equation \eq{eq:gcsf1eq}
with $\tilde{N}(t)=\Ntott$, and also from models such as that of Krzyzanski~\cite{Krzyzanski2010} that take account
of the G-CSF receptors in marrow, but still obtain renal dominated kinetics.


The study of congenital diseases like cyclical neutropenia (CN)--an inherently oscillatory and dynamic disease--
and exogenous dosing regimens (such as during chemotherapy)
necessitate that the dynamics of G-CSF be well-characterised.
Hence we use the more realistic model \eqref{eq:FreeGCSF},\eqref{eq:BoundGCSF} for G-CSF pharmacokinetics
rather than the single equation reduction \eq{eq:gcsf1eq}.

\subsection{Modelling Granulopoiesis}
\label{sec:PDEDerivation}

The first term on the right hand side of \eq{eq:Reservoir} gives the rate that mature neutrophils enter the bone marrow reservoir at time $t$, and is obtained by modelling the differentiation of stem cells at time $t-\tau_N(t)$
through mitosis of neutrophil precursors to time $t-\tau_N(t)+\tauNP=t-\tauNM(t)$ followed by maturation of the cells
until time $t$. The time variation of $\tau_N(t)$ and $\tauNM(t)$ is solution dependent so this term involves state-dependent delays. Granulopoiesis models incorporating state-dependent delay have been employed before
in \cite{Foley:09,Foley:09b,Brooks2012}, but the derivation of those models was inaccurate and they missed the important
$\VN(G_1(t))/\VN(G_1(t-\tauNM(t)))$ term. Here we will show in detail how the
mitotic and maturation stages of the neutrophil precursors
can be modelled by age-structured PDE models, whose solution by the method of characteristics leads to the
state-dependent delay terms in \eq{eq:Reservoir}.

We do not model the cell-cycle process during mitosis, nor do we differentiate between the different maturation stages
of dividing cells (myeloblasts, promyelocytes, myelocytes). Rather, to simplify the modelling and the
resulting differential equations we model mitosis as an exponential process from the moment the HSC
commits to differentiate to the end of the mitosis. The proliferation rate $\etaNP$ is assumed to
be independent of which stage in mitosis the cell has reached.
There is evidence  that the cytokine G-CSF affects the
differentiation of HSCs and the effective proliferation rate during mitosis, as
explained in \cite{Endele2014}, and so we allow both the differentiation rate $\kappaN$ and
the proliferation rate $\etaNP$ to vary with $G_1$, the circulating G-CSF, as seen in
equations \eqref{eq:NewKappa},\eqref{eq:etaNP}, and explained in Section~\ref{sec:PDUpdates}.

We let $n_p(t,a)$ be the cell density as a function of time $t$ and age $a$ during proliferation.
We assume that cells age at a constant rate, $\dot{a}=1$, from age $0$ to age $\tauNP$, so $\tauNP$
is also the time period that cells spend in proliferation, and the proliferation rate is $\tauNP(G_1(t))$.
Then, differentiating,
$$\etaNP(G_1(t))n_p(t,a)=\frac{dn_p}{dt}
=\frac{\partial n_p}{\partial t}+\frac{da}{dt}\frac{\partial n_p}{\partial a}
=\frac{\partial n_p}{\partial t}+\frac{\partial n_p}{\partial a}$$
so the age-structured PDE model for proliferation is
\be \label{eq:PDEProl}
\frac{\partial n_p}{\partial t}+\frac{\partial n_p}{\partial a}=\etaNP(G_1(t))n_p(t,a), \qquad t\geq0, \quad a\in[0,\tauNP],
\ee
which, by the method of characteristics has solution
\begin{equation} \label{eq:Proliferation}
\mbox{}\hspace{-0.5em}n_p(t,a)=n_p(t-a,0)\exp\left[\int_{t-a}^t \hspace{-1em}\etaNP(G_1(s)) ds \right]\!, \; t\geq0, \; a\in[0,\min\{t,\tauNP\}].
\end{equation}
If $\tauNP\geq a>t>0$ the solution depends on the initial condition $n_p(0,a-t)$, but a similar expression applies. Here we have taken homeostasis as the initial condition throughout and so the solution in \eq{eq:Proliferation} is all that is required.

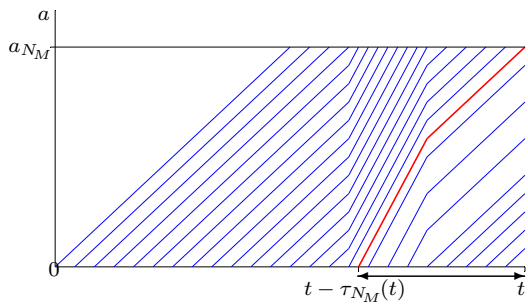
\begin{figure}[t]
\begin{center}
\includegraphics[scale=0.3]{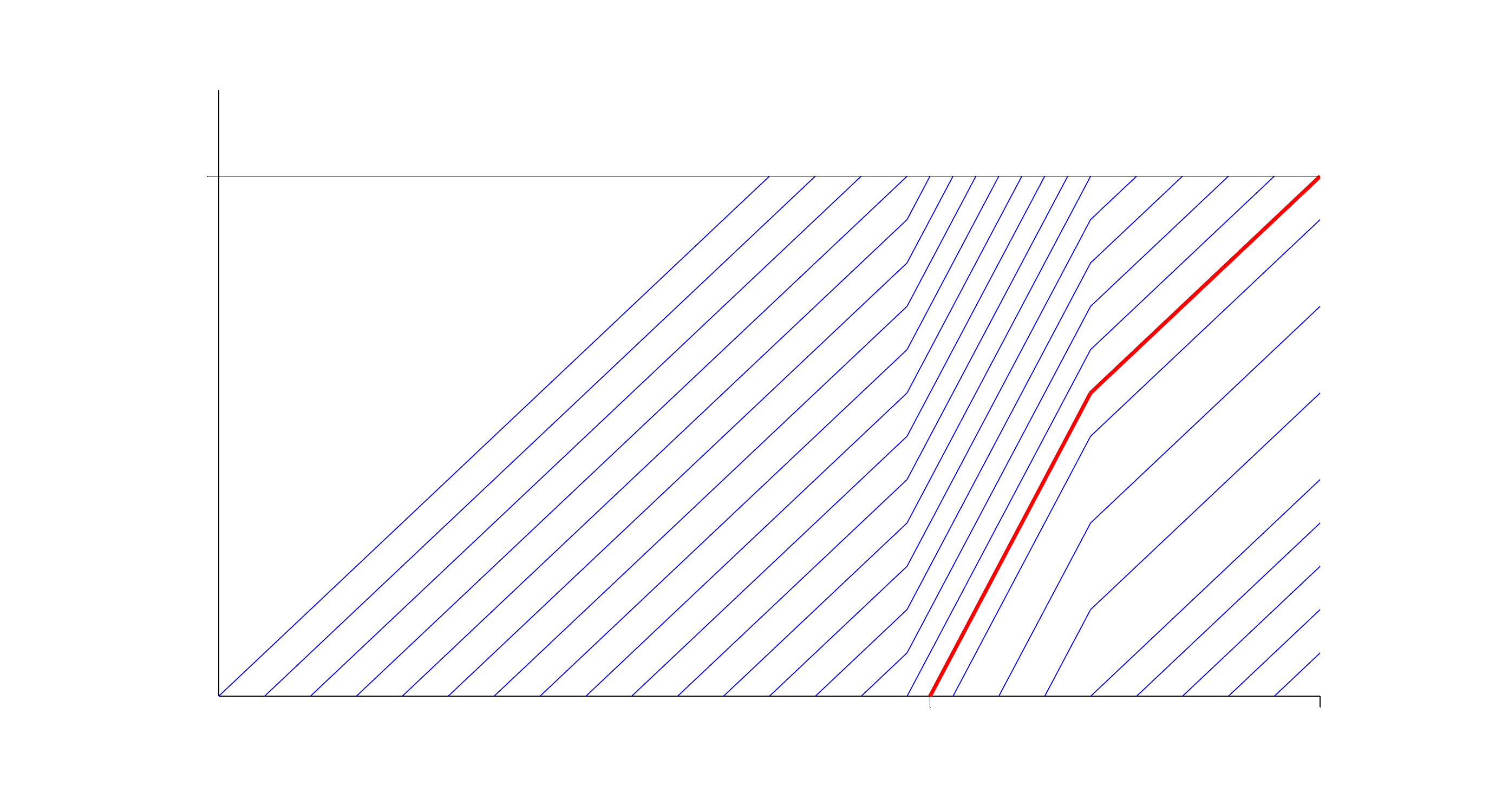}
\put(-206,15){$0$}
\put(-221,100){$\aNM$}
\put(-30,8){$t$}
\put(-110,8){$t-\tauNM(t)$}
\put(-210,112){$a$}
\put(-90,15){\vector(1,0){63}}
\put(-27,15){\vector(-1,0){63}}
\caption{During maturation the aging rate is variable with $\dot{a}(t)=\VN(G_1(t))$,
so age is not trivially related to time, and the maturation time $\tauNM(t)$ is variable.} \label{fig:matpde}
\end{center}
\end{figure}

We model the maturing neutrophil precursors (metamyelocytes and bands) as a single homogeneous compartment.
There is evidence that G-CSF affects the time that cells spend in maturation \cite{Spiekermann1997,Basu2002} and the speed up in maturation has been
measured experimentally \cite{Price}. Since the exact mechanism by which G-CSF affects maturation time
is unknown, we will model this process by decoupling time from age and demanding that cells age by an amount $\aNM$,
but allowing them to mature at a variable
aging rate $\dot{a}(t)=\VN(G_1(t))$ where $\VN(G_1)$ is a monotonically increasing function with
$\VN(0)>0$ and $\lim_{G_1\to\infty}\VN(G_1)=V_{max}<\infty$.

See Section~\ref{sec:PDUpdates} for further discussion
of the function $\VN(G_1)$.
We assume that the rate of cell death, $\gammaNM$, during maturation is constant independent of the concentration
of G-CSF. 

We let $n_m(t,a)$ be the cell density as a function of time $t$ and age $a$ during maturation for
$t\geq0$ and $a\in[0,\aNM]$.
Then the age-structured maturation model is
\be \label{eq:PDEMat}
\frac{\partial n_m}{\partial t}+\VN(G_1(t))\frac{\partial n_m}{\partial a}=
\frac{\partial n_m}{\partial t}+\frac{da}{dt}\frac{\partial n_m}{\partial a}=
\frac{dn_m}{dt}=-\gammaNM n_m(t,a).
\ee
The characteristics are defined by $\dot a=\VN(G_1(t))$, and along characteristics
for $t\geq\tauNM(t)$ we obtain
\be \label{eq:Maturation}
n_m(t,\aNM)  
= n_m(t-\tauNM(t),0)e^{-\gammaNM\tauNM(t)}.
\ee

Age-structured PDE models have been used in hematopoiesis models many
times previously \cite{Lei:10,Foley:09,Colijn:2,Craig2015}, but special care needs to be taken to
interpret $n_m(t,a)$ when the maturation has variable velocity, or an incorrect
solution will be obtained.

Cells which mature at time $t$ enter maturation at time $t-\tauNM(t)$ and so differentiated from
HSCs at time $t-\tauNM(t)-\tauNP=t-\tau_N(t)$. The rate at which cells differentiate at time
$t-\tau_N(t)$ is $\kappaN(G_1(t-\tau_N(t)))Q(t-\tau_N(t))$, and hence
$$n_p(t-\tau_N(t),0)=\kappaN(G_1(t-\tau_N(t)))Q(t-\tau_N(t)).$$
Then by \eq{eq:Proliferation}
\begin{align} \notag
n_p(t-&\tauNM(t),\aNM)=n_p(t-\tau_N(t),0)\exp\left[\int_{t-\aNM}^t \hspace{-0.7em}\etaNP(G_1(s)) ds \right]\\
&=\kappaN(G_1(t-\tau_N(t)))Q(t-\tau_N(t))\exp\left[\int_{t-\aNM}^t \hspace{-0.7em}\etaNP(G_1(s)) ds \right].
\label{eq:npa}
\end{align}

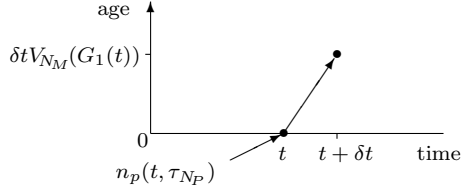
\begin{figure}[t]
\begin{center}
\begin{picture}(165,70)(-35,-20)
\put(10,0){\line(1,0){110}}
\put(80,-2){\line(0,1){2}}
\put(60,0){\circle*{3}}
\put(80,30){\circle*{3}}
\put(-10,45){age}
\put(10,0){\vector(0,1){50}}
\put(5,-5){$0$}
\put(40,-10){\vector(2,1){19}}
\put(60,0){\vector(2,3){19}}
\put(58,-10){$t$}
\put(73,-10){$t+\delta t$}
\put(110,-10){time}
\put(-43,27){$\delta t \VN(G_1(t))$}
\put(10,30){\line(-1,0){2}}
\put(-3,-17){$n_p(t,\tauNP)$}
\end{picture}
\caption{Transition from proliferation to maturation.}
\end{center}
\end{figure}

To obtain the boundary condition for the maturation phase, note that $n_p(t,\tauNP)$ is the rate
at which cells leave proliferation and enter maturation. Hence, to leading order,
$n_p(t,\tauNP)\delta t$ cells enter maturation in the time interval $[t,t+\delta t]$.
Cells that enter maturation at time $t$ will already have age $\VN(G_1(t))\delta t$ by time
$t+\delta t$. Since $n_p(t,a)$ and $n_m(t,a)$ describe the density of cells in the
proliferation and maturation phases, to avoid the spontaneous creation or destruction of cells
at the transition between proliferation and maturation we require
$$\int_{0}^{\VN(G_1(t))\delta t}\hspace{-1.5em} n_m(t+\delta t,a)da-
\int_{t}^{t+\delta t}\hspace{-1.5em}n_p(t,\tauNP)dt=\cO(\delta t^2).$$
Then
\begin{align} \notag
\VN(G_1(t))n_m(t,0)
&=\lim_{\delta t\to 0}\frac{1}{\delta t}\int_{0}^{\VN(G_1(t))\delta t}\hspace{-1.5em} n_m(t+\delta t,a)da\\
&=\lim_{\delta t\to 0}\frac{1}{\delta t}\int_{t}^{t+\delta t}\hspace{-1.5em}n_p(t,\tauNP)dt=n_p(t,\tauNP),
\label{eq:nnn}
\end{align}
and hence the boundary condition for the maturation compartment is
\be \label{eq:MatBC}
n_m(t-\tauNM(t),0)=n_p(t-\tauNM(t),\tauNP)/\VN(G_1(t-\tauNM(t))).
\ee
Combining \eq{eq:Maturation}, \eq{eq:npa}, \eq{eq:MatBC} and \eq{eq:AN} we obtain
\begin{align} \notag
n_m(t,&\aNM) = \frac{n_p(t-\tauNM(t),\tauNP)}{\VN(G_1(t-\tauNM(t)))}e^{-\gammaNM\tauNM(t)}\\ \notag
&=\frac{\kappaN(G_1(t-\tauN(t))Q(t-\tauN(t))}{\VN(G_1(t-\tauNM(t)))}
\exp\left[\int_{t-\tauN(t)}^{t-\tauNM(t)}\hspace{-2em}\etaNP(G_1(s)) ds-\gammaNM\tauNM(t) \right]\\
&=\frac{\kappaN(G_1(t-\tau_N(t)))Q(t-\tau_N(t))}{\VN(G_1(t-\tauNM(t)))} A_N(t).  \label{eq:PDEmat2}
\end{align}

Again because of the variable aging-rate there is a correction factor to apply to $n_m(t,\aNM)$
to obtain the rate that cells leave maturation.
To calculate this rate notice that cells which reach age $\aNM$ at time $t$
have age $\aNM-\VN(G_1(t))\delta t+\cO(\delta t^2)$ at time $t-\delta t$. Thus the number of
neutrophils that mature in the time interval $[t-\delta t,t]$ is
$$\int\limits_{\aNM- \VN(G_1(t))\delta t}^{\aNM}\hspace{-2.5em} n_m(t-\delta t,a)da+\cO(\delta t^2)
=\VN(G_1(t)) n_m(t,\aNM)\delta t+\cO(\delta t^2).$$
Hence, the rate that cells leave maturation is $\VN(G_1(t)) n_m(t,\aNM)$,
which using \eq{eq:PDEmat2} can be written as
\be \label{eq:PDEmat3}
\kappaN(G_1(t-\tau_N(t)))Q(t-\tau_N(t))A_N(t)\frac{\VN(G_1(t))}{\VN(G_1(t-\tauNM(t)))},
\ee
which is the first term on the right-hand side of \eq{eq:Reservoir}. The correction factor
${\VN(G_1(t))}/{\VN(G_1(t-\tauNM(t)))}$ was omitted from the state-dependent DDE
models in \cite{Foley:09,Brooks2012}.


\subsection{G-CSF Pharmacodynamics}
\label{sec:PDUpdates}

G-CSF in concert with many other cytokines regulates important parts of granulopoiesis.
The precise mechanisms by which it does this are not fully understood (and would probably be
beyond the level of detail that we would want to model mathematically even if they were)
but it is known that G-CSF acts along several signalling pathways in complex processes which activate and generate secondary signals that regulate neutrophil production \cite{Greenbaum2011,Semerad2002,Ward1998}.

The initiation of signalling pathways and the transfer of the resulting signals responsible for the various effects of a given drug may be driven directly by receptor binding and/or the internalisation of the drug. Assuming the rate at which
a drug is internalised is proportional to its bound concentration, we do not need to distinguish between the different possible pathways and will use the concentration of the bound drug
to drive the pharmacodynamics and produce the effects in the body.

Many previous models applied the cytokine paradigm mentioned in the introduction to model cytokine effects directly from the circulating neutrophil concentrations. For example in  \cite{Colijn:1,Colijn:2,Foley:09,Lei:10,Brooks2012,Craig2015}, the differentiation function was taken to be a monotonically decreasing function of the circulating neutrophil concentration. Some authors preferred instead to introduce simplified pharmacodynamic models using direct and indirect PD effects related to the concentration of unbound G-CSF \cite{Wang2001,Shochat2007} while other more detailed approaches have also been studied \cite{Scholz2012,Vainstein2005,Vainas2012}.

The cytokine paradigm breaks down when G-CSF is given exogenously. Immediate
responses of the hematopoietic system to G-CSF administration include releasing neutrophils from the marrow
reservoir into circulation, and increasing the maturation speed of neutrophils, so the circulating concentration
of neutrophils and the total number of neutrophils in the reservoir and circulation both increase, which results
in G-CSF and neutrophil concentrations being high concurrently.
Consequently we will use G-CSF concentrations
from \eq{eq:FreeGCSF},\eq{eq:BoundGCSF} to directly model the pharmacodynamic effects
of G-CSF
on the differentiation rate of HSCs $\kappaN$, the effective proliferation
rate of neutrophil precursors in mitosis $\etaNP$, the aging rate of maturing neutrophils $\VN$,
and the release rate of neutrophils from the bone marrow reservoir $\ftrans$.

We use Hill and Michaelis-Menten functions to model the G-CSF dependency of these effects.
There is some disagreement in the literature over exactly which cytokines are important
in different parts of the process, and we may be assigning some effects to G-CSF that are actually
due to GM-CSF or one of the other myriad of cytokines that regulate granulopoiesis. If these other cytokines are
mostly in quasi-equilibrium with G-CSF, using G-CSF as a cipher for all the cytokines should produce
very similar effects without the extraordinary complexity that would be inherent in modelling
each one of the cytokines.

Mammalian studies \cite{Dale91,Bugl2012,Lui2013}
reveal that neutrophils are still produced even in the absence of G-CSF, presumably
because other cytokines are acting. Accordingly, we will construct our effects functions to
have non-zero activity even in the complete absence of G-CSF. Moreover, in Section~\ref{sec:KO} we
will consider the case of G-CSF knockout mathematically with our model to derive a parameter constraint
to reduce the number of unknown parameters.

Recall that the concentration of G-CSF bound to mature neutrophils satisfies the inequality
$G_2(t)\leq V\Ntott$  with equality only if every G-CSF receptor were bound to two G-CSF molecules.
We suppose that the rate that mature neutrophils are released from the marrow reservoir into circulation is dependent
on the fraction $G_{BF}(t)=G_2(t)/(V\Ntott)$ of their receptors which are bound to G-CSF. The
rate is then given by the
Michaelis-Menten function $\ftrans(G_1)$ defined by \eqref{eq:nu}. 
Letting
\be \label{eq:ftransratio}
\ftrans^\textit{ratio}=\frac{\ftrans^\textit{max}}{\ftranshomeo}>1,
\ee
this function is also similar to the one used by
Shochat \cite{Shochat2007} that was adapted in Craig \cite{Craig2015} except that we use the
fraction of bound receptors
to drive the function.
At homeostasis \eq{eq:GBF} and \eq{eq:nu} imply that
$$\ftrans(G_{BF}^*)=\ftrans(\Gtwohomeo/[V(\Nhomeo+\NRhomeo)])=\ftranshomeo.$$
The parameter $b_G$ defines the half-effect concentration with
$$\ftrans(G_{BF}^*+b_G)=\frac12(\ftranshomeo+\ftrans^\textit{max}),$$
while the condition
$\ftrans(0)>0$ implies the constraint
\begin{equation}
\label{eq:Constraint1}
b_G>\ftrans^\textit{ratio}G_{BF}^*=\frac{\Gtwohomeo\ftrans^\textit{ratio}}{V(\NRhomeo+\Nhomeo)}.
\end{equation}

To model the effects of G-CSF on the differentiation, proliferation and maturation some care
must be taken. We posit that it is cytokine signalling that drives
these processes, and $G_2(t)$ denotes the concentration of bound G-CSF, which is proportional to the rate
that G-CSF is internalised. So it would be tempting
to use $G_2(t)$ to govern these processes, and indeed initially we tried this without
success. The problem is that $G_2(t)$ models the concentration of G-CSF bound to mature neutrophils
in the marrow reservoir and circulation. Through \eq{eq:FreeGCSF} and \eq{eq:BoundGCSF}
this gives a very good model of the removal of G-CSF from circulation because
although the neutrophil progenitor cells also have G-CSF receptors, these cells are relatively few
in number and have relatively few receptors, hence they can be ignored when modelling the
G-CSF kinetics. However, when modelling the pharmacodynamic effects of G-CSF it appears
to be crucial to take account of the binding of G-CSF to the neutrophil precursors, and it
is the freely circulating G-CSF which is available to bind to the G-CSF receptors on
the immature neutrophils and precursors. Consequently, we should use $G_1(t)$ to govern the cytokine
dependent differentiation, proliferation, and maturation.

Another way to see that it should be the circulating
G-CSF $G_1(t)$, and not the G-CSF bound to mature neutrophils $G_2(t)$ that should govern these processes is
as follows. If the concentration of mature neutrophils is decreased then the concentration of bound G-CSF
will also decrease because the number of receptors available to bind to will be decreased, but the concentration
of unbound G-CSF will increase because the rate the G-CSF is removed by internalisation is reduced. However, with a
reduced concentration of neutrophils, an elevated cytokine concentration is needed to
increase differentiation, proliferation and maturation speed.

We model the differentiation rate from HSCs to neutrophil precursors using the Hill function
\eq{eq:NewKappa}. Very little is known about how the differentiation rate changes in function of G-CSF, but we suppose that
it will not vary by orders of magnitude, since this would lead to instability in the HSC population,
while the HSC population is observed to be very stable in healthy subjects \cite{Riether2015}.
It is then convenient to assume that
the homeostatic rate is at the midpoint of the range of possible differentiation rates so
\be \label{eq:KKK}
\kappaNhomeo=\frac12(\kappaN^\textit{min}+\kappaN^\textit{max}).
\ee
With this assumption \eq{eq:NewKappa} is a standard sigmoidal Hill function with minimum
differentiation rate $\kappaN(0)=\kappaN^\textit{min}$, and with $\kappaN(G_1)$ increasing monotonically with $G_1$
and such that at homeostasis $\kappaN(\Gonehomeo)=\kappaNhomeo$, while for large concentrations
$\lim_{G_1\to\infty}\kappaN(G_1)=\kappaNhomeo+(\kappaNhomeo-\kappaN^\textit{min})=\kappaN^\textit{max}$.
To ensure that neutrophils are still produced in the complete absence of G-CSF we will require that
$\kappaN^\textit{min}>0$.


G-CSF is believed to increase the effective rate of mitosis during proliferation by reducing apoptosis.
Thus we
use a monotonically increasing Michaelis-Menten function $\etaNP(G_1(t))$ defined by \eq{eq:etaNP} to describe the G-CSF dependent
effective proliferation rate (which measures the difference between actual proliferation and apoptosis).
This function looks a little different than the other Michaelis-Menten functions we will use, but this is simply because it has been
scaled to give the correct minimal and homeostasis effects with $\etaNP(0)=\etaNP^\textit{min}>0$
and $\etaNP(\Gonehomeo)=\etaNPhomeo$, with $\etaNP(G_1)$ a monotonically increasing function of $G_1$.

Letting
$$\etaNP^\textit{max}=\lim_{G_1\to\infty}\etaNP(G_1)=\etaNPhomeo+\frac{\bNP}{\Gonehomeo}(\etaNPhomeo-\etaNP^\textit{min}),$$
we see that
$$\frac{\bNP}{\Gonehomeo}=\frac{\etaNP^\textit{max}-\etaNPhomeo}{\etaNPhomeo-\etaNP^\textit{min}},$$
so the parameter $\bNP>0$ determines the relative position of $\etaNPhomeo\in[\etaNP^\textit{min},\etaNP^\textit{max}]$
with $\etaNPhomeo>(\etaNP^\textit{min}+\etaNP^\textit{max})/2$ when $\bNP\in(0,\Gonehomeo)$ and
$\etaNPhomeo<(\etaNP^\textit{min}+\etaNP^\textit{max})/2$ when $\bNP>\Gonehomeo$.



G-CSF is known to affect the time that neutrophils spend in maturation \cite{Spiekermann1997,Basu2002}, an acceleration in maturation that Price \cite{Price} measured experimentally, but the mechanism by which G-CSF
speeds up maturation is not well understood. We choose to model this process by decoupling
time from age and demanding that cells age by an amount $\aNM$,
but allowing them to mature at a variable
aging rate $\dot{a}(t)=\VN(G_1(t))$ where $\VN(G_1)$ is a monotonically increasing Michaelis-Menten function
given in \eq{eq:Vn}.
This is similar to the form used in Craig \cite{Craig2015} which was adopted
from Foley~\cite{Foley:09}, and is also functionally equivalent to \eq{eq:nu}.

$b_V$ is the half effect parameter for the aging velocity with $\VN(\Gonehomeo+b_V)=(1+V_{max})/2$.
We require that $\VN(0)>0$, which from \eq{eq:Vn} is equivalent to
\begin{equation} \label{eq:Constraint3}
b_V>\Gonehomeo V_{max}.
\end{equation}
This constraint ensures that the aging velocity $\VN(G_1)$ is strictly positive for all $G_1\geq0$.
The function $\VN(G_1)$ also satisfies the homeostasis condition $\VN(\Gonehomeo)=1$, so that at
homeostasis the aging rate is $1$. The aging rate saturates with $\lim_{G_1\to\infty}\VN(G_1)=V_{max}<\infty$.

Notice that, using \eq{eq:DerivTaus}
\be \label{eq:VVfac}
\tfrac{d}{dt}(t-\tauNM(t))=1-\tfrac{d}{dt}\tauNM(t)=\frac{\VN(G_1(t))}{\VN(G_1(t-\tauNM(t)))},
\ee
and positivity of $\VN(G_1)$ assures that $t-\tauNM(t)$, and similarly $t-\tauN(t)$,
are monotonically increasing functions of $t$.
This is important in state-dependent DDE theory for existence and uniqueness of solutions.
Physiologically, it assures that cells which have exited proliferation or maturation never re-enter those
phases.



The responses of our new model and the model
of Craig~\cite{Craig2015} to exogenous administration of G-CSF are very different. With our new
model both differentiation and proliferation are increased with increased G-CSF so that after some time delay
the marrow reservoir gets replenished. In the previous model, the G-CSF triggered an immediate release of
neutrophils from the marrow reservoir into circulation and the resulting high circulating
neutrophil count would cause differentiation and proliferation to be decreased. This meant the the marrow reservoir would
suffer a double depletion with increased release into circulation combined with reduced production of new mature
neutrophils, which could lead to instabilities in the model that ought not to be occurring in the granulopoiesis
of healthy subjects.

Since the four functions \eq{eq:NewKappa},\eq{eq:etaNP},\eq{eq:Vn} and \eq{eq:nu}
describe the effects of G-CSF on granulopoiesis, rather than modelling
the processes that lead to the effects, the parameters in these functions do not correspond to physiological quantities that can be measured directly. Nevertheless these parameters can be determined by fitting the response of
the system to experimental data as described in Section~\ref{sec:NeutFit}.

\subsection{Modelling exogenous drug administration}
\label{sec:exogcsf}

As noted following \eqref{eq:FreeGCSF}, $I_G(t)$ denotes the input of exogenous G-CSF. The administration of rhG-CSF (in our case filgrastim) typically takes two forms: IV infusion (where the drug is given intravenously over a period of time) or subcutaneously (injection under the skin). In the former case, the drug passes directly into the bloodstream meaning the bioavailable fraction (the percentage of the administered dose that enters the blood) is 100\%. In this case, we express the single exogenous administration as
\begin{equation}
\label{eq:IVadmin}
I_G(t)= \begin{cases}
      \frac{Do}{t_\textit{inf}V_d}, & t_0 \leq t\leq t_{inf} \\
       0 & \textrm{otherwise},
   \end{cases}
\end{equation}
where $Do$ is the administered dose, $t_0$ is the start of the infusion, $t_\textit{inf}$ is the time of infusion and $V_d$ is the volume of distribution.
The volume of distribution is a pharmacokinetic parameter which relates the hypothetical volume a drug would occupy to the concentration it is observed in the plasma.
It is typically calculated for a drug
by dividing the administered dose by the concentration in the blood immediately following an
administration for the simplest case of IV bolus administration (instantaneous administration into the blood).
Drugs given subcutaneously do not immediately reach the bloodstream. Instead, a certain proportion of the medication remains in the subcutaneous tissue pool before diffusing into the plasma. Some previous studies, notably \cite{Foley:09,Brooks2012} used an extra transition compartment to model the administered G-CSF concentration in the tissues before reaching the blood and allowed for the free exchange between this central (blood) compartment and the tissue compartment. Owing to the specifics of the pharmacokinetics of filgrastim, we will instead use the following direct input functions from \cite{Krzyzanski2010} and \cite{Craig2015} to model subcutaneous administration as
\begin{equation}
\label{eq:SCadmin}
I_G(t)=\begin{cases}
\frac{k_aDoF}{V_d}e^{k_a t}, & t \geq t_0\\
0 & t < t_0,
\end{cases}
\end{equation}
where $k_a$ is the constant of absorption, and $F$ is the bioavailable fraction (the fraction of non-metabolised dose which enters the system).
This direct form is preferred over the two compartment method previously employed in \cite{Foley:09,Brooks2012} because of the relatively small volume of distribution exhibited by filgrastim (the bio-similar exogenous form of G-CSF), which is to say that $V_d$ is less than the standard 70L measure of highly distributed drugs \cite{Craig2015} and that the drug does not have a strong tendency to redistribute into the tissues.

The pharmacokinetic model of the chemotherapeutic drug (Zalypsis\textsuperscript{\textregistered}) used in this paper is the same as in \cite{Craig2015}. Briefly, the concentration of chemotherapeutic drug in the system is modelled using a set of four ordinary differential equations which was determined to be suitable through population pharmacokinetic analysis \cite{PerezRuixo2012}. The PK model of Zalypsis\textsuperscript{\textregistered} is given by
\begin{gather} \notag
\tfrac{d}{dt}C_{p}(t)=I_C (t)+k_{fp}\Cf(t)+k_{sl_{1}p}C_{sl_1}(t)-(k_{pf}+k_{psl_{1}}+k_{el_C})C_p(t)\\
\tfrac{d}{dt}C_f(t)=k_{pf}C_{p}(t)+k_{sl_{2}f}C_{sl_{2}}(t)-(k_{fp}+k_{fsl_{2}})\Cf(t)  \label{eq:ChemoModel1} \\
\tfrac{d}{dt}C_{sl_{1}}(t)=k_{psl_{1}}C_{p}(t)-k_{sl_{1}p}C_{sl_{1}}(t), \quad
\tfrac{d}{dt}C_{sl_{2}}(t)=k_{fsl_{2}}\Cf(t)-k_{sl_{2}f}C_{sl_{2}}(t), \notag
\end{gather}
where $C_p$ is the concentration in the central (blood) compartment, $\Cf$ is the concentration in the fast-exchange tissues, and $C_{sl_{1}}$ and $C_{sl_{2}}$ are the concentrations in the slow-exchange tissues, $k_{ij}$ are traditional rate constants between the $i^{th}$ and $j^{th}$ compartments ($i,j=p,f,sl_1,sl_2$), and $k_{el_C}$ is the rate of elimination from the central compartment. 
We consider the chemotherapeutic drug to be administered by IV infusion, so $I_C(t)=Dose_{Zal}/\Delta_{t}$, where $Dose_{Zal}$ is the administered dose and $\Delta_t$ is the time of infusion.

In contrast to the pharmacodynamic effects of G-CSF, chemotherapy has negative effects on the neutrophil (and other blood) lineages. Chemotherapy (and radiotherapy) works by disrupting the cell-cycle of tumours \cite{Maholtra2003} but this interference also affects all cells which are dividing, including the neutrophil
progenitors. The cytotoxic side effects chemotherapeutic treatment has on the neutrophils is called myelosuppression and it is a leading cause of treatment adaptation and/or cessation for patients undergoing chemotherapy \cite{Craig2015}. Since chemotherapy's myelosuppressive action only affects 
cells capable of division, we model the pharmacodynamic effects of chemotherapy on the HSCs, which rarely divide, and the neutrophil progenitors in the proliferative phase, which divide regularly
until they exit the mitotic phase.

Since the effects of chemotherapy on the HSCs are not clear, we model the antiproliferative effect as
a simple linear decrease
of the rate of apoptosis experienced by these cells by replacing $\gamma_Q$ in equation \eq{eq:AQCraig}
by $\gamma_Q+h_QC_p(t)$ where $C_p(t)$ is the concentration of the chemotherapeutic drug in the central blood compartment given by \eqref{eq:ChemoModel1}, and $h_Q$ is a factor to be determined
(as outlined in Section~\ref{sec:ChemoEstimation}). Then \eq{eq:AQCraig} gives
\begin{equation}
\label{eq:HSCchemo}
A_Q(t)=2e^{-\gamma_Q\tau_Q-h_Q\int_{t-\tau_Q}^t C_p(s)ds}.
\end{equation}
It is convenient to numerically implement \eq{eq:HSCchemo} as a differential equation,
and applying Leibniz's Rule to \eq{eq:HSCchemo},
similar to the derivation of \eqref{eq:dANdt}, we obtain
\begin{equation}
\label{eq:dAQdt}
\tfrac{d}{dt}A_Q(t)=(h_Q(C_p(t-\tau_Q)-C_p(t)))A_Q(t),
\end{equation}
and we replace \eq{eq:AQ} by \eq{eq:dAQdt} when chemotherapy is administered.

%
The second effect of chemotherapeutic drugs is to reduce the effective proliferation rate of the mitotic neutrophil progenitors. We model this by replacing $\etaNP$ of \eqref{eq:etaNP} by
\begin{equation}
\label{eq:etaNPchemo}
\etaNP^{chemo}(G_1(t),C_p(t))=\etaNPinf+\frac{\etaNP(G_1(t))-\etaNPinf}{1+(C_p(t)/EC_{50})^{s_c}},
\end{equation}
which is a modification of the model used in \cite{Craig2015}. Here $\etaNPinf$ corresponds to the effective proliferation rate in the presence of an infinite dose of the drug. We require $\etaNPinf<\etaNP^\textit{min}$ to ensure that effective proliferation is reduced, so $\etaNP^{chemo}(G_1(t),C_p(t))<\etaNP^{chemo}(G_1(t))$ whenever
$C_p(t)>0$. We will allow the possibility of $\etaNPinf<0$, which would correspond to negative
effective proliferation (more death than division in the mitotic phase) in the presence of very large
concentrations of the chemotherapeutic drug, though we note that because the drug is cleared from
circulation relatively quickly we will have $\etaNP^{chemo}(G_1(t),C_p(t))>0$ most of
the time even if $\etaNPinf<0$. If $\etaNPinf\in(0,\etaNP^\textit{min})$ then effective cell division is
reduced but never completely halted however large the concentration of the chemotherapeutic drug.
$EC_{50}$ is the concentration of chemotherapeutic drug which gives
the half-maximal effect, and $s_c$ is a Hill coefficient.
The parameters $h_Q$, $\etaNPinf$, $EC_{50}$, and $s_c$ will all be estimated using fitting techniques described in Section~\ref{sec:ChemoEstimation}.

\section{Parameter Estimation and Equation Constraints}
\label{sec:ParEstim}

In this section we show how our mathematical model imposes constraints on its own parameters to be self-consistent,
and how experimental data can be used to determine model parameters. We begin in Section~\ref{sec:NeutConstr}
by studying the model at homeostasis and deriving inequalities that the parameters must satisfy, as well
as showing how experimentally measured quantities can be used to directly determine some parameters in the model.
In Section~\ref{sec:GCSFParameters} we show how the G-CSF pharmacokinetic parameters can be determined using a
combination of model equation constraints and parameter fitting to experimental data
from single administrations of G-CSF. In Section~\ref{sec:KO}, G-CSF knockout is used to derive further parameter
constraints and relationships.
Finally in Section~\ref{sec:NeutFit} we show how the pharmacodynamic parameters in the neutrophil equations can
be determined by fitting the model to experimental data for the circulating neutrophil concentrations
after a single IV or subcutaneous administration  of G-CSF.

\subsection{Neutrophil Steady-State Parameter Determination and Constraints}
\label{sec:NeutConstr}

At homeostasis let $\Qhomeo$ be the stem cell concentration and
denote the sizes of the four neutrophil compartments at homeostasis by
$N_P^*$ (proliferation) , $N_M^*$ (maturation), $\NRhomeo$ (marrow reservoir), $\Nhomeo$ (total blood neutrophil pool),
and the average time that a cell spends in one of these stages at homeostasis by
$\tauNP$, $\aNM$, $\tauNRhomeo$ and $\tauNChomeo$, respectively. With the exception of $\tauNP$, all of
these quantities have been determined experimentally, but unfortunately only
$\tauNP$ and $\aNM$ actually appear in our model.
In this section we show that our model imposes some constraints on
the values of these parameters, and also how the values of $\kappaNhomeo$,
$N_P^*$, $N_M^*$, $\NRhomeo$, $\Nhomeo$, $\aNM$, $\tauNRhomeo$ and $\tauNChomeo$ can be used through the model
to determine values for the parameters
$\tauNP$, $\etaNPhomeo$, $\gammaNM$, $\gammaNR$, $\gamma_N$ and $\ftranshomeo$ which do appear in the model
in Section~\ref{sec:Model}.

At homeostasis equations \eq{eq:HSCs}--\eq{eq:Neutrophils} become
\begin{gather} \label{eq:HSCHomeo}
0 = -\bigl(\kappaNhomeo + \kappa_\delta + \beta(\Qhomeo)\bigr)\Qhomeo + A_Q^*\beta(\Qhomeo)\Qhomeo,\\
\label{eq:NMatBal}
\kappaNhomeo\Qhomeo A_N^*=(\ftranshomeo+\gammaNR)\NRhomeo, \\
\label{eq:Neutrohomeo}
\ftranshomeo\NRhomeo=\gamma_N \Nhomeo .
\end{gather}
Equation \eq{eq:HSCHomeo}
has the trivial solution $\Qhomeo =0$ with other solutions given by
\be \label{eq:Ksum}
\kappaNhomeo + \kappa_\delta= (A_Q^*-1)\beta(\Qhomeo )
\ee
To the best of our knowledge, there is no experimental data to determine the
relative rates of differentiation to the three cell lines (erythrocytes, neutrophils, thrombocytes)
at homeostasis. In the absence of any evidence to the
contrary, we will assume that these are all equal. Since $\kappaNhomeo$ denotes the differentiation
rate to the neutrophil line and $\kappa_\delta$ differentiation to erythrocyte and thrombocyte precursors
we obtain
\be \label{eq:kappaVals}
\kappaNhomeo=\tfrac12\kappa_\delta=\tfrac13(A_Q^*-1)\beta(Q^*).
\ee

At homeostasis neutrophil precursors are assumed to enter the mitotic phase at rate $\kappaNhomeo\Qhomeo$.
They then proliferate at a rate $\etaNPhomeo$ for a time $\tauNP$.
The total number of cells in the proliferative phase at homeostasis is thus
\be \label{eq:NProlifTot}
N_P^*=\int_0^{\tauNP}\kappaNhomeo\Qhomeo e^{\etaNPhomeo s}ds
=\kappaNhomeo\Qhomeo \frac{e^{\etaNPhomeo\tauNP}-1}{\etaNPhomeo},
\ee
and cells leave proliferation and enter maturation at a rate $R_P^*$
given by
\be \label{eq:RPstar}
R_P^*=\kappaNhomeo\Qhomeo e^{\etaNPhomeo\tauNP}.
\ee

At homeostasis from \eq{eq:Vn} we have $\VN(\Gtwohomeo)=1$, and thus
from \eqref{eq:tauNM1}, the time spent in maturation at homeostasis is $\aNM$.
The number of cells of age $s$ for $s\in[0,\aNM]$ in the maturation phase
is then $\kappaNhomeo\Qhomeo\exp(\etaNPhomeo\tauNP-\gammaNM s)$, and the total number
of cells in the maturation phase is
\be \label{eq:NMatTot}
N_M^*=\int_0^{\aNM}\kappaNhomeo\Qhomeo e^{\etaNPhomeo\tauNP-\gammaNM s}ds
=\kappaNhomeo\Qhomeo e^{\etaNPhomeo\tauNP}\frac{1-e^{-\gammaNM\aNM}}{\gammaNM}.
\ee
Writing
\be \label{eq:ANstar}
A_N^* = \exp\bigl(\etaNPhomeo\tauNP - \gammaNM\aNM\bigr),
\ee
which corresponds to \eq{eq:AN} at homeostasis, we
can rewrite \eq{eq:NMatTot} as
\be \label{eq:NMatTotANstar}
N_M^*=\kappaNhomeo\Qhomeo A_N^*\frac{e^{\gammaNM\aNM}-1}{\gammaNM}.
\ee
Now the rate at which cells leave the maturation phase is
$$\kappaNhomeo\Qhomeo e^{\etaNPhomeo\tauNP-\gammaNM\aNM}
=\kappaNhomeo\Qhomeo A_N^*.$$

The average time, $\tauNChomeo$,
that neutrophils spend in circulation in the blood (in the total blood neutrophil pool)
has been measured a number of times. However, what is actually measured is the half removal time, $\tauNChalf$, which gives
$\gamma_N$, the removal rate from circulation by
\be \label{eq:tauNC}
\gamma_N =\frac{1}{\tauNChomeo}=\frac{\ln 2}{\tauNChalf}.
\ee

Equation \eq{eq:Neutrohomeo} ensures that at homeostasis the rate neutrophils leave the reservoir and
enter circulation equals the rate at which they are removed from circulation.
From this we obtain
\be \label{eq:ftransGstar}
\ftranshomeo=\frac{\gamma_N \Nhomeo }{\NRhomeo}.
\ee

The rate at which neutrophils exit the mature marrow reservoir
is given by
$(\ftranshomeo+\gammaNR)\NRhomeo$ where $\ftranshomeo$ is the transition rate constant for cells
entering circulation and $\gammaNR$ is the random death rate. Thus the average time that cells spend in
the reservoir at homeostasis is
\be \label{eq:tauNR}
\tauNRhomeo=\frac{1}{\ftranshomeo+\gammaNR}.
\ee
Hence the random death rate in the
reservoir, $\gammaNR\geq0$, is given by
\be \label{eq:gammaNR}
\gammaNR=\frac{1}{\tauNRhomeo}-\ftranshomeo,
\ee
and we require that
\be \label{eq:NdelConstr}
\tauNRhomeo\ftranshomeo\leq 1
\ee
to ensure that $\gammaNR\geq0$. That said, using \eq{eq:tauNC} and \eq{eq:ftransGstar},
we can rewrite \eq{eq:NdelConstr} as
\be \label{eq:NdelRatConstr}
\frac{\tauNRhomeo}{\tauNChomeo}\leq \frac{\NRhomeo}{\Nhomeo }.
\ee

The apoptosis rate during the maturation phase, $\gammaNM\geq0$, is calculated
by eliminating $\kappaNhomeo\Qhomeo A_N^*$ from
\eq{eq:NMatBal} and \eq{eq:NMatTotANstar}. Also making use of \eq{eq:gammaNR}, we obtain
\be \label{eq:gammaNMeq}
F_M(\gammaNM):=\NRhomeo(e^{\gammaNM\aNM}-1)-\gammaNM\tauNRhomeo N_M^*=0.
\ee
It is easy to see that $F_M(0)=0$ and hence $\gammaNM=0$ is one solution of \eq{eq:gammaNMeq}.
Since $F_M''(\gamma)>0$ for all $\gamma\geq0$, if $F_M'(0)<0$
there is a unique $\gammaNM>0$ such that $F_M(\gammaNM)=0$,
and no positive value of $\gamma$ such that $F_M(\gamma)=0$ if $F_M'(0)\geq0$.
Since cell death is known to occur in the maturation compartment (see \cite{Mackey:03}),
we should choose our parameters  so that \eq{eq:gammaNMeq} admits a solution $\gammaNM>0$.
The condition
$F_M'(0)>0$ is equivalent to
\be \label{eq:MatApopRatCond}
\frac{\NRhomeo}{N_M^*} < \frac{\tauNRhomeo}{\aNM},
\ee
and to include apoptosis in the maturation compartment
our parameters must be chosen to satisfy \eq{eq:MatApopRatCond}.

Equation \eq{eq:NdelRatConstr} can be interpreted as a lower bound on
$\tauNRhomeo$, and \eq{eq:MatApopRatCond} as an upper bound.
Eliminating $\tauNRhomeo$ from these two bounds we find that
the parameters must satisfy
\be \label{eq:NdelConsist}
\frac{\aNM}{\tauNChomeo}<\frac{N_M^*}{\Nhomeo }
\ee
for the constraints \eq{eq:NdelRatConstr} and \eq{eq:MatApopRatCond}
to be consistent. Then $\tauNRhomeo$ must satisfy
\be \label{eq:tauNRstarConstr}
\tauNRhomeo \in \left( \aNM\frac{\NRhomeo}{N_M^*} , \tauNChomeo\frac{\NRhomeo}{\Nhomeo} \right)
\ee
for
both \eq{eq:NdelRatConstr} and \eq{eq:MatApopRatCond} to be satisfied as strict inequalities.
All the quantities in \eq{eq:tauNRstarConstr} have been estimated experimentally. To be consistent with
our model the values must satisfy both \eq{eq:NdelConsist} and \eq{eq:tauNRstarConstr}.
In Section~\ref{sec:ParVals} we state parameters that satisfy these constraints.
With those parameters we take $\gammaNM>0$ to be the unique strictly positive solution to \eq{eq:gammaNMeq}.


Equation \eq{eq:NMatBal} ensures that the rate cells enter and leave the reservoir are
equal at homeostasis. Rearranging and using \eq{eq:ftransGstar}
we obtain
\be \label{eq:ANstar2}
A_N^*=\frac{\NRhomeo}{\kappaNhomeo\Qhomeo\tauNRhomeo},
\ee
which determines $A_N^*$. Now from \eq{eq:ANstar} we have
\be \label{eq:expetatau}
e^{\etaNPhomeo\tauNP}=A_N^*e^{\gammaNM\aNM},
\ee
which determines $e^{\etaNPhomeo\tauNP}$, and it remains to determine one of
$\etaNPhomeo$ or $\tauNP$ in order to be able to find the other. However \eq{eq:NProlifTot}
implies that
\be \label{eq:etaNPval}
\etaNPhomeo
=\kappaNhomeo\Qhomeo \frac{e^{\etaNPhomeo\tauNP}-1}{N_P^*}=\kappaNhomeo\Qhomeo \frac{A_N^*e^{\gammaNM\aNM}-1}{N_P^*}
\ee
and now from \eq{eq:expetatau} we have
\be \label{eq:tauNP2}
\tauNP=\frac{1}{\etaNPhomeo}\ln\bigl(A_N^*e^{\gammaNM\aNM}\bigr).
\ee
In Section~\ref{sec:ParVals} we use the equations of this section to determine parameter values for our model.

\subsection{Estimation of G-CSF Pharmacokinetic Parameters}
\label{sec:GCSFParameters}

Following \cite{Watari1989,Kawakami1990,Barreda2004,Krzyzanski2010} we take the homeostasis
concentration of the free circulating G-CSF to be
$\Gonehomeo=0.025\unit{ng/mL}$.
The parameter $V$ in \eq{eq:BoundGCSF} is the same parameter $V$ as appears in \eq{eq:gcsf1eq}.
But $V$ is difficult to interpret directly from \eq{eq:gcsf1eq}, and although
published values are available, they vary widely
between sources. For the pharmacokinetic G-CSF model
\eq{eq:FreeGCSF},\eq{eq:BoundGCSF} the meaning of $V$ is clear; its simply the conversion factor
that converts a neutrophil concentration $N$ in units of $10^9$ cells per kilogram
of body mass, into the corresponding G-CSF concentration $VN$ in units of nanograms per millilitre
when every receptor on the neutrophils is bound.

To compute $V$, we first note that
the molecular mass of G-CSF is $18.8\unit{kDa}=18800\unit{g/mol}$ \cite{Krzyzanski2010}
or dividing by Avogadro's constant, the equivalent weight of G-CSF
is $G_{mw}=3.12\times 10^{-11}\unit{ng/molecule}$.
We take the number of receptors per neutrophil to be $R=600$, which is in the middle of the range that
Barreda \cite{Barreda2004} cites, though we note that both smaller and larger numbers can be found in the literature.
Then given $N$, the number of receptors per millilitre is
$$R\times\frac{70}{5000}\times10^9\times N,$$
where we assume body mass of $70\unit{kg}$ and $5000\unit{mL}$ of blood. Since two molecules
bind to each receptor the maximum concentration of bound G-CSF is
$$VN=2\times G_{mw}\times R\times \frac{70}{5000}\times10^9\times N
=0.525 N \unit{ng/mL}$$
and hence
\be \label{eq:Vval}
V=0.525 \unit{(ng/mL)/(10^9 cells/kg)}.
\ee

Values have been published for several of the other parameters in the G-CSF equations
\eq{eq:FreeGCSF},\eq{eq:BoundGCSF}, but these have been largely based on \textit{in vitro} experiments and/or
simpler G-CSF models using mixed-effects estimation techniques, and so are
not directly applicable to our model \cite{Krzyzanski2010,Wang2001,Scholz2012,Sarkar2003}.

At homeostasis, equations \eq{eq:FreeGCSF},\eq{eq:BoundGCSF} give
\be \label{eq:Gtwohomeo}
\Gtwohomeo=\frac{(\Gonehomeo)^\Pow}{(\Gonehomeo)^\Pow+(k_{int}+k_{21})/k_{12}}V \Ntothomeo,
\ee
and
\begin{align} \notag
\Gprod  & = k_{ren}\Gonehomeo+k_{int}\Gtwohomeo \\
& = k_{ren}\Gonehomeo+k_{int}V \Ntothomeo\frac{(\Gonehomeo)^\Pow}{(\Gonehomeo)^\Pow+(k_{int}+k_{21})/k_{12}}.
\label{eq:Gprod}
\end{align}

Once values of $k_{int}$, $k_{12}$, $k_{21}$,
$k_{ren}$ and $\Pow$ are determined as we describe below, \eq{eq:Gtwohomeo} and \eq{eq:Gprod}
determine values for $\Gtwohomeo$ and $\Gprod$. 


The remaining parameters might be determined by simulating the full model with exogenous G-CSF administration
and fitting the response of the model to published data for such experiments. However, that would involve
also fitting the as yet undetermined pharmacodynamic parameters in equations \eq{eq:HSCs}--\eq{eq:nu}
which would create a very large optimisation problem, with the potential for interactions between the
pharmacokinetic and pharmacodynamic parameters to create a complicated functional with many local minima.
To avoid this, we prefer to determine the pharmacokinetic and pharmacodynamic parameters separately. Here we
determine the PK parameters by decoupling the G-CSF equations \eq{eq:FreeGCSF}-\eq{eq:BoundGCSF}
from the neutrophil dynamics.

There have been a number of studies tracking the response of the hematopoietic system to a single administration of exogenous G-CSF including Wang~\cite{Wang2001} and Krzyzanski~\cite{Krzyzanski2010}.
If data were available for circulating neutrophil and marrow reservoir neutrophil concentrations
as functions of time
it would be possible to treat equations \eq{eq:FreeGCSF}-\eq{eq:BoundGCSF} separately from the rest of the model
as a system of two ordinary differential equations with $\Ntott$ treated as a known non-autonomous forcing term determined by the data. But unfortunately it is not known how to directly measure either
marrow neutrophil reservoir or bound G-CSF concentrations, and such values are not reported
in the literature.
%
%

In the absence of marrow neutrophil data we will decouple the G-CSF kinetic equations
\eq{eq:FreeGCSF}-\eq{eq:BoundGCSF} from the rest of the model by replacing the time dependent term
$\Ntott$ by the constant $\Ntot$ to obtain
\begin{align}\notag
\tfrac{d}{dt}G_{1\hspace{-0.1em}}(t) & = I_G(t) 
+\Gprod-k_{ren}G_{1\hspace{-0.1em}}(t)\\ \label{eq:UnboundGCSFsimp}
& \mbox{}\qquad\quad-k_{12}(\Ntot V-G_{2\hspace{-0.05em}}(t))G_{1\hspace{-0.1em}}(t)^{\Pow}
+k_{21}G_{2\hspace{-0.05em}}(t) \\ \label{eq:BoundGCSFsimp}
\tfrac{d}{dt}G_{2\hspace{-0.05em}}(t) & = -k_{int}G_{2\hspace{-0.05em}}(t)+k_{12}\bigl(\Ntot V\hspace{-0.4em}-G_{2\hspace{-0.05em}}(t)\hspace{-0.1em}\bigr)G_{1\hspace{-0.1em}}(t)^{\Pow}-k_{21}G_{2\hspace{-0.05em}}(t).
\end{align}
In \eq{eq:UnboundGCSFsimp} and \eq{eq:BoundGCSFsimp} the constant
$\Ntot$ represents the constant total number of neutrophils available for G-CSF binding,
and will be treated as an extra parameter to be determined during the fitting. It should
correspond approximately to an average value of $\Ntott$ across the time course of the data.

With data for bound G-CSF unavailable we are constrained to fit \eq{eq:UnboundGCSFsimp},\eq{eq:BoundGCSFsimp}
to data for the unbound G-CSF. To do this we use digitisations of
two sets of data from Wang~\cite{Wang2001} from a 750\unit{\mu g}
intravenous (IV) administration of G-CSF and from a subcutaneous (SC) administration of the same dose.
SC administrations necessarily include the absorption kinetics of a drug, as outlined in equation \eqref{eq:SCadmin},
whereas IV administrations reach the blood directly and can be modelled more simply as in \eqref{eq:IVadmin}.
For these reasons, both IV and SC data were used simultaneously during the fitting procedure to best characterise the 
parameters. Rather than fitting directly to the data from Wang~\cite{Wang2001}, to obtain robust parameter fits
we took the G-CSF data from the SC and IV administrations and fit a spline through each to define functions
$G_{dat}^{SC}(t)$ and $G_{dat}^{IV}(t)$ over the time intervals $0\leq t\leq 2\unit{days}$ for which the data were taken. With postulated parameter values we then use the Matlab \cite{Mathworks} ordinary differential equation solver
\textit{ode45} to simulate \eq{eq:UnboundGCSFsimp},\eq{eq:BoundGCSFsimp} over the same time interval to define
functions $G_{1}^{SC}(t)$ and $G_{1}^{IV}(t)$. We measure the error between the simulated solutions
and the data using the $L^2$ function norm defined by
\be \label{eq:L2}
\|G\|_2^2=\int_0^T \hspace{-0.5em}G(t)^2dt.
\ee
For the IV data which varies over orders of magnitude, as seen in Figure~\ref{fig:gcsffails}, we use a log scale, while for the SC data a linear scale is appropriate. We define a combined error function for both simulations by
\be \label{eq:PKErr}
\textit{Err} = \|\log(G_1^{IV})-\log(G_{dat}^{IV})\|_2^2 + \chi^{0.95}\|G_1^{SC}-G_{dat}^{SC}\|_2^2,
\ee
where the scale factor $\chi$ defined by
\be \label{eq:chi}
\chi=\frac{\max_{t\in[0,T]}\log(G_{dat}^{IV}(t))-\min_{t\in[0,T]}\log(G_{dat}^{IV}(t))}{\max_{t\in[0,T]}G_{dat}^{SC}(t)-\min_{t\in[0,T]}G_{dat}^{SC}(t)},
\ee
effectively rescales the data so that both data sets have equal weight. (Since $\chi<1$ the power $0.95$ in \eq{eq:PKErr} works to give slightly more weight to the SC data).

Fitting was performed using the Matlab \cite{Mathworks} \textit{lsqcurvefit}
least squares solver, with the error function
$\textit{Err}$ evaluated numerically by sampling the functions at a thousand equally spaced points.
It is convenient to define the constant
\begin{equation} \label{eq:Nelim}
N_{elim}=1-\frac{k_{ren}G_1^*}{G_{prod}}
\end{equation}
where $N_{elim}$ is the fraction of G-CSF clearance performed through internalisation at homeostasis
(obtained in \eq{eq:Nelim} as one minus the fraction of renal clearance at homeostasis).
The estimation was performed for the G-CSF parameters: $k_{12}$, $k_{21}$, $\Pow$, $k_{int}$,
the neutrophil constant $N_{elim}$,
and the pharmacokinetic drug parameters $k_a$, and $F$. The elimination fraction $N_{elim}$ was
either fixed ($N_{elim}=0.6$ and $0.8$ in Table~\ref{tab:PKParams}) or fitted (the other
entries in Table~\ref{tab:PKParams}). At each step of the optimisation the candidate
$k_{12}$, $k_{21}$, $\Pow$, $k_{int}$ and $N_{elim}$ are used to determine the dependent parameters
$G_2^*$, $k_{ren}$, and $G_{prod}$, which from \eq{eq:UnboundGCSFsimp},\eq{eq:BoundGCSFsimp} and
\eq{eq:Nelim} are given by
\begin{align}
\label{eq:G2Calculation}
G_2^*&=VN_{tot}\frac{(G_1^*)^\Pow}{(G_1^*)^\Pow+(k_{21}+k_{int})/k_{12}}\\
\label{eq:krenCalculation}
k_{ren}&=\left(-1+\frac{1}{N_{elim}}\right)Vk_{int}(G_1^*)^{(\Pow-1)}\frac{N_{tot}}{(G_1^*)^\Pow+(k_{21}+k_{int})/k_{12}}\\
\label{eq:GprodCalculation}
G_{prod}&=k_{int}G_2^*+k_{ren}G_1^*.
\end{align}

\begin{table}[ht]
\begin{center}
\begin{tabular}{|c|c|c|c|c|c|c|}\hline
Name & Value 1 & Value 2 & Value 3 & Value 4 & Value 5 &Units  \\ \hline
$N_{elim}$ & 0.097478 & 0.6 & 0.71678 & 0.8 & 0.87358  & $-$ \\ \hline
$k_{ren}$ & 1.3142  & 0.45064 & 0.2456 & 0.16139 & 0.094597 & days$^{-1}$\\
$k_{12}$ & 2.3004  & 2.2519 & 2.1342 & 2.2423 & 2.878 & days$^{-1}$\\
$k_{21}$ & 407.1641  & 198.2403 & 168.2588 & 184.8658 & 259.8087 & days$^{-1}$\\
$k_{int}$ & 394.5111  & 459.2721 & 275.2744 & 462.4209 & 632.0636 & days$^{-1}$\\
 $\Pow$ & 1.7355 & 1.4418 & 1.4631 & 1.4608 & 1.4815 & $-$ \\
 $\Ntot$ & 3.9496  & 4.1767 & 4.1457 &4.2009 & 3.606 & $10^9$ cells/kg\\ \hline
 \multicolumn{7}{|c|}{$Do=750\unit{\mu g}$, $V_d=2178.0$ \unit{mL}} \\ \hline
 $F$ & 0.99752& 0.75 & 0.75 & 0.75 & 0.98271 & $-$\\
 $k_a$ & 3.8154 & 5.2142 & 5.0574 & 5.143 & 4.1931 & days$^{-1}$\\
 $Err$ & 0.16352  & 0.15716 & 0.17901 & 0.18543 & 0.21130 & $-$ \\ \hline
 \multicolumn{7}{|c|}{$Do=300\unit{\mu g}$, $V_d=4754.7$ \unit{mL}} \\ \hline
 $F$ & 1& 0.63361 & 0.62299 & 0.64466 & 0.71424 & $-$\\
 $k_a$ & 6.3783 & 8.0804 & 8.0628 & 8.0236 & 7.4367 & days$^{-1}$\\ \hline
 \multicolumn{7}{|c|}{$Do=375\unit{\mu g}$, $V_d= 2322.9$ \unit{mL}} \\ \hline
 $F$ & 0.89831& 0.4801 & 0.48549 & 0.49964 & 0.57618 & $-$\\
 $k_a$ & 4.18161 & 6.7326 & 6.6324 & 6.6133 & 6.1259 & days$^{-1}$\\
\hline
\end{tabular}
\end{center}
\caption{Pharmacokinetic parameter estimates from the simplified
G-CSF model \eq{eq:UnboundGCSFsimp},\eq{eq:BoundGCSFsimp} for different homeostasis
elimination fractions of $N_{elim}$. $Err$ is defined by \eq{eq:PKErr} for the $750\unit{\mu g}$ dose.
As described in the text, dose-dependent drug parameters were only recalculated for the lower doses.}
\label{tab:PKParams}
\end{table}%

\begin{figure}[ht]
\begin{subfigure}[b]{0.5\linewidth}
    \begin{center}
    \includegraphics[width=0.9\linewidth]{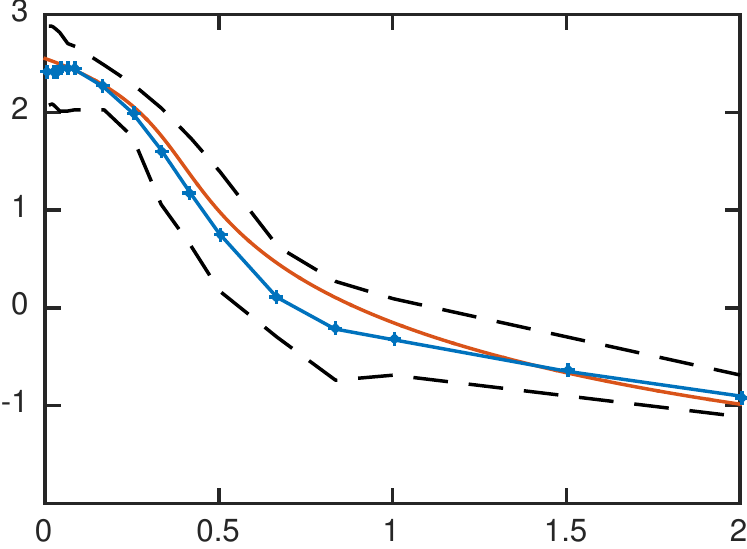}
     \put(-159,88){\rotatebox[origin=c]{90}{\scriptsize$\log(G_1(t))$}}
     \put(-20,-6){\scriptsize Days}
     \end{center}
    \caption{25 minute IV infusion}
    \vspace{4ex}
  \end{subfigure}
  \begin{subfigure}[b]{0.5\linewidth}
    \begin{center}
    \includegraphics[width=0.9\linewidth]{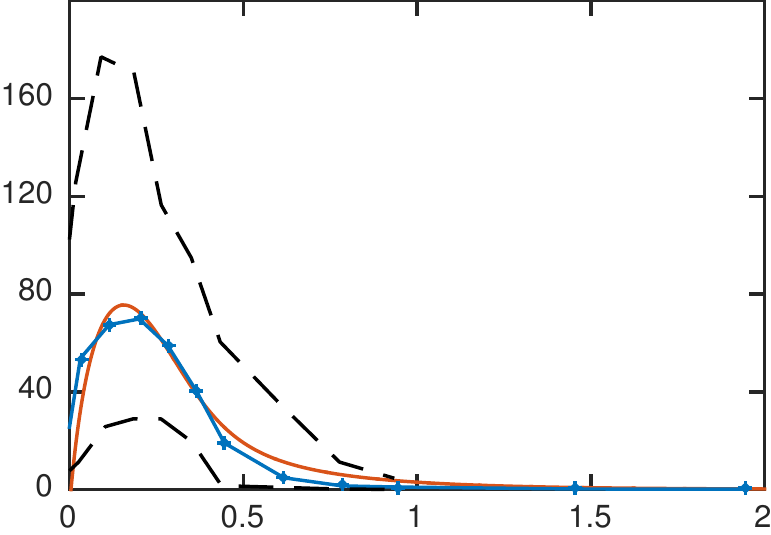}
          \put(-159,95){\rotatebox[origin=c]{90}{\scriptsize$G_1(t)$}}
     \put(-20,-6){\scriptsize Days}
     \end{center}
    \caption{Subcutaneous injection}
    \vspace{4ex}
  \end{subfigure}
\caption{G-CSF PK parameter fitting results of \eq{eq:UnboundGCSFsimp},\eq{eq:BoundGCSFsimp} with parameter values taken from Table~\ref{tab:PKParams} with $N_{elim}=0.80$. In both panels, a 750 $\mu$g dose is administered following the protocol described in Wang \cite{Wang2001}. Blue lines with data: digitised data median values, red solid lines: model solution with estimated parameters, black dashed lines: maximum and minimum values of the digitised data.}
 \label{fig:GoodPKFits}
\end{figure}

\begin{figure}[ht]
  \begin{subfigure}[b]{0.5\linewidth}
    \begin{center}
    \includegraphics[width=0.9\linewidth]{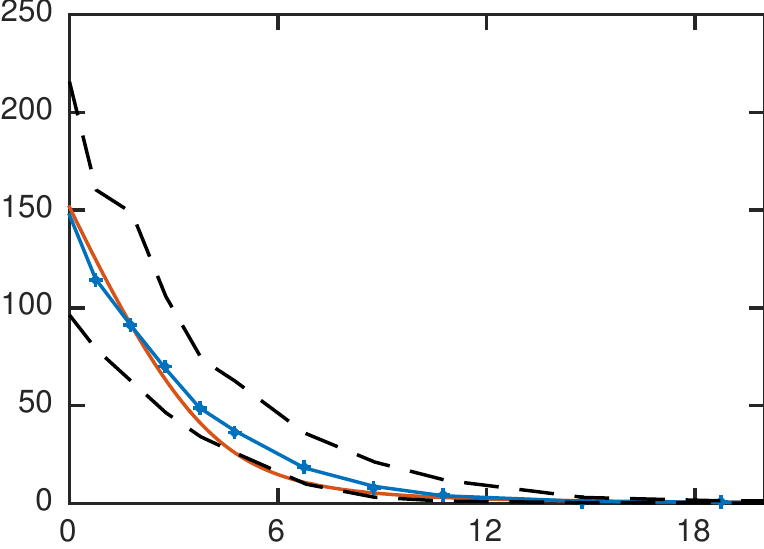}
      \put(-161,95){\rotatebox[origin=c]{90}{\scriptsize$G_1(t)$}}
     \put(-22,-6){\scriptsize Hours}
     \end{center}
    \caption{$5\unit{\mu g/kg}$ ($350\mug$ total) IV infusion}
    \vspace{4ex}
  \end{subfigure}
  \begin{subfigure}[b]{0.5\linewidth}
    \begin{center}
    \includegraphics[width=0.9\linewidth]{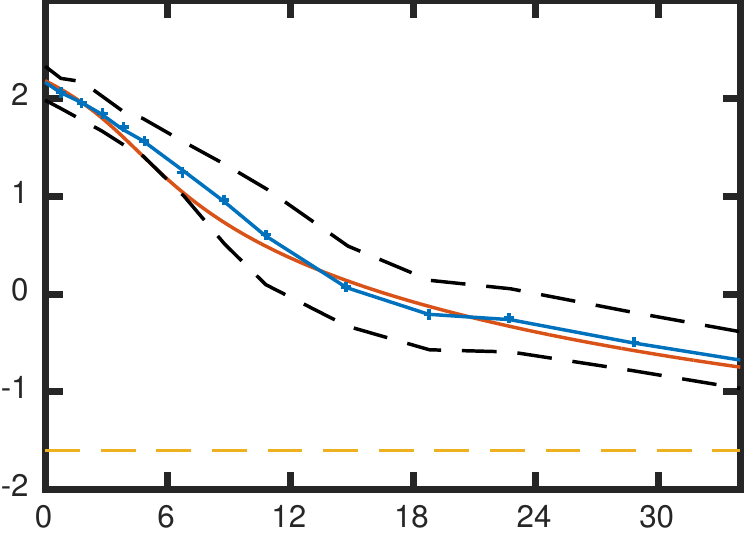}
         \put(-159,87){\rotatebox[origin=c]{90}{\scriptsize$\log(G_1(t))$}}
     \put(-22,-6){\scriptsize Hours}
     \put(-100,20){\scriptsize$\log(G_1(t))$}
     \end{center}
    \caption{$5\unit{\mu g/kg}$ ($350\mug$ total) IV infusion}
    \vspace{4ex}
  \end{subfigure}
  \begin{subfigure}[b]{0.5\linewidth}
    \begin{center}
    \includegraphics[width=0.9\linewidth]{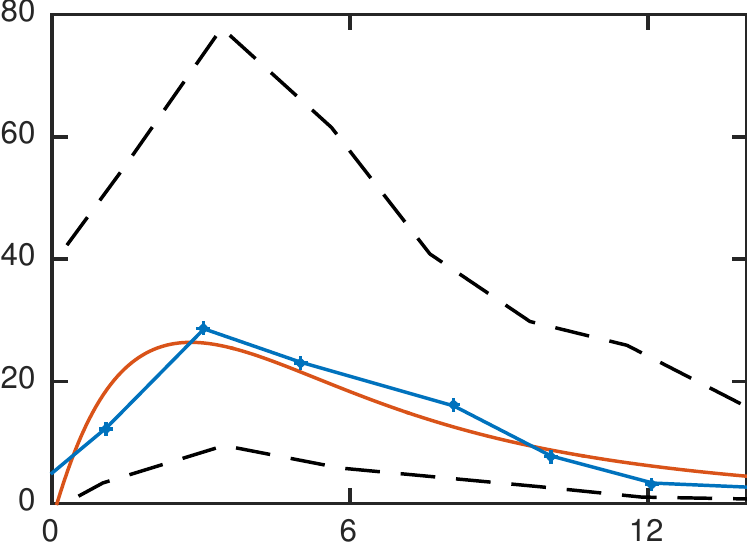}
          \put(-161,95){\rotatebox[origin=c]{90}{\scriptsize $G_1(t)$}}
     \put(-22,-6){\scriptsize Hours}
     \end{center}
    \caption{$375\mug$ subcutaneous administration}
  \end{subfigure}
  \begin{subfigure}[b]{0.5\linewidth}
    \begin{center}
    \includegraphics[width=0.9\linewidth]{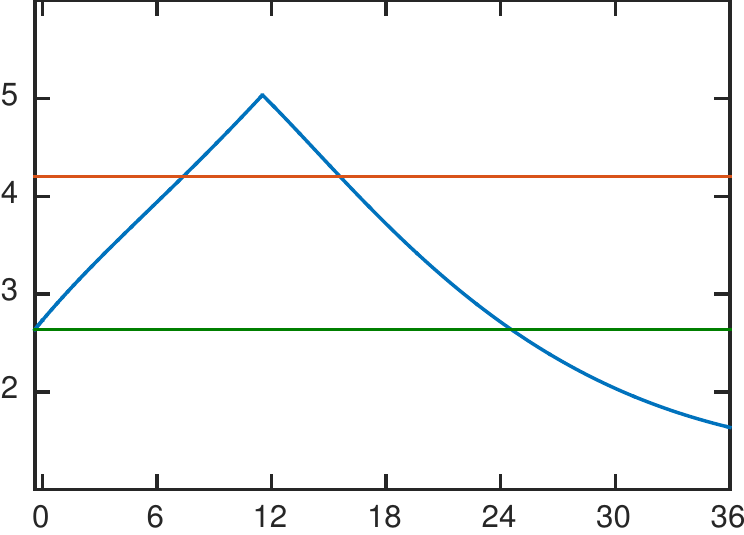}
          \put(-161,82){\rotatebox[origin=c]{90}{\scriptsize$N(t)+N_R(t) $}}
     \put(-42,77){\scriptsize$\Ntot$}
     \put(-112,46){\scriptsize$\NRhomeo+\Nhomeo$}
     \put(-22,-6){\scriptsize Hours}
     \end{center}
    \caption{$750\mug$ IV infusion}
  \end{subfigure}
   \caption{G-CSF pharmacokinetic parameter fitting results of \eq{eq:UnboundGCSFsimp},\eq{eq:BoundGCSFsimp} with parameter values taken from Table~\ref{tab:PKParams} with $N_{elim}=0.80$ compared for different administration types, doses, and datasets.
   a) \& b) A simulation of \eq{eq:UnboundGCSFsimp},\eq{eq:BoundGCSFsimp}
   is compared to data from \cite{Krzyzanski2010} in linear and log scales.
   c) A simulation compared to data from \cite{Wang2001}. d) Neutrophil concentrations (blue line) of the full
   neutrophil model \eq{eq:HSCs}-\eq{eq:nu} compared to the value of $N_{tot}$ 
   and $N_R^*+N^*$. 
   For a to c: blue line with data: digitised data median values, red solid line: model solution from estimated parameters, black dashed lines--digitised data maximum and minimum values. 
   }
  \label{fig:PKParameters}
\end{figure}

The following fitting procedure was employed. First parameters were fit from IV data for a 750\unit{\mu g} administration \cite{Wang2001} on the log scale to ensure that behaviour at both high and low concentrations were properly characterised. Next initial SC parameters were fit from 750\unit{\mu g} SC data in linear scale.
Using the parameters from these two fits as seed values, we next obtain final parameter values by fitting both log-concentration IV and linear SC data simultaneously using the norm defined in \eq{eq:PKErr}.
Finally, as the pharmacokinetic parameters related to the SC administration have been shown to be dose-dependent \cite{Scholz2012}, we re-estimate $F$ and $k_a$ for lower doses of $300\unit{\mu g}$
and $375\unit{\mu g}$ \cite{Krzyzanski2010,Wang2001}. Since $V_d$ is typically calculated by the ratio of the dose to the initial concentration in the blood for IV administrations \cite{DiPiro2010}, we have applied the same calculation here to scale the G-CSF prediction to the first measured data point. Accordingly, the volume of distribution was recalculated to fit the administered dose.
The resulting parameters are reported in Table~\ref{tab:PKParams}.

Figure~\ref{fig:GoodPKFits} compares the solutions from the fitting procedure of the simplified model \eq{eq:UnboundGCSFsimp} \eq{eq:BoundGCSFsimp}
for the parameter set with $N_{elim}=0.80$ from Table~\ref{tab:PKParams} to the Wang \cite{Wang2001} data
for $750\unit{\mu g}$ IV and SC doses in log and linear scales, respectively.

Figure~\ref{fig:PKParameters}(a-b) gives linear and log scale plots of the simulations of \eq{eq:UnboundGCSFsimp},\eq{eq:BoundGCSFsimp} with the $N_{elim}=0.80$ parameter set from Table~\ref{tab:PKParams}
for an IV administration
from Krzyzanski \cite{Krzyzanski2010}. In this case no fitting was performed; the Krzyzanski \cite{Krzyzanski2010}
protocol is simulated using parameters obtained from fitting to the Wang data, and a good fit to the data
is still obtained. Figure~\ref{fig:PKParameters}(c) shows another simulation for a slightly larger SC dose, with the
same G-CSF parameters (only the the dose-dependent drug parameters $k_a$ and $F$ were fit, as already noted), and we
again obtain good agreement with the data.

Figure~\ref{fig:PKParameters}(d) validates the use of the $N_{tot}$ simplification used for \eq{eq:UnboundGCSFsimp},\eq{eq:BoundGCSFsimp} by comparing $N_{tot}$ to $N_R(t)+N(t)$ from the solution of the full model \eq{eq:HSCs}-\eq{eq:nu} and to $\NRhomeo+\Nhomeo$. This demonstrates how $N_{tot}$ effectively averages $N_R(t)+N(t)$ over most of the simulation.


We characterize the parameter sets found for the simplified
G-CSF model \eq{eq:UnboundGCSFsimp},\eq{eq:BoundGCSFsimp}
by the fraction $N_{elim}$ of the G-CSF that is cleared by binding and internalisation at homeostasis.
For $0\leq N_{elim}<1/2$ the elimination is renal dominated at homeostasis, while for
$1/2< N_{elim}\leq1$ the pharmacokinetics are internalisation dominant. As already mentioned in Section~\ref{sec:GCSFModel},
from a clinical standpoint, it is believed that $N_{elim}>1/2$, while a number of previously published models including
\cite{Craig2015,Krzyzanski2010,Wang2001} have $N_{elim}$ close to zero.

When we included $N_{elim}$ as a parameter to be fit the results were very sensitive to the seed values
used to start the optimisation and had a tendency to produce parameter sets with very low or very high $N_{elim}$
(see the $N_{elim}=0.097$ and $N_{elim}=0.87358$ parameter sets in Table~\ref{tab:PKParams}), but we also found
a good fit with $N_{elim}=0.71678$ and were able to find good fits for any fixed value of $N_{elim}$, as seen in Figure~\ref{fig:GoodPKFits}
(see Table~\ref{tab:PKParams} for parameter
sets with $N_{elim}=0.6$ and $0.8$). Our results seem to indicate that there is at least a one parameter family of plausible parameter sets with each set characterised by the  value of $N_{elim}$. 
This arises
because we are fitting the simplified model \eq{eq:UnboundGCSFsimp},\eq{eq:BoundGCSFsimp} without any data for
the bound G-CSF concentrations. If the model \eq{eq:UnboundGCSFsimp},\eq{eq:BoundGCSFsimp} were linear then
parameter identifiability theory would require data from both components of the solution in order to identify unique
parameters in the model. Even though \eq{eq:UnboundGCSFsimp},\eq{eq:BoundGCSFsimp} is nonlinear, the lack of
any bound G-CSF data allows us to fit the unbound G-CSF concentrations with different parameter sets, which will
result in different solutions for the unmeasured bound G-CSF concentrations.  In Section~\ref{sec:NeutFit} we will
see that different G-CSF kinetic parameter sets will result in
similar G-CSF responses, but markedly different neutrophil dynamics. The small differences in the reported
errors $Err$ in Table~\ref{tab:PKParams} are not sufficient alone to make a definitive judgement of
which is the optimal parameter set. In the following sections we will study the response
of the full system \eq{eq:HSCs}-\eq{eq:nu} not just to exogenous G-CSF but also chemotherapeutic treatment (both alone and with prophylactic exogenous G-CSF) which will lead us to conclude that the PK parameters from Table~\ref{tab:PKParams} with
$N_{elim}=0.80$ produce the best model responses to a variety of scenarios.

As seen in Table~\ref{tab:PKParams}, the estimates obtained for $N_{tot}$ are significantly larger than $\Ntothomeo$.
However as Figure~\ref{fig:PKParameters}(d) shows for a 750\unit{\mu g} dose administered by a 25 minute IV infusion, $N_{tot}$ is an approximate average
for $\Ntott$ over the initial part of the simulation. This, along with the similarity between the results
given by \eq{eq:FreeGCSF}-\eq{eq:BoundGCSF} and the full model (as illustrated in Figure~\ref{fig:gcsffails})
gives us confidence not only in the simplified model \eq{eq:UnboundGCSFsimp}-\eq{eq:BoundGCSFsimp} for estimating the
G-CSF kinetic parameters, but also provides additional confirmation that the marrow reservoir
neutrophils $N_R(t)$ must be included along with the total blood neutrophil pool $N(t)$ in the full
kinetic G-CSF model \eq{eq:FreeGCSF}-\eq{eq:BoundGCSF} in order to reproduce the observed physiological
response.


\subsection{Parameter estimates from G-CSF knockout}
\label{sec:KO}

Several murine studies \cite{Bugl2012,Lui2013}
have looked at the effects of G-CSF knockout by producing mice lacking G-CSF receptors and measuring the differences in circulating neutrophil counts compared to wild type mice. The conclusion of these studies is that even in the case of
complete incapacity of the neutrophils to bind with G-CSF, neutrophil counts were still between 20 and 30\% of normal levels.
This is consistent with G-CSF not being the sole cytokine to regulate
neutrophil production.
Consequently we will ensure that our model produces reduced but non-zero circulating neutrophil concentrations
in the complete absence of G-CSF, and so in this section we consider the behaviour of the equations
defining neutrophil production when $G_1\equiv G_2\equiv0$. In that case the four G-CSF dependent functions take
values $\kappaN(0)=\kappaN^\textit{min}$, $\etaNP(0)=\etaNP^\textit{min}$, $\VN(0)\in(0,1)$ (by \eq{eq:Constraint3}), and
$\ftrans(0)\in(0,\ftranshomeo)$ (by \eq{eq:Constraint1}).

We let $\Nkohomeo$ denote the total blood neutrophil pool under G-CSF knockout and define the ratio
\be \label{eq:Cko}
C_{ko}=\Nkohomeo/\Nhomeo.
\ee
Let $\theta=R_{Pko}/R_P^*$ be
the ratio of the rate of cells leaving proliferation in the absence of G-CSF to the rate of cells leaving proliferation at homeostasis.
Using \eq{eq:RPstar} and a similar calculation for $R_{Pko}$ we obtain
\begin{equation}
\label{eq:theta}
\theta=\frac{R_{Pko}}{R_P^*}=\frac{\kappaN^\textit{min}\Qhomeo e^{\tauNP\etaNP^\textit{min}}}{\kappaNhomeo\Qhomeo e^{\tauNP\etaNPhomeo}}
=\frac{\kappaN^\textit{min}}{\kappaNhomeo}e^{\tauNP\etaNPhomeo(\mu-1)},
\end{equation}
where we also introduce the second auxiliary parameter
\begin{equation} \label{eq:mu}
\mu={\etaNP^\textit{min}}/{\etaNPhomeo}\leq 1,
\end{equation}
which measures the fractional reduction in the proliferation rate at knockout. In \eq{eq:theta} we have assumed that
the number of stem cells is unchanged at knockout. Since the differentiation rate to neutrophils will be decreased
from $\kappaNhomeo$ to $\kappaN^\textit{min}$ in the absence of G-CSF, the number of stem cells will actually increase, but
using \eq{eq:Ksum} and \eq{eq:betaQ} this increase can be calculated and is found to be less than $1\%$ for our
model parameters. 


For given values of $\theta$, $\mu$ and $e^{\tauNP\etaNPhomeo}$ we will use \eq{eq:theta} to determine
the ratio $\kappaN^\textit{min}/\kappaNhomeo$.
Since $\kappaN^\textit{min}\leq\kappaNhomeo$ (see \eq{eq:KKK}), \eq{eq:theta} implies that
$\theta\leq e^{\tauNP\etaNPhomeo(\mu-1)}$. Rearranging this gives a lower bound for $\mu$, from which obtain
the constraint
\be \label{eq:muinterval}
\mu \in \left(1+\frac{\ln(\theta)}{\tauNP\etaNPhomeo}, 1\right).
\ee
Here $\mu=1$ corresponds to a constant
proliferation rate independent of G-CSF, with the reduced production of neutrophils at
knockout caused by a reduction of the differentiation rate $\kappaN$. If $\mu$ is equal to its
lower bound then $\kappaN$ is constant independent of G-CSF concentration, and the reduced production
of neutrophils is caused by the reduced effective proliferation rate $\etaNP$.
For intermediate values of $\mu$, both $\kappaN^\textit{min}$ and $\etaNP^\textit{min}$ are reduced from their homeostasis values,
and $\mu$ acts as a tuning parameter to
weight the relative contribution of each mechanism with $\kappaN^\textit{min}/\kappaNhomeo$ a monotonically
decreasing function of $\mu={\etaNP^\textit{min}}/{\etaNPhomeo}$. 


A value for $\theta$ can be computed by studying the dynamics in the absence of G-CSF after the proliferation stage.
Letting $\Nkohomeo$ and $\NRkohomeo$ denote the number of neutrophils at knockout in the total blood pool and
in the marrow reservoir respectively, the rate that cells enter and leave circulation should be equal
implying that $\gamma_N \Nkohomeo=\ftrans(0)\NRkohomeo$, or
\be \label{eq:ReservoirKO}
\NRkohomeo =\frac{1}{\ftrans(0)}\gamma_N \Nkohomeo.
\ee
The rate $R_{Mko}$ that mature neutrophils are created at knockout is then equal to the rate that
neutrophils enter and leave the marrow reservoir, and hence
\be \label{eq:KOCons2}
R_{Mko}=(\ftrans(0)+\gammaNR)\NRkohomeo  = \gamma_N \Nkohomeo \left (1 + \frac{\gammaNR}{\ftrans(0)}\right).
\ee
During G-CSF knockout, the maturation time is given by $\aNM/\VN(0)$, during which cells die at a constant rate
$\gammaNM$ (which is not affected by G-CSF). Hence the rate $R_{Pko}$ that cells exit proliferation in the absence of G-CSF is related to $R_{Mko}$ by
$$R_{Pko}e^{-\gammaNM\frac{\aNM}{\VN(0)}}=R_{Mko}.$$
Thus,
\be \label{eq:KOCons5}
R_{Pko}=e^{\gammaNM\frac{\aNM}{\VN(0)}}R_{Mko}=
\gamma_N \Nkohomeo\left (1+\frac{\gammaNR}{\ftrans(0)}\right )e^{\gammaNM\frac{\aNM}{\VN(0)}}.
\ee
A similar calculation yields  $R_P^*$, the rate that cells leave proliferation at homeostasis (with G-CSF), as
\begin{equation}
\label{eq:ProlRateHomeo}
R_P^*=\gamma_N \Nhomeo \left (1+\frac{\gammaNR}{\ftranshomeo}\right)e^{\gammaNM \aNM}.
\end{equation}
Then
\begin{equation} \label{eq:thetaval}
\theta=\frac{R_{Pko}}{R_P^*}=
C_{ko}\frac{\ftrans(0)+\gammaNR}{\ftranshomeo+\gammaNR}\exp\Bigl[\aNM\gammaNM\Bigl(\frac{1}{\VN(0)}-1\Bigr)\Bigr],
\end{equation}
where $C_{ko}$ is defined by \eq{eq:Cko}.





\subsection{Estimating the Pharmacodynamic Parameters}
\label{sec:NeutFit}

We still require estimates for six parameters, $\mu$, $\bNP$, $V_{max}$, $b_V$, $b_G$ and $\ftrans^\textit{max}$
in the functions defining the pharmacodynamic effects of G-CSF on the neutrophil production and mobilisation.

We digitised data from Wang~\cite{Wang2001} for average circulating neutrophil concentrations for three days following a
$375\mug$ and a $750\mug$ 25-minute IV infusion. The data also contained circulating G-CSF concentrations, but we did
not use the G-CSF concentrations for fitting. As in Section~\ref{sec:GCSFParameters}, instead of fitting directly to the
data points we used it to to define two continuous functions $N_{dat}^{375}(t)$ and $N_{dat}^{750}(t)$, one for each dose,
and fit the response of the full model \eq{eq:HSCs}-\eq{eq:nu} to these functions.

The fitting is difficult because no data is available for reservoir or stem cell concentrations, and the
circulating neutrophil concentrations are only measured for
three days after the infusion. Since the proliferation time for neutrophil precursors is about a week, this data cannot
be used to fit any stem cell parameters, as no cells that commit to differentiate to the neutrophil line after the infusion will
reach circulation during this time (which is why we do not re-estimate any stem cell parameters in the current work).
Although at homeostasis it also takes about a week for cells to traverse maturation and the marrow reservoir, these
processes are greatly sped up after G-CSF administration, and cells that are in proliferation at the time of the infusion can
reach circulation within a day, enabling us to estimate relevant parameters.

After three days the
neutrophil concentrations have not returned to their homeostatic values. If parameters are fit just using this short
interval of data, we found parameters which gave good fits to the circulating neutrophil concentrations over the first three
days, but for which the neutrophil concentrations then under went very large deviations from homeostasis values lasting months or more.
There is no evidence of a single G-CSF administration destabilising granulopoiesis \cite{Molineux2001}. Accordingly, we will require that the fit
parameters result in stable dynamics. We do this by adding artificial data points for $7 \leq t\leq 21\unit{days}$.
Accordingly we construct $N_{dat}^{375}(t)$ and $N_{dat}^{750}(t)$ over two disjoint time intervals as splines through the data points
for $t\in[0,3]$    
and as  constant functions $N_{dat}^{dose}(t)=\Nhomeo$ for $t\in[7,21]$.  
Since we have no data for $t$ between $3$ and $7$ days describing how the neutrophils return to homeostasis, we do not define values
for $N_{dat}^{dose}(t)$ for this time interval.

For candidate parameter values, we then used Matlab's \cite{Mathworks} delay differential equation solver \textit{ddesd} to simulate \eqref{eq:HSCs}-\eqref{eq:nu} over the full 21-day time interval. This defined the functions $N^{375}(t)$ and $N^{750}(t)$, from which we were able to measure the error between the data and the simulated solutions using an
$L^2$ function norm similar to the one defined in \eqref{eq:L2}. 
For the disjoint time intervals, we have two integrals to perform, and rescale them to carry equal weight and hence define
\begin{equation} \label{eq:L2Neutrophils}
\|N\|^2_2=\frac13\int_{0}^3N(t)^2dt+\frac{1}{14}\int_{7}^{21}N(t)^2dt,
\end{equation}
with corresponding fitting error
\begin{equation} \label{eq:NeutrophilError}
Err=\|N_{dat}^{375}(t)-N^{375}(t)\|_2^2+\|N_{dat}^{750}(t)-N^{750}(t)\|_2^2.
\end{equation}

Parameter estimation was performed using the \textit{fmincon} function in Matlab \cite{Mathworks}. As in the G-CSF fitting described in Section~\ref{sec:GCSFParameters}, the error was evaluated by sampling the functions at one thousand points (with $500$ in each
time interval because of the scaling in \eq{eq:L2Neutrophils}).

Instead of directly fitting the six parameters specified at the start of this section, we let $\tilde{b}_V=b_V/V_{max}$ and
fit to the six
parameters $\mu$, $\bNP$, $V_{max}$, $\tilde{b}_V$, $\ftrans(0)$ and $\ftrans^\textit{ratio}$. This set of parameters is easier to
fit to because the constraints \eq{eq:Constraint1} and \eq{eq:Constraint3} then become simply $\ftrans(0)>0$ and $\tilde{b}_V>\Gonehomeo$, while the original constraints both involve more than one of the unkown parameters.
From \eq{eq:nu},\eq{eq:ftransratio} and \eq{eq:theta}, at each step of the optimisation the six fitting parameters
define the remaining parameters via
%
%
\begin{equation}\label{eq:PDParamsDdet}
\begin{split}
&\etaNP^\textit{min}=\mu \, \etaNPhomeo, \qquad
\ftrans^\textit{max}=\ftrans^\textit{ratio}\ftranshomeo, \qquad
b_V=\tilde{b}_V V_{max},\\
&\kappaN^\textit{min}=\theta\kappaNhomeo e^{(\tauNP\etaNPhomeo(1-\mu))},\qquad
b_G=G_{BF}^*\frac{\ftrans^\textit{max}-\ftrans(0)}{\ftranshomeo-\ftrans(0)}.
\end{split}
\end{equation}
where $\theta$ itself is calculated from \eqref{eq:thetaval}. The Hill coefficient of \eqref{eq:NewKappa} was set to be $s_1=1.5$, midway within its plausible range of values, as explained in Section~\ref{sec:ParVals}.

The estimation of $\mu$ requires some caution as its lower bound in \eqref{eq:muinterval} changes at each iteration of the optimisation as $\theta$ varies, and we see from \eq{eq:thetaval} that $\theta$ itself depends on three of the
parameters to which we are fitting.  Nonsensical results are obtained if the model is simulated with $\mu$ outside its bounds. Since the constraint is difficult to apply, to ensure that \eqref{eq:muinterval} is respected we use a penalty method.
Consequently, \eqref{eq:muinterval} is checked at each iteration of the optimisation
and if $\mu$ is outside of its bounds
$\mu$ is reset to the bound and after the simulation is computed $Err$ is multiplied by
the penalty factor
$e^{|\mu-\mu_{bound}|^{1/2}}$ which is larger than $1$ when $\mu\ne\mu_{bound}$. The error function thus
penalised cannot have a minimum with $\mu$ outside of its bounds, and so the
optimisation routine is forced to find values for $\mu$ within the permissible range.

A family of G-CSF kinetic parameter sets was reported in
Table~\ref{tab:PKParams} in Section~\ref{sec:GCSFParameters}.
Estimates for the pharmacodynamic parameters were performed for
every parameter set in Table~\ref{tab:PKParams}. 
The resulting pharmacodynamic parameters are reported in Table~\ref{tab:NeutrophilValues}.

\begin{table}[ht]
\setlength{\tabcolsep}{2pt}
\begin{center}
\begin{tabular}{|c|c|c|c|c|c|c|c|}\hline
Name & Value 1 & Value 2 & Value 3 & Value 4 & Value 5  & Units \\ \hline
$N_{elim}^{simp}$ &0.097478 & 0.6 & 0.71678 & 0.8 & 0.87358 & $-$  \\ \hline
$N_{elim}$ &0.3631 & 0.4508 & 0.6204 & 0.7033& 0.8153 & $-$  \\ \hline
$\mu$ & 0.96381 & 0.86303& 0.85482 & 0.84458 & 0.90768  & $-$ \\
$\bNP$ & 0.125 & 0.026182 & 0.025994 & 0.022868 & 0.024908 & \unit{ng/mL} \\
$V_{max}$ & 7.9932 & 7.9881 & 7.9697 & 7.867 & 7.994 & $-$\\
$\tilde{b}_V$ & 0.031250 &0.031251 & 0.031255 & 0.031283 & 0.031261 & \unit{ng/mL} \\
$\ftrans(0)$ & 0.072801 & 0.026753 & 0.023154 & 0.020056 & 0.049852  & \unit{days^{-1}} \\
$\ftrans^\textit{ratio}$ & 10.9606 & 11.7257 & 11.9442 & 11.3556 & 11.9706 &  $-$\\ \hline
$\etaNP^\textit{min}$ & 1.6045 & 1.4367 & 1.4231 & 1.406 & 1.5111 & \unit{days^{-1}} \\
$\ftrans^\textit{max}$ & 3.9897 & 4.2682 & 4.3478 & 4.1335 & 4.3574 & \unit{days^{-1}}\\
$b_V$ & 0.24979 & 0.24964 & 0.24909 & 0.24611 & 0.2499 & \unit{ng/mL} \\
$b_G$ & 6.3999$\times 10^{-5}$ & 0.0002107 & 0.00019058 & 0.00018924 &0.00018725& $-$   \\
$\theta$ & 0.45978 & 0.18895 & 0.17099 & 0.15096 & 0.32529 & $-$  \\
$\kappaN^\textit{min}$ &0.0052359 & 0.0073325 & 0.0073325 & 0.0073325 & 0.0073325 & days$^{-1}$ \\ \hline
 $Err$ & 0.3482 & 0.3153 & 0.2928 & 0.2843 &  0.3762 & $-$ \\ 
\hline
\end{tabular}
\end{center}
\caption{Parameter estimation results for the pharmacodynamic parameters. $N_{elim}^{simp}$ refers to $N_{elim}$ value of the
corresponding kinetic parameters for the simplified G-CSF model given in Table~\ref{tab:PKParams}. $N_{elim}$
is the corresponding value for the full model, then stated are the six fit parameters, followed by the dependent
parameters. The approximation error to the data is found by integrating \eq{eq:Neutrophils} as in \eq{eq:NeutrophilError} and comparing to Wang's data \cite{Wang2001} for a 375 $\mu$g and 750 $\mu$g IV infusion of 25 minutes. }
\label{tab:NeutrophilValues}
\end{table}%


Since $\Gtwohomeo$ in the full model \eq{eq:HSCs}-\eq{eq:nu} is given by \eq{eq:Gtwohomeo} which differs from the
value given by \eq{eq:G2Calculation} for the simplified model \eq{eq:UnboundGCSFsimp},\eq{eq:BoundGCSFsimp},
the values of $G_{prod}$ and $N_{elim}$ derived for the two models will also be different.
In Table~\ref{tab:NeutrophilValues} the values from Section~\ref{sec:GCSFParameters} for the simplified model
are referred to as $N_{elim}^{simp}$, and we also state the corresponding value of $N_{elim}$ for the
full model from \eq{eq:Nelim} using \eq{eq:Gprod}.


It is important to note that if $\mu$ were close to $1$ and far from its lower bound, then
$\kappaN^\textit{min}/\kappaN^*\ll1$, and the wide variation in possible differentiation rates
could have potentially destabilising effects on the stem cells. However, for most of the investigated parameter sets (except for $N_{elim}^{simp}=0.097478$)
with the added `stabilising' data, $\mu$ was found to be essentially equal to its lower bound.
In this case $\kappaN^{min}$ is almost equal to $\kappaN^*$, and  the rate of differentiation out of the stem cell 
compartment is essentially constant and \eq{eq:NewKappa} is virtually independent of the influence of G-CSF. For the current model with the imposed stabilising data, this implies that any change in production is produced by variations in the effective proliferation rate of \eq{eq:etaNP}. Without the additional data points, we found parameter estimates where $\mu$ was far from its lower bound and $\kappaN^{min}$ was similarly lower than $\kappaN^*$ but these led to unstable dynamics. As seen in Sections~\ref{sec:ChemoEstimation} and \ref{sec:ModelEvaluation}, the parameter estimates obtained are shown to successfully reproduce protocols for chemotherapy-alone and chemotherapy with adjuvant G-CSF. Accordingly, differentiation from the hematopoietic stem cells is likely close to constant in reality but from our results, we cannot conclude that differentiation is independent of G-CSF.

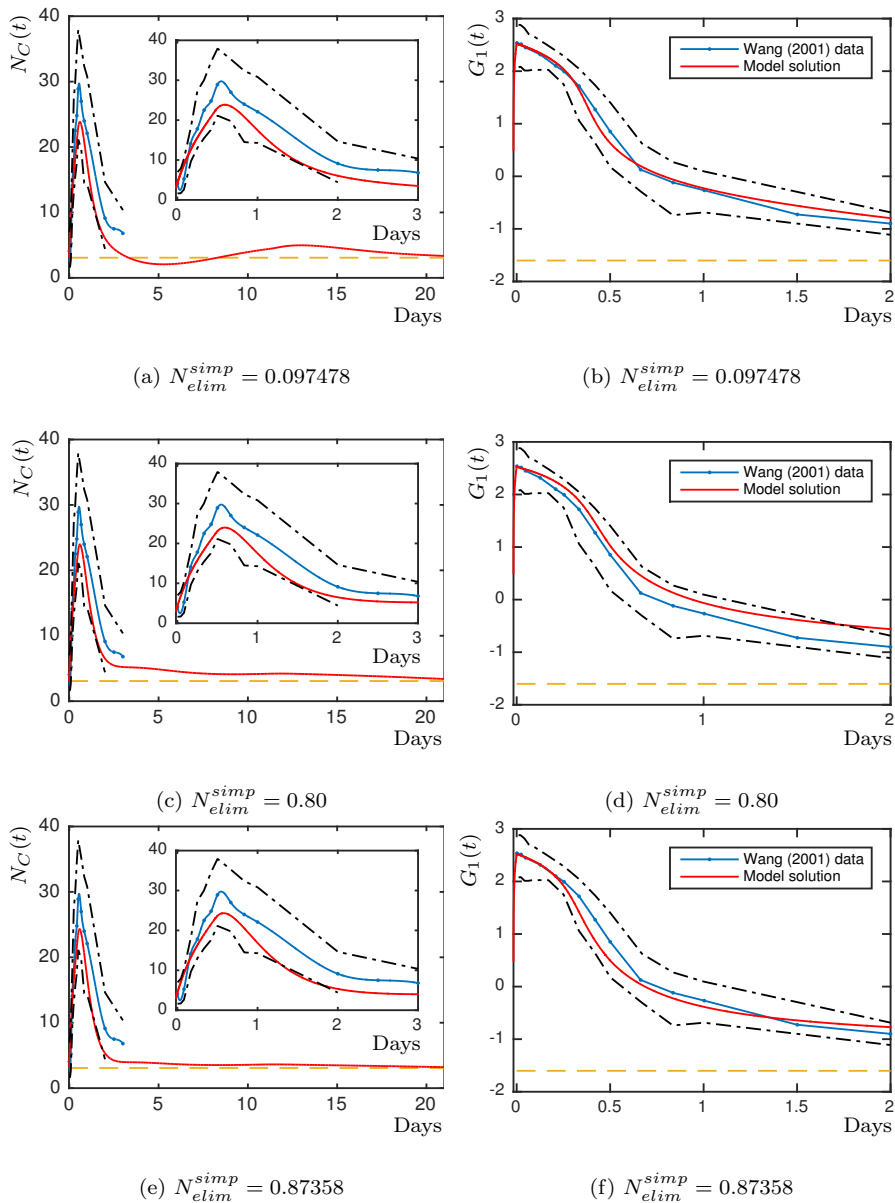
\begin{figure}[tp]
  \begin{subfigure}[b]{0.5\linewidth}
    \begin{center}
    \includegraphics[width=0.9\linewidth]{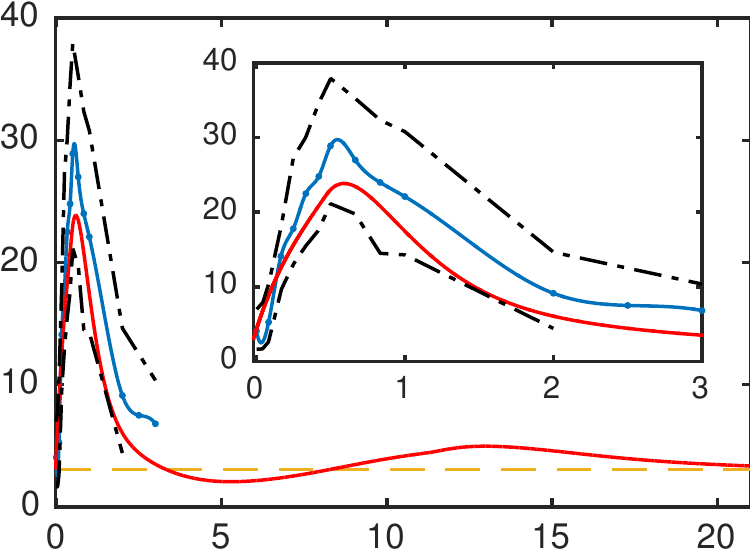}
         \put(-163,97){\rotatebox[origin=c]{90}{$N_C(t)$}}
     \put(-19,-6){Days}
      \put(-27,25){Days}
     \end{center}
    \caption{$N_{elim}^{simp}=0.097478$}
    \label{fig:NeutrophilsNelim09}
    \vspace{4ex}
  \end{subfigure}
  \begin{subfigure}[b]{0.5\linewidth}
    \begin{center}
    \includegraphics[width=0.9\linewidth]{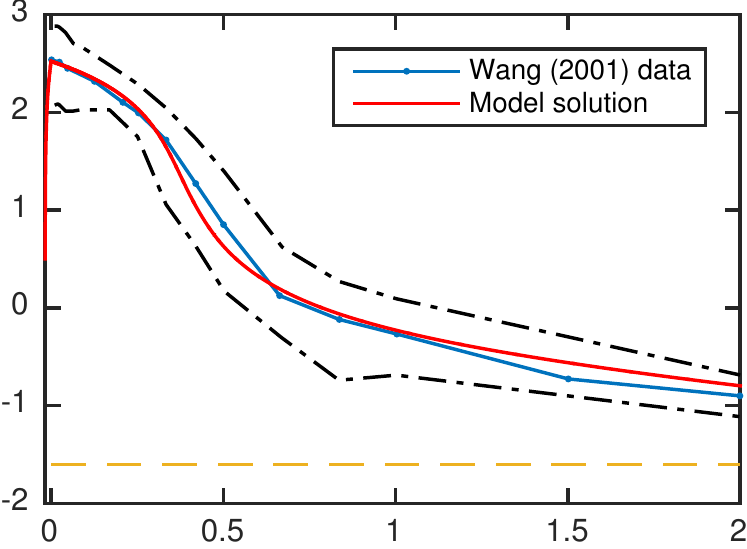}
     \put(-160,95){\rotatebox[origin=c]{90}{$G_1(t)$}}
     \put(-19,-6){Days}
     \end{center}
    \caption{$N_{elim}^{simp}=0.097478$}
    \label{fig:GCSFNelim09}
    \vspace{4ex}
  \end{subfigure}
  \begin{subfigure}[b]{0.5\linewidth}
    \begin{center}
    \includegraphics[width=0.9\linewidth]{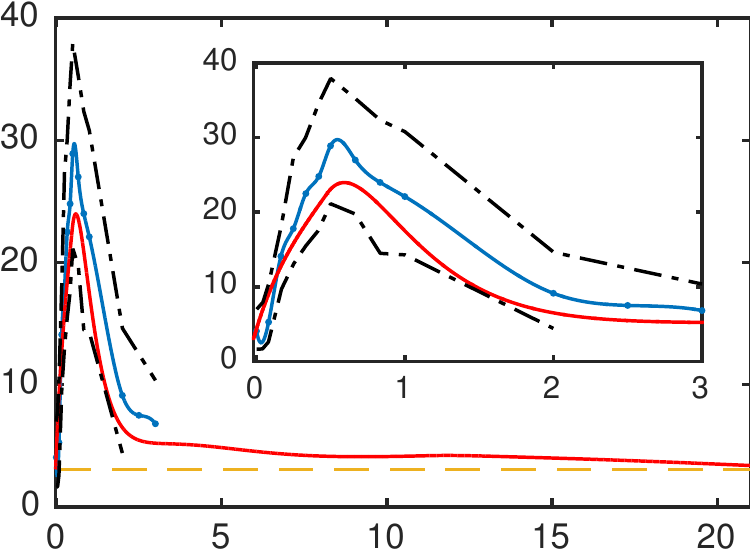}
         \put(-162,97){\rotatebox[origin=c]{90}{$N_C(t)$}}
    \put(-19,-6){Days}
          \put(-27,25){Days}
     \end{center}
    \caption{$N_{elim}^{simp}=0.80$}
    \label{fig:NeutrophilsNelim72}
  \end{subfigure}
  \begin{subfigure}[b]{0.5\linewidth}
    \begin{center}
    \includegraphics[width=0.9\linewidth]{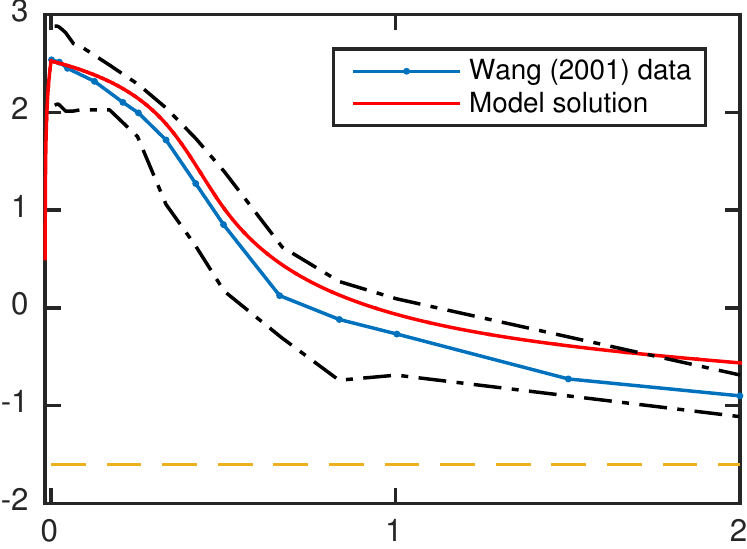}
         \put(-160,95){\rotatebox[origin=c]{90}{$G_1(t)$}}
    \put(-19,-6){Days}
     \end{center}
    \caption{$N_{elim}^{simp}=0.80$}
    \label{fig:GCSFNelim72}
  \end{subfigure}
    \begin{subfigure}[b]{0.5\linewidth}
    \begin{center}
    \includegraphics[width=0.9\linewidth]{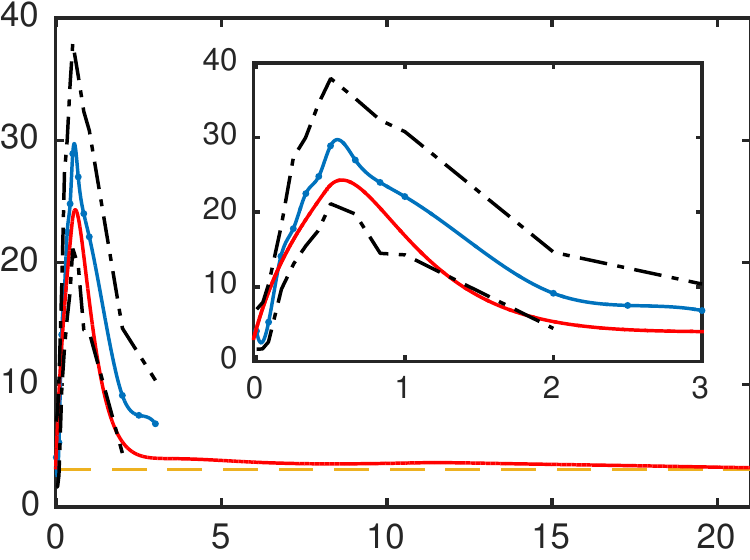}
         \put(-163,97){\rotatebox[origin=c]{90}{$N_C(t)$}}
    \put(-19,-6){Days}
       \put(-27,25){Days}
     \end{center}
    \caption{$N_{elim}^{simp}=0.87358$}
    \label{fig:NeutrophilsNelim87}
  \end{subfigure}
  \begin{subfigure}[b]{0.5\linewidth}
    \begin{center}
    \includegraphics[width=0.9\linewidth]{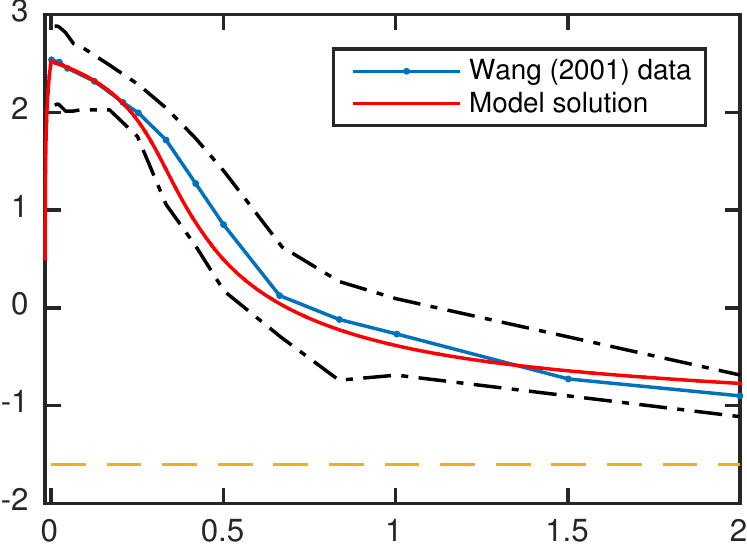}
         \put(-163,95){\rotatebox[origin=c]{90}{$G_1(t)$}}
     \put(-19,-6){Days}
     \end{center}
    \caption{$N_{elim}^{simp}=0.87358$}
    \label{fig:GCSFNelim87}
  \end{subfigure}
     \caption{Simulations of the full model \eq{eq:HSCs}-\eq{eq:nu} for various parameter sets with
     different $N_{elim}$ values. Left: Circulating neutrophil concentrations in \unit{10^9 cells/kg}
     over 21 days, with the first three days shown as an inset. Right: The corresponding circulating G-CSF
     concentrations.
     Blue lines with data: digitised data from Figure~7 (neutrophil concentrations) and Figure~6 (G-CSF concentrations) of Wang~\cite{Wang2001}, red solid lines: model solution, black dashed lines: maximum and minimum digitised data values from Figure~7 and Figure~6 of \cite{Wang2001}, yellow dashed lines: respective homeostatic values.}
  \label{fig:NeutrophilParameters}
\end{figure}

Figure~\ref{fig:NeutrophilParameters}, compares the resulting model solutions for three different values of $N_{elim}$, two of which are shown to be less optimal. Also included are the corresponding G-CSF predictions without any re-estimation from the values obtained in Section~\ref{sec:GCSFParameters}. For $N_{elim}=0.097478$, the G-CSF response is well predicted as seen in Figure~\ref{fig:GCSFNelim09} but because of the renal domination of these parameters, the cytokine paradigm fails in the endogenous-only case. Moreover, repeated administrations of exogenous G-CSF will not accumulate per clinical observations. The G-CSF response seems to be well characterised by the $N_{elim}^{simp}=0.87358$ parameters in Figure~\ref{fig:GCSFNelim87} however the dynamics of the neutrophil response in Figure~\ref{fig:NeutrophilsNelim87} do not stay within the data bounds, and so are not a good fit. Using $N_{elim}^{simp}=0.80$, both the neutrophil and G-CSF responses are successfully predicted in Figures~\ref{fig:NeutrophilsNelim72} and \ref{fig:GCSFNelim72}. The two sets with the lowest errors ($N_{elim}^{simp}=0.71678$ and $N_{elim}^{simp}=0.8$) were used to determine parameters relating to the pharmacodynamic effects of chemotherapy, which is discussed in Section~\ref{sec:ChemoEstimation}.

\subsection{Estimation of Chemotherapy Related Parameters}
\label{sec:ChemoEstimation}

To estimate parameters in \eqref{eq:HSCchemo} and \eqref{eq:etaNPchemo}, data from the results of the Phase I clinical trial of Zalypsis\textsuperscript{\textregistered} were digitised from Gonz\'alez-Sales \cite{GonzalezSales2012}. Unlike the data used for fitting in Sections~\ref{sec:GCSFParameters} and \ref{sec:NeutFit}, here the protocols differ from one subject to the next and are reported per patient. All dosing regimens were as stated \cite{GonzalezSales2012} with doses scaled by body surface area (BSA). Since the subjects were patients undergoing anti-cancer treatments, deviations from the prescribed protocols were frequent. Thus only subjects in the top row (A, B) and bottom row (D, E) of Figure~3 in \cite{GonzalezSales2012} were retained for our analyses.

As with the parameter estimation of the two previous sections, we define the function $N^{ch_{j}}_{dat}(t)$ from a spline fit to the data, where $j=A,B,D,E$ corresponds to each of the retained subjects. Likewise, the function $N^{ch_{j}}(t)$ was defined from the solution from the DDE solver \textit{ddesd} in Matlab \cite{Mathworks} for each patient. When the subject was administered two or more cycles of chemotherapy, we took time intervals corresponding to the first two cycles. Thus, the time spans differed for each subject-specific fitting procedure and were: $t_{span_{A}}=[0,43]$, $t_{span_{B}}=[0,41]$, $t_{span_{C}}=[0,47]$, and $t_{span_{D}}=[0,61]$. As explained in Section~\ref{sec:ParVals}, to account for each subject's baseline ANC, we adjust a scaling factor so our homeostasis $N^*$ value matches each individual's. We have previously shown the robustness of a similar model to pharmacokinetic interindividual and interoccasion variability which substantiates this adjustment and the use of average values in physiological models \cite{Craig2015b}. For each of the four patients, we define the integrals

\begin{equation}
\label{eq:L2chemo}
\frac{1}{| t_{span_{j}}|}\int_{\min(t_{span_{j}})}^{\max(t_{span_{j}})} N(t)^2dt,
\end{equation}
where $j=A,B,D,E$. To find average parameter values which fit to all four patients together, we further defined the average error in the $L^2$ function norm of \eq{eq:L2} between the simulated solutions and the data by

\begin{equation}
\label{eq:ChemoErr}
Err=\frac{1}{4}\sum_{j}\|N^{ch_{j}}(t)-N_{dat}^{ch_{j}(t)}\|_2^2.
\end{equation}

\begin{table}[!h]
\begin{center}
\begin{tabular}{|c|c|c|c|}\hline
Name & Value 1 & Value 2  & Units  \\ \hline
$N_{elim}^{simp}$  & 0.71678 & 0.8  & $-$ \\
\hline
$N_{elim}$ & 0.6204 & 0.7033  & $-$ \\
\hline
$h_Q$ & 0.0071122 & 0.0079657  & $-$\\
$EC_{50}$ & 0.78235 & 0.72545 & ng/mL \\
$s_c$  & 0.90568 & 0.89816 & $-$ \\
$\etaNP^{inf}$ & 0 & 0 & days$^{-1}$\\
\hline
$Err$ & 0.17068 & 0.16965 & $-$\\
\hline
\end{tabular}
\end{center}
\caption{Results of the parameter estimation of chemotherapy effects values.}
\label{tab:ChemoValues}
\end{table}%


\begin{figure}[ht]
  \begin{subfigure}[b]{0.5\linewidth}
    \begin{center}
    \includegraphics[width=0.9\linewidth]{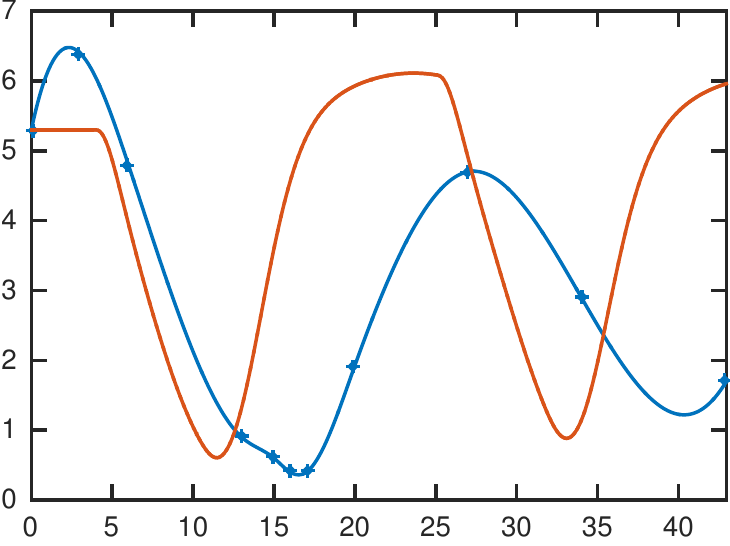}
        \put(-163,100){\rotatebox[origin=c]{90}{$N_C(t)$}}
	\put(-19,-6){Days}
	\end{center}
    \caption{Subject A}
    \label{fig:ChemoParametersA}
    \vspace{4ex}
  \end{subfigure}
  \begin{subfigure}[b]{0.5\linewidth}
    \begin{center}
    \includegraphics[width=0.9\linewidth]{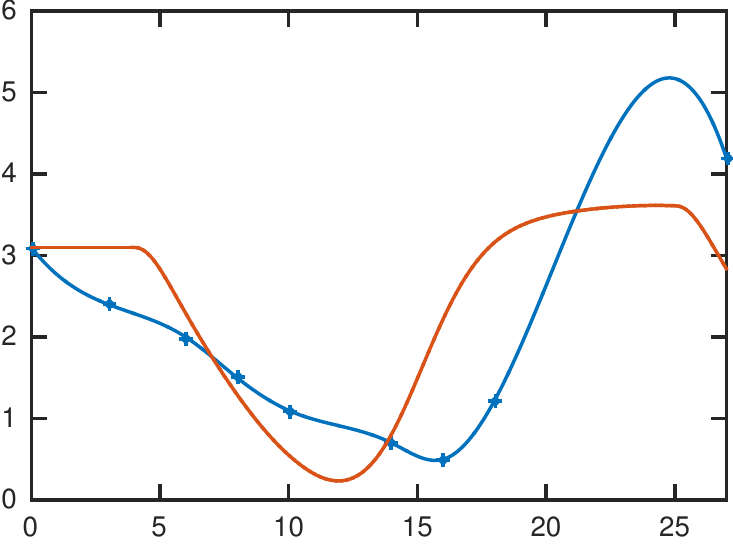}
            \put(-163,100){\rotatebox[origin=c]{90}{$N_C(t)$}}
	\put(-19,-6){Days}
	\end{center}
    \caption{Subject B}
    \label{fig:ChemoParametersB}
    \vspace{4ex}
  \end{subfigure}
  \begin{subfigure}[b]{0.5\linewidth}
    \begin{center}
    \includegraphics[width=0.9\linewidth]{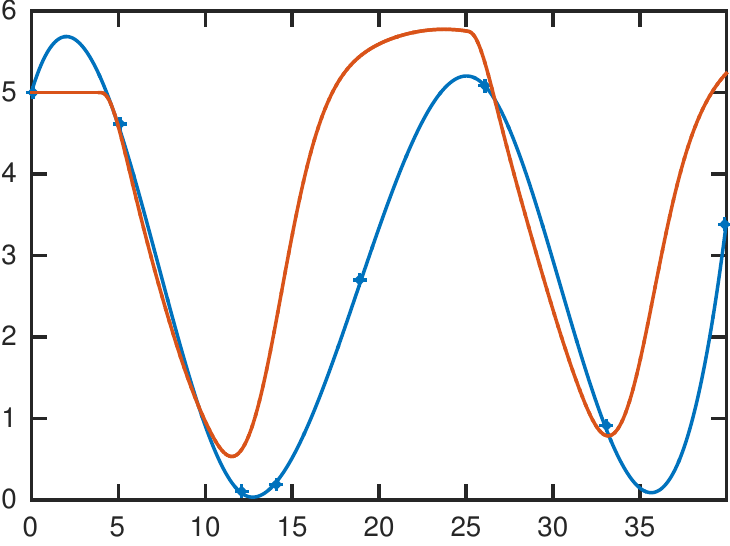}
            \put(-163,100){\rotatebox[origin=c]{90}{$N_C(t)$}}
	\put(-19,-6){Days}
	\end{center}
    \caption{Subject D}
    \label{fig:ChemoParametersC}
  \end{subfigure}
  \begin{subfigure}[b]{0.5\linewidth}
    \begin{center}
    \includegraphics[width=0.9\linewidth]{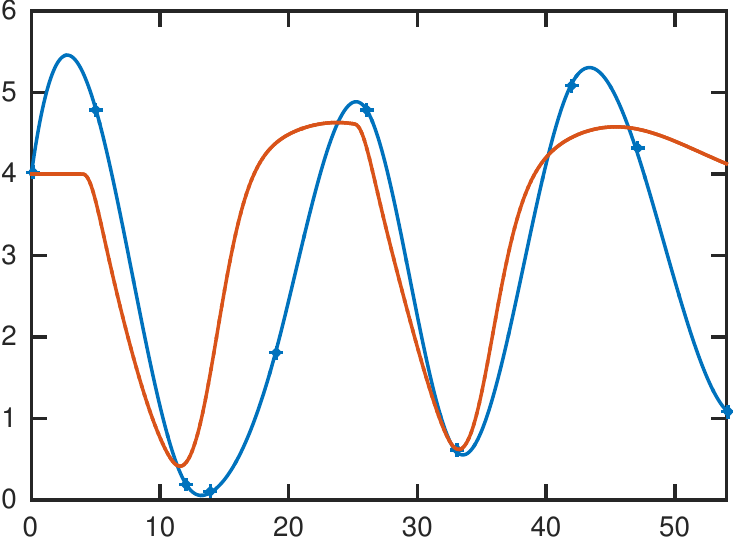}
            \put(-163,100){\rotatebox[origin=c]{90}{$N_C(t)$}}
	\put(-19,-6){Days}
	\end{center}
    \caption{Subject E}
    \label{fig:ChemoParametersD}
  \end{subfigure}
   \caption{Results from the chemotherapy parameter fitting for $N_{elim}^{simp}=0.80$ parameters over two chemotherapy cycles. Model solutions were obtained using the parameters given in Table~\ref{tab:NeutrophilValues} and by simulating the full model \eq{eq:HSCs}-\eq{eq:nu}. Chemotherapeutic concentrations are obtained via \eq{eq:ChemoModel1} and \eq{eq:etaNPchemo}. Equation~\eq{eq:AQCraig} is replaced by \eq{eq:HSCchemo} and solved by using \eq{eq:dAQdt}  Data and experimental protocols from Figure~3 of Gonz\`alez-Sales \cite{GonzalezSales2012}. Blue lines with data: digitised data, red solid lines: model solution.}
  \label{fig:ChemoParameters}
\end{figure}

Parameters $h_Q$, $\etaNP^{inf}$, $s_c$, and $EC_{50}$  were then estimated using the \textit{lsqcurvefit} optimisation routine in Matlab \cite{Mathworks} and similarly averaged. These values are reported in Table~\ref{tab:ChemoValues} and the results of Figure~\ref{fig:ChemoParameters} were obtained from simulations using these parameters. For each of $h_Q$, $EC_{50}$, $s_c$, and $\etaNP^{inf}$, similar estimates were obtained for $N_{elim}^{simp}=0.71578$ and $N_{elim}^{simp}=0.8$, although the average error of $N_{elim}^{simp}=0.8$ is slightly smaller and was accordingly retained as optimal.

\clearpage

\section{Parameter Values}
\label{sec:ParVals}

Here we summarise the parameter values we use in the full model taken from experimental results and the fitting procedures described in Section~\ref{sec:ParEstim}. For the model to be self-consistent these parameters must be positive and satisfy the parameter constraints that we derived above, namely:  \eq{eq:ftransratio}, \eq{eq:Constraint1},
\eq{eq:Constraint3}, \eq{eq:NdelConsist}, \eq{eq:tauNRstarConstr} and \eq{eq:muinterval}.

The main model parameters are stated in Table~\ref{tab:modelparams}.
For the stem cells we reuse parameter values for $\Qhomeo$, $\gamma_Q$, $\tau_Q$,
$f_Q$, $s_2$ and $\beta(\Qhomeo)$ from previous modelling (sometimes rounding them to fewer significant figures).
The value of $\theta_2$ is obtained by evaluating \eq{eq:betaQ} at homeostasis and rearranging to
obtain
\be \label{eq:theta2}
\theta_2=\left[\frac{(\Qhomeo)^{s_2}\beta(\Qhomeo)}{f_Q-\beta(\Qhomeo)}\right]^\frac{1}{s_2}.
\ee
In Table~\ref{tab:modelparams} we quote a value of $\theta_2$ to five significant figures, but in our computations all parameters defined by formulae are evaluated to full machine precision. This ensures that
our differential equation model has a steady state exactly at the stated homeostasis values.


\begin{table}[ht!]
\noindent\mbox{}\hspace{-1cm}
\setlength{\tabcolsep}{2pt}
\begin{tabular}{|c|c|c|c|c|}
\hline
\!\!Name\!\! & Interpretation & Value & Units & Source \\ \hline
$\gamma_Q$ & HSC apoptosis rate  & $0.1$ & \unit{days^{-1}} & \cite{Brooks2012,Craig2015} \\
$\tau_Q$ & Time for HSC re-entry & $2.8$ & \unit{days} & \!\cite{Mackey:01,Bernard2003,Lei:10,Craig2015}\! \\
$A_Q^*$ & HSC Amplification Factor & $1.5116^\dagger$ & $-$ & Eq.~\eq{eq:AQ} \\
$f_Q$ & Maximal HSC re-entry rate & $8$ & \unit{days^{-1}} & \cite{Bernard2003,Brooks2012,Craig2015} \\
$s_2$ & HSC re-entry Hill coefficient & $2$ & $-$ & \cite{Bernard2003,Brooks2012,Craig2015} \\
$\theta_2$ & Half-effect HSC concentration & $0.080863^\dagger$ & \unit{10^6 cells/kg} & Eq.~\eq{eq:theta2} \\   \hline
$\kappa_\delta$ & HSC differentiation rate to other lines & $0.014665^\dagger$ & \unit{days^{-1}} & Eq.~\eq{eq:kappaVals} \\
$\kappa^\textit{min}$ & HSC-neutrophil minimal differentiation rate & $0.0073325^\dagger$ & \unit{days^{-1}} & Eq.~\eq{eq:PDParamsDdet} \\
$\kappaNhomeo$ & HSC-neutrophil homeo differentiation rate & \!\!$0.0073325^\dagger$\!\! & \unit{days^{-1}} & Eq.~\eq{eq:kappaVals}\\
$s_1$ & HSC-neutrophil differentiation Hill coefficient & $1.5$ & $-$ & Eq.~\eq{eq:s1} \\
\hline
$\etaNPhomeo$ & Neutrophil homeostasis effective proliferation rate & $1.6647^\dagger$ & \unit{days^{-1}} & Eq.~\eq{eq:etaNPval} \\
$\bNP$ & Neutrophil proliferation M-M constant & $0.022868$ & \unit{ng/mL} & Fit Table~\ref{tab:NeutrophilValues} \\
$\etaNP^\textit{min}$ & Neutrophil minimal proliferation rate & $1.4060$ & \unit{days^{-1}} & Eq.~\eq{eq:PDParamsDdet} \\
$\tauNP$ & Neutrophil proliferation time & $7.3074^\dagger$ & \unit{days} & Eq.~\eq{eq:tauNP2} \\
\hline
$V_{max}$ & Maximal neutrophil maturation velocity & $7.8670$ & $-$ & Fit Table~\ref{tab:NeutrophilValues} \\
$b_V$ & maturation velocity half-effect concentration & $0.24611$ & \unit{ng/mL} & Eq.~\eq{eq:PDParamsDdet} \\
$\aNM$ & Homeostasis neutrophil maturation time & $3.9$ & \unit{days} &  \cite{Dancey1976,Hearn1998} \& \eq{eq:aNMtauNR} \\
$\gammaNM$ & Neutrophil death rate in maturation & $0.15769^\dagger$ & \unit{days^{-1}} & Eq. \eq{eq:gammaNMeq} \\
\hline
$\ftranshomeo$ & Homeostasis Reservoir Release rate & $0.36400^\dagger$ & \unit{days^{-1}} & Eq. \eq{eq:ftransGstar} \\
$\ftrans^\textit{max}$ & Maximal Reservoir Release rate & $4.1335^\dagger$ & \unit{days^{-1}} & Eq.~\eq{eq:PDParamsDdet} \\
$b_G$ & Reservoir Release half-effect concentration & $1.8924\times10^{-4}$ & $-$ & Eq.~\eq{eq:PDParamsDdet} \\
$\gammaNR$ & Neutrophil death rate in reservoir & \!\!$0.0063661^\dagger$\!\! & \unit{days^{-1}} &  Eq. \eq{eq:gammaNR} \\
\hline
$\gamma_N$ & Neutrophil Removal Rate from Circulation & $35/16$ & \unit{days^{-1}} & Eq. \eq{eq:tauNC}  \\
\hline
$\Gonehomeo$ & Homeostasis Free G-CSF Concentration & $0.025$ & \unit{ng/mL} & \cite{Watari1989,Kawakami1990,Barreda2004,Krzyzanski2010} \\
$G_{BF}^*$ & Homeostasis neutrophil receptor bound fraction & $1.5823\times10^{-5}$ & $-$ & Eq. \eq{eq:GBF} \\
$\Gprod$ & Endogenous G-CSF production rate & $0.014161^\dagger$ & \!\!\unit{ng/mL/day}\!\! & Eq. \eq{eq:Gprod}\\
$V$ & Bound G-CSF conversion factor & 0.525 & \!\!\unit{\frac{ng/mL}{10^9 cells/kg}}\!\! & Eq. \eq{eq:Vval}\\
$k_{ren}$  & G-CSF renal elimination rate & 0.16139 & \unit{days^{-1}} & \!\!Fit Table~\ref{tab:PKParams}\!\! \\
$k_{int}$  & G-CSF effective internalisation rate  & 462.42 &
                                      \unit{days^{-1}} & \!\!Fit Table~\ref{tab:PKParams}\!\! \\
$k_{12}$  & G-CSF Receptor binding coefficient & 2.2423 & {\tiny\unit{(ng/mL)^{-Pow}days^{-1}}} & \!\!Fit Table~\ref{tab:PKParams}\!\! \\
$k_{21}$  & G-CSF Receptor unbinding rate & 184.87 & \unit{days^{-1}} & \!\!Fit Table~\ref{tab:PKParams}\!\! \\
$\Pow$  & Effective G-CSF binding coefficient & 1.4608 & $-$ & \!\!Fit Table~\ref{tab:PKParams}\!\! \\
\hline
\end{tabular}
\caption{Model Parameters. $^\dagger$ -- these parameters are displayed to 5 significant figures here, but
the value is actually defined by the stated equation, and in simulations/computations we use the
stated formula to define the value to machine precision.}
\label{tab:modelparams}
\end{table}

\begin{table}[ht!]
\noindent\mbox{}\hspace{-2.25cm}
\setlength{\tabcolsep}{2pt}
\begin{tabular}{|c|c|c|c|c|}
\hline
Name & Interpretation & Value & Units & Source \\
\hline
$\Qhomeo$ & HSC homeostasis concentration  & $1.1$ & \unit{10^6 cells/kg} & \cite{Bernard2003,Lei:10,Craig2015} \\
$\beta(\Qhomeo)$ & HSC re-entry rate & $0.043$ & \unit{days^{-1}} & \cite{Mackey:01,Craig2015}  \\
\hline
$\Nhomeo$ & Homeostasis Total Blood Neutrophil Pool & $0.22/0.585$ & \unit{10^9 cells/kg} & Eq.~\eq{eq:NhomeoVal} \\
$\NRhomeo$ & Homeostasis Neutrophil Reservoir Concentration & $2.26$ & \unit{10^9 cells/kg} & \cite{Dancey1976} \\
$N_P^*$ & Homeostasis Neutrophil Proliferation Concentration & $0.93$ & \unit{10^9 cells/kg} & Eq.~\eq{eq:NPNMVals} \\
$N_M^*$ & Homeostasis Neutrophil Maturation Concentration & $4.51$ & \unit{10^9 cells/kg} & Eq.~\eq{eq:NPNMVals} \\
\hline
$\Gtwohomeo$ & Homeostasis bound G-CSF concentration & $2.1899\times10^{-5}$ & \unit{ng/mL} & Eq.~\eq{eq:Gtwohomeo}\\ \hline
$\tauNRhomeo$ & Homoeostasis Neutrophil mean time in reservoir & $2.7$ & \unit{days} & \cite{Dancey1976,Hearn1998} \& \eq{eq:aNMtauNR} \\
$\tauNChomeo$ & Homoeostasis Neutrophil mean time in circulation & $16/35$ & \unit{days} & \cite{Dancey1976} \\
$\tau_{1/2}$ & Circulating Neutrophil half-removal time & $7.6$ & \unit{hours} & \cite{Dancey1976} \\
\hline
$A_N^*$ & Homeostasis neutrophil proliferation+maturation amplification & $1.0378\times10^{5\dagger}$  & $-$ & Eq. \eq{eq:ANstar2} \\ \hline
$\tilde{b}_V$ & scaled maturation half-effect concentration & $0.031283$ & \unit{ng/mL} & Fit Table~\ref{tab:NeutrophilValues} \\ \hline
$\ftrans^\textit{ratio}$ & Ratio of maximal and homeostasis reservoir release rates & $11.356$ & $-$ & Fit Table~\ref{tab:NeutrophilValues} \\
$\ftrans(0)$ & Minimal reservoir release rate & $0.020056$  & \unit{days^{-1}} & Fit Table~\ref{tab:NeutrophilValues} \\ \hline
$\theta$ & Ratio of rate cells leave proliferation at knockout to homeostasis & $0.15096$ & $-$ & Eq.~\eq{eq:thetaval} \\
$C_{ko}$ & Knockout total blood neutrophil pool fraction & $0.25$ & $-$ & \cite{Bugl2012,Lui2013} \\
$\mu$ & Ratio of minimal and homeostasis proliferation rates & $0.84458$ & $-$ & Fit Table~\ref{tab:NeutrophilValues}\\ \hline
\end{tabular}
\caption{Auxiliary Parameters which are not in the model in Section~\ref{sec:Model}, but whose values are used to define other parameters. $^\dagger$ -- these parameters are displayed to 5 significant figures here, but
the value is actually defined by the stated equation, and in simulations/computations we use the
stated formula to define the value to machine precision.}
\label{tab:auxparams}
\end{table}

For the neutrophil parameters we mainly take experimental values from the work of Dancey~\cite{Dancey1976}
and use the formulae of Section~\ref{sec:NeutConstr} to determine the related model parameter values. However,
some choices and adjustments need to be made to ensure that the values are consistent with the model.
Dancey~\cite{Dancey1976} measured the circulating neutrophil pool to be $0.22\times\unit{10^9cells/kg}$
and the recovery rate to be $0.585$ from which we obtain the total blood neutrophil pool $\Nhomeo$ (including the
marginated pool) to be
\be \label{eq:NhomeoVal}
\Nhomeo=\frac{0.22}{0.585}\approx 0.3761\times\unit{10^9cells/kg}.
\ee
Since $N(t)$ measures the total blood neutrophil pool in units of \unit{10^9cells/kg} some care needs to
be taken when comparing to data, where absolute neutrophil counts (ANC) measure the circulating neutrophil
pool in units of
\unit{cell/\mu L}. Based on $70\unit{kg}$ of body mass and $5\unit{litres}$ of blood we have the default conversion
factor for healthy subjects of
\be \label{eq:ANC}
ANC=0.585\times\frac{70}{5}\times1000\times N(t) = 8190\, N(t)\unit{cell/\mu L}.
\ee
This gives a baseline homeostasis ANC of $8190\Nhomeo=3080\unit{cell/\mu L}$,
well within the accepted normal range of $1800-7000 \unit{cells/\mu L}$ \cite{Ryan2016}.
When comparing our model to data for individuals with different
baseline ANCs (as in Section~\ref{sec:ChemoEstimation}) we adjust the conversion factor \eq{eq:ANC}, but not the parameter values in our
model, so that $\Nhomeo$ gives the homeostasis ANC of the data.

Dancey~\cite{Dancey1976} measures the proliferation and maturation phases at homeostasis to
be $N_P^*=2.11\unit{\times10^9cells/kg}$ (mainly promyelocytes and myelocytes)
and $N_M^*=3.33\unit{\times10^9cells/kg}$ (metamyelocytes and bands). Using these numbers in the
calculations in Section~\ref{sec:NeutConstr} results in a proliferation time $\tauNP$ defined by
\eq{eq:tauNP2} of about $26$ days. In our model $\tauNP$ is the time from when the HSC first commits to
differentiate to the neutrophil line to the end of proliferation of the neutrophil
precursors. Although this time has never been definitively measured, $26$ days seems to be too long. This
is confirmed by the time to neutrophil replenishment in the blood after both allogenic and autologous stem cell transplantation \cite{Baiocchi1993,Cairo1992}, where circulating neutrophils are seen two weeks after the transplant.
We suspect that this overly long proliferation time results from the simplification in our model
of considering proliferation as a single homogenous process as detailed in Section~\ref{sec:PDEDerivation}.

\sloppy{To obtain a more realistic proliferation time of around a week, close to the $6.3 \unit{days}$
that Smith \cite{Smith2016} states, we keep the total of
$N_P^*+N_M^*=5.44\unit{\times10^9cells/kg}$ as found by Dancey~\cite{Dancey1976}, but redistribute cells between
proliferation and maturation and set}
\be \label{eq:NPNMVals}
N_P^*=0.93\unit{\times10^9cells/kg}, \qquad N_M^*=4.51\unit{\times10^9cells/kg}.
\ee

Dancey~\cite{Dancey1976} measured the half removal time of neutrophils from circulation to be $t_{1/2}=7.6\unit{hrs}$.
Accordingly, using \eq{eq:tauNC} and rounding, we set $\gamma_N=35/16=2.1875\unit{days^{-1}}$ and obtain
$\tauNChomeo$ as the reciprocal of this. Then equation \eq{eq:MatApopRatCond} imposes the constraint that
$\aNM<5.4823\unit{days}$. If we set $\aNM=3.9\unit{days}$
close to the value of $3.8\unit{days}$
found by Hearn~\cite{Hearn1998}, then \eq{eq:tauNRstarConstr} imposes the constraint that
$\tauNRhomeo\in(1.9543,2.7472)$.
Hence we take
\be \label{eq:aNMtauNR}
\aNM=3.9\unit{days}, \qquad \tauNRhomeo=2.7\unit{days},
\ee
so that both constraints are satisfied, and
$\aNM+\tauNRhomeo=6.6\unit{days}$, the value given in \cite{Dancey1976}. The rest of the neutrophil homeostasis
parameters are calculated using the formulae of Section~\ref{sec:NeutConstr}, paying attention in \eq{eq:ANstar2} to
multiply $\Qhomeo$ by $10^{-3}$ to convert it to the same units as $\NRhomeo$.


The G-CSF pharmacokinetic parameters are fit using the simplified G-CSF
model \eq{eq:UnboundGCSFsimp},\eq{eq:BoundGCSFsimp} as described in Section~\ref{sec:GCSFParameters}.
This produces multiple, but equally plausible, parameter sets but as described in subsequent sections not all of these result in good fits to data when we consider the neutrophil response of the full model \eq{eq:HSCs}-\eq{eq:nu}
to administrations of G-CSF or of chemotherapy. Consequently as stated in Section~\ref{sec:ChemoEstimation},
to obtain the best responses of the system to these scenarios we use the fourth
set of pharmacokinetic parameters from Table~\ref{tab:PKParams} which for the simplified G-CSF
model have $N_{elim}^{simp}=0.8$ to define $k_{ren}$, $k_{12}$, $k_{21}$, $k_{int}$ and $Pow$, as
well as the exogenous G-CSF parameters $V_d$, $F$, $k_a$ (where the last three are dose dependent).
Equations \eq{eq:Gtwohomeo}, \eq{eq:Gprod} and \eq{eq:Nelim} then define $\Gtwohomeo$, $G_{prod}$ and $N_{elim}=0.7033$
for the full model.

At G-CSF knockout, from \cite{Bugl2012,Lui2013} we have $C_{ko}\in[0.2, 0.3]$, so it is natural to set $C_{ko}=0.25$.

For the pharmacodynamic parameters, similar to $Pow$, arguments could be made for choosing $s_1=1$ or $s_1=2$,
but having fit $Pow$ and finding it close to $1.5$, we will simply set
\be \label{eq:s1}
s_1=1.5
\ee
to reduce the number of parameters that need to be fit by one. The remaining pharmacodynamic
parameters $\mu$, $\bNP$, $V_{max}$, $\tilde{b}_V$, $\ftrans(0)$ and $\ftrans^{ratio}$ were
then fit as described in Section~\ref{sec:NeutFit}, with these parameters defining values of
the dependent parameters
$\etaNP^{min}$, $\ftrans^{max}$, $b_V$ and $b_G$ via \eq{eq:PDParamsDdet}.  From Section~\ref{sec:KO} we also obtain values for $\theta$ from \eq{eq:thetaval}
and $\kappaN^\textit{min}$ from \eq{eq:PDParamsDdet}. Each set of kinetic parameters from Table~\ref{tab:PKParams}
defines a different set of pharmacodynamic parameters as reported in Table~\ref{tab:NeutrophilValues},
but as noted already we prefer the parameter set for
$N_{elim}^{simp}=0.80$ which corresponds to $N_{elim}=0.7033$.

The full set of parameter values for our combined neutrophil and G-CSF model \eq{eq:HSCs}-\eq{eq:nu} are given in Table~\ref{tab:modelparams}, along with their units, interpretation and source.
Since some of these parameters are defined by equations involving auxiliary parameters that do not
explicitly appear in the full model we state these parameters and their source in Table~\ref{tab:auxparams}. Parameters related to the pharmacokinetics and pharmacodynamics of both of the exogenous drugs which have not previously been stated are given in Table~\ref{tab:exodrugs}.

\begin{table}[ht!]
\begin{center}
\setlength{\tabcolsep}{2pt}
\begin{tabular}{|c|c|c|c|c|}
\hline
Name & Interpretation & Value & Units & Source \\
\hline
\multicolumn{5}{|c|}{Filgrastim} \\ \hline
\multicolumn{5}{|c|}{300 \unit{mcg} dose} \\ \hline
$V_d$ & Volume of distribution  & 4754.7  & \unit{mL} & Fit Table~\ref{tab:PKParams} \\
$F$ & Bioavailable fraction & 0.64466 & $-$ & Fit Table~\ref{tab:PKParams}\\
$k_a$ & Subcutaneous rate of absorption & 8.0236 & \unit{days^{-1}}  & Fit Table~\ref{tab:PKParams}\\\hline
\multicolumn{5}{|c|}{375 \unit{mcg} dose} \\ \hline
$V_d$ & Volume of distribution  & 2322.9  & \unit{mL} & Fit Table~\ref{tab:PKParams} \\
$F$ & Bioavailable fraction & 0.49964 & $-$ & Fit Table~\ref{tab:PKParams}\\
$k_a$ & Subcutaneous rate of absorption & 6.6133 & \unit{days^{-1}}  & Fit Table~\ref{tab:PKParams}\\
\hline
\multicolumn{5}{|c|}{750 \unit{mcg} dose} \\ \hline
$V_d$ & Volume of distribution  & 2178.0  & \unit{mL} & Fit Table~\ref{tab:PKParams} \\
$F$ & Bioavailable fraction & 0.75 & $-$ & Fit Table~\ref{tab:PKParams}\\
$k_a$ & Subcutaneous rate of absorption & 5.143 & \unit{days^{-1}}  & Fit Table~\ref{tab:PKParams}\\
\hline
\multicolumn{5}{|c|}{Zalypsis\textsuperscript{\textregistered}} \\ \hline
$k_{fp}$ & Rate of exchange from compartment $f$ to $p$ & 18.222& \unit{days^{-1}} & \cite{PerezRuixo2012}\\
$k_{sl_{1}p}$ & Rate of exchange from compartment $sl_1$ to $p$ & 0.6990 & \unit{days^{-1}} & \cite{PerezRuixo2012}\\
$k_{pf}$ & Rate of exchange from compartment $p$ to $f$ & 90.2752 & \unit{days^{-1}} & \cite{PerezRuixo2012}\\
$k_{psl_{1}}$ & Rate of exchange from compartment $p$ to $sl_1$ & 8.2936 & \unit{days^{-1}} & \cite{PerezRuixo2012}\\
$k_{el_{C}}$ & Rate of elimination & 132.0734 & \unit{days^{-1}} & \cite{PerezRuixo2012}\\
$k_{sl_{2}f}$  & Rate of exchange from compartment $sl_2$ to $f$ & 62.5607 & \unit{days^{-1}} & \cite{PerezRuixo2012}\\
$k_{fsl_{2}}$ & Rate of exchange from compartment $f$ to $sl_2$ & 9.2296 & \unit{days^{-1}} & \cite{PerezRuixo2012}\\
BSA & Body surface area & 1.723 & \unit{m^2} & \cite{PerezRuixo2012}\\
$h_Q$ & Effect of chemotherapy on $Q(t)$ & 0.0079657 & $-$ & Fit Table~\ref{tab:ChemoValues} \\
$EC_{50}$ & Half-maximal effect of chemotherapy on $\etaNP$ &  0.75390 & $-$ & Fit Table~\ref{tab:ChemoValues} \\
$s_c$ & Chemotherapy effect Hill coefficient &  0.89816 & $-$ & Fit Table~\ref{tab:ChemoValues} \\
$\etaNP^{inf}$ & Proliferation rate with infinite chemotherapy dose & 0 & \unit{days^{-1}}& Fit Table~\ref{tab:ChemoValues}\\
\hline
\end{tabular}
\caption{Exogenous drug administration parameters determined by parameter fitting as explained in Sections~\ref{sec:GCSFParameters} and \ref{sec:ChemoEstimation}. For Zalpysis\textsuperscript{\textregistered}, $p$: plasma/central compartment, $f$: fast-exchange tissues, $sl_1$: first slow-exchange tissues, $sl_2$: second slow-exchange tissues.}
\label{tab:exodrugs}
\end{center}
\end{table}

\section{Model evaluation and functional responses}
\label{sec:ModelEvaluation}

\sloppy{Having estimated the G-CSF pharmacokinetic, homeostasis related, and chemotherapy pharmacodynamic parameters individually as described in Sections~\ref{sec:GCSFParameters}, \ref{sec:NeutFit}, and \ref{sec:ChemoEstimation}, a convincing evaluation of the ability of the model is to successfully predict data obtained during the concurrent administration of both exogenous drugs. For this, as in \cite{Craig2015}, we simulated the CHOP14 protocol described in \cite{Pfreundschuh2004a} and \cite{Pfreundschuh2004b} which includes the administration of both chemotherapy and exogenous G-CSF. Although the chemotherapeutic drug modelled in Section~\ref{sec:exogcsf} is not part of the combination therapy of the CHOP14 regimen, the cytotoxic effects of the anticancer drugs are presumed to be similar. To compare to the CHOP14 data published in \cite{Krinner2013}, we simulated a regimen of six cycles of 14-day periodic chemotherapeutic treatment with rhG-CSF treatment beginning four days after the administration of chemotherapy and continuing for ten administrations per cycle. As in \cite{Craig2015}, the simulated dose of 4 $\mu$g of Zalypsis\textsuperscript{\textregistered} was selected from the optimal regimens identified in \cite{GonzalezSales2012} and per the CHOP14 protocol outlined in \cite{Pfreundschuh2004a,Pfreundschuh2004b}, ten 300 $\mu$g doses of subcutaneous G-CSF were simulated per cycle. The lower dose of 300 $\mu$g was selected since we assumed an average weight of 70kg per patient throughout.}

\begin{figure}[ht]
\begin{center}
\includegraphics[scale=1]{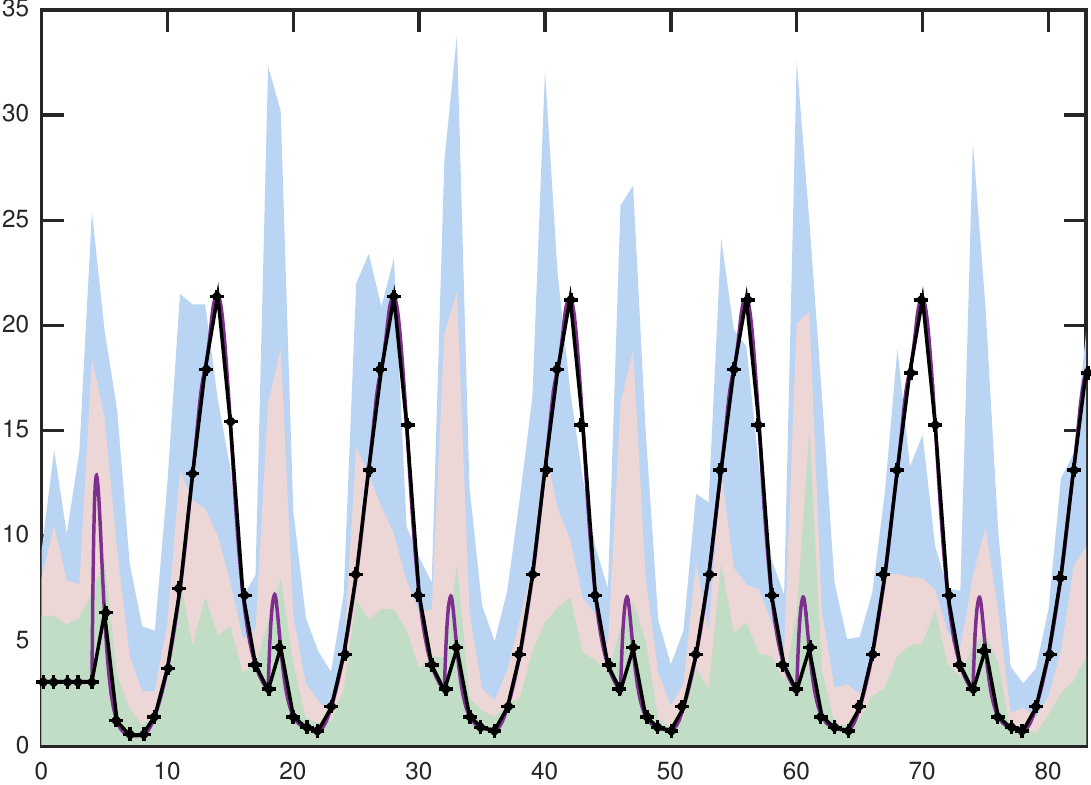}
\put(-325,210){\rotatebox[origin=c]{90}{$N_C(t)$}}
\put(-20,-7){Days}
\end{center}
\caption{Comparison of the predicted neutrophil response to the CHOP14 protocol \cite{Pfreundschuh2004a,Pfreundschuh2004b} for $N_{elim}^{simp}=0.80$. In this regimen, 4 $\mu$g of Zalypsis\textsuperscript{\textregistered} given by a 1 hour IV infusion is administered 14 days apart, beginning on day 0, for 6 cycles (84 days total). Per cycle, ten administrations of 300 $\mu$g subcutaneous doses of filgrastim are given beginning four days after the start of the chemotherapeutic cycle and continuing to day 13 post-chemotherapy. The simulation is compared to data from \cite{Krinner2013}, presented in quartiles. In pale green: the first quartile, in pale pink: median range, in pale blue: third quartile. Black line with sampling points: model prediction sampled every day at clinical sampling points, solid purple line: full model prediction.}
\label{fig:CHOP}
\end{figure}

\begin{figure}[ht]
  \begin{subfigure}[b]{0.5\linewidth}
    \begin{center}
    \includegraphics[width=0.9\linewidth]{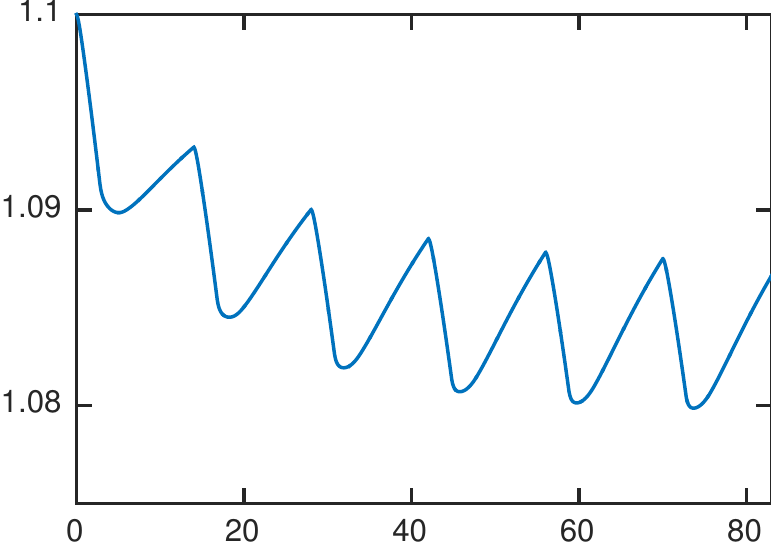}
    \put(-163,96){\rotatebox[origin=c]{90}{$Q(t)$}}
	\put(-20,-7){Days}
    \end{center}
    \caption{}
    \label{fig:OtherPopsCHOPA}
    \vspace{4ex}
  \end{subfigure}
    \begin{subfigure}[b]{0.5\linewidth}
    \begin{center}
    \includegraphics[width=0.9\linewidth]{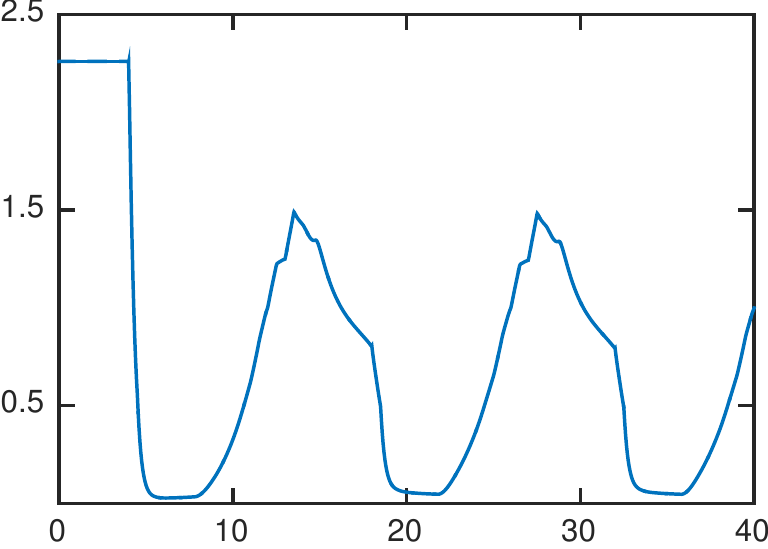}
        \put(-163,94){\rotatebox[origin=c]{90}{$N_R(t)$}}
	\put(-20,-7){Days}
	\end{center}
    \caption{}
    \label{fig:OtherPopsCHOPB}
        \vspace{4ex}
  \end{subfigure}
  \begin{subfigure}[b]{0.5\linewidth}
    \begin{center}
    \includegraphics[width=0.9\linewidth]{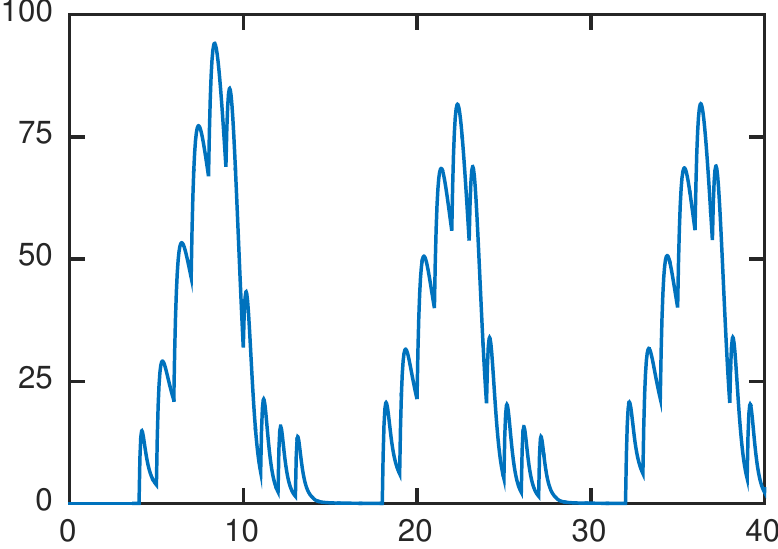}
        \put(-163,93){\rotatebox[origin=c]{90}{$G_1(t)$}}
	\put(-20,-7){Days}
	\end{center}
    \caption{}
  \end{subfigure}
  \begin{subfigure}[b]{0.5\linewidth}
    \begin{center}
    \includegraphics[width=0.9\linewidth]{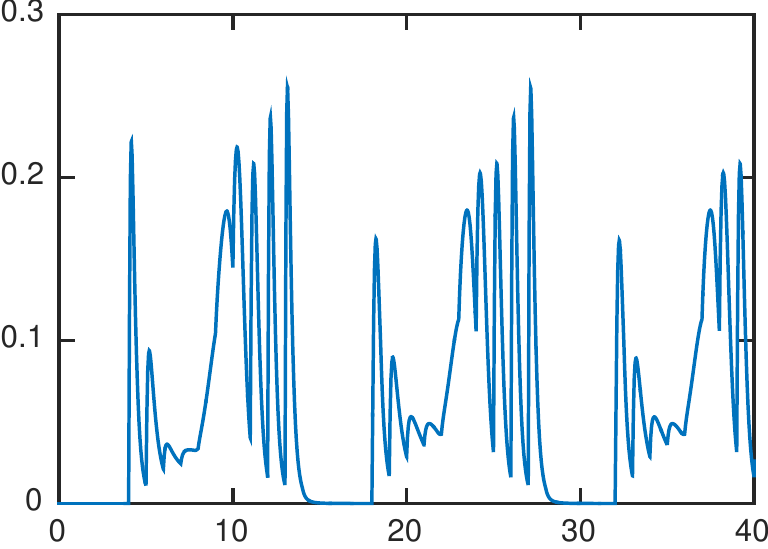}
        \put(-163,93){\rotatebox[origin=c]{90}{$G_2(t)$}}
	\put(-20,-7){Days}
    \end{center}
    \caption{}
  \end{subfigure}
   \caption{Model responses to the CHOP14 protocol as described in Section~\ref{sec:ModelEvaluation}. In a) $Q(t)$ over the six CHOP cycles detailed above, b), c), and d) $N_R(t)$, $G_1(t)$, and $G_2(t)$ over three CHOP cycles.}
  \label{fig:OtherPopsCHOP}
\end{figure}

 \begin{figure}[ht!]
  \begin{subfigure}[b]{0.5\linewidth}
    \begin{center}
    \includegraphics[width=0.9\linewidth]{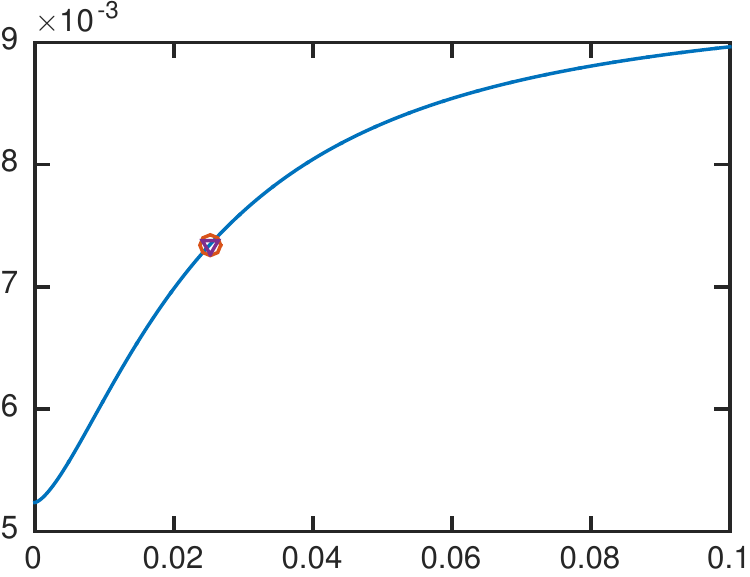}
    \put(-163,90){\rotatebox[origin=c]{90}{$\kappaN(G_1)$}}
     \put(-22,-8){$G_1$}
     \end{center}
    \caption{}
    \label{fig:FunctionResponsesA}
    \vspace{4ex}
  \end{subfigure}
  \begin{subfigure}[b]{0.5\linewidth}
    \begin{center}
    \includegraphics[width=0.9\linewidth]{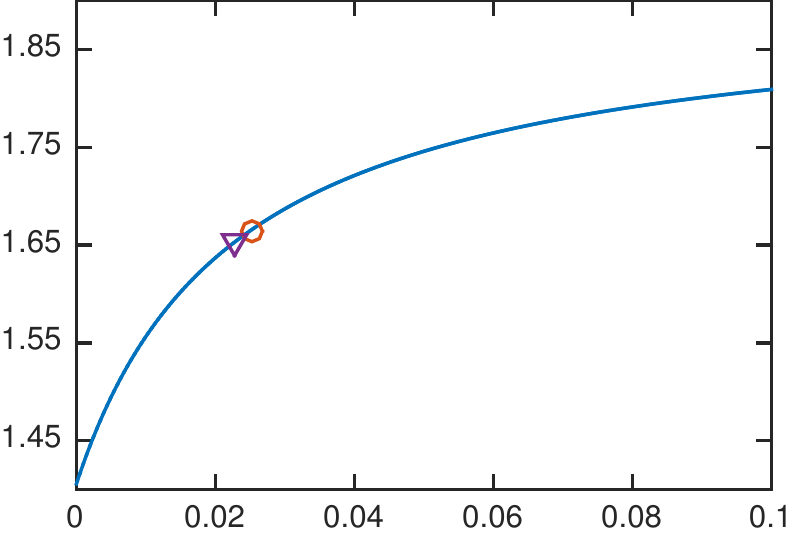}
        \put(-163,85){\rotatebox[origin=c]{90}{$\etaNP(G_1)$}}
     \put(-22,-8){$G_1$}
     \end{center}
    \caption{}
    \label{fig:FunctionResponsesB}
    \vspace{4ex}
  \end{subfigure}
    \begin{subfigure}[b]{0.5\linewidth}
    \begin{center}
    \includegraphics[width=0.933\linewidth]{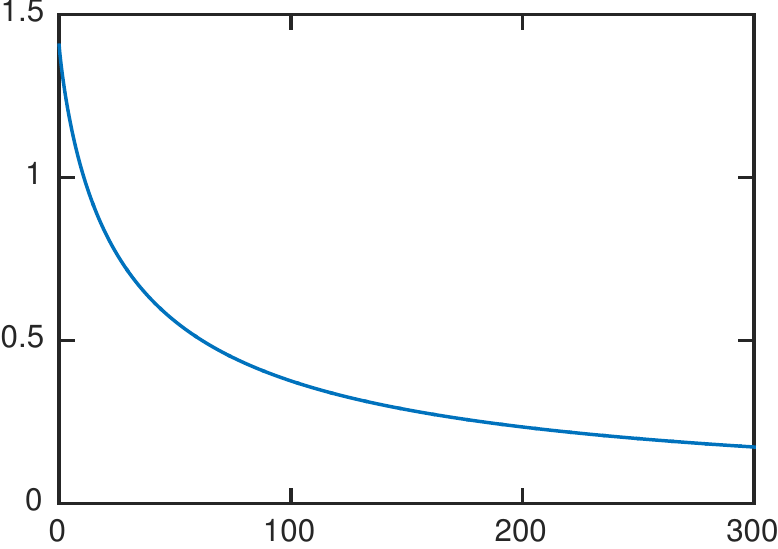}
        \put(-171,70){\rotatebox[origin=c]{90}{$\etaNP^{chemo}(G_1^*,C_p)$}}
     \put(-22,-8){$C_p$}
     \end{center}
    \caption{}
    \label{fig:FunctionResponsesC}
        \vspace{4ex}
  \end{subfigure}
  \begin{subfigure}[b]{0.5\linewidth}
    \begin{center}
    \includegraphics[width=0.9\linewidth]{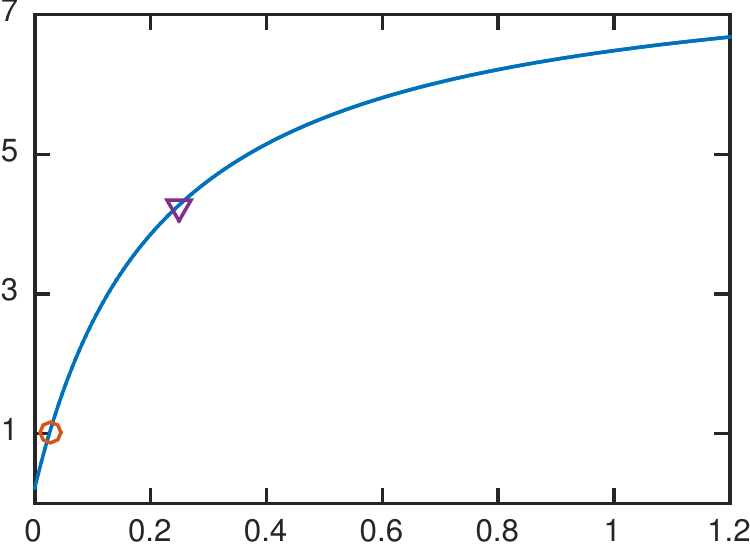}
   \put(-163,85){\rotatebox[origin=c]{90}{$\VN(G_1)$}}
     \put(-22,-8){$G_1$}
     \end{center}
    \caption{}
    \label{fig:FunctionResponsesD}
        \vspace{4ex}
  \end{subfigure}

  \begin{subfigure}[b]{0.5\linewidth}
    \begin{center}
    \includegraphics[width=0.9\linewidth]{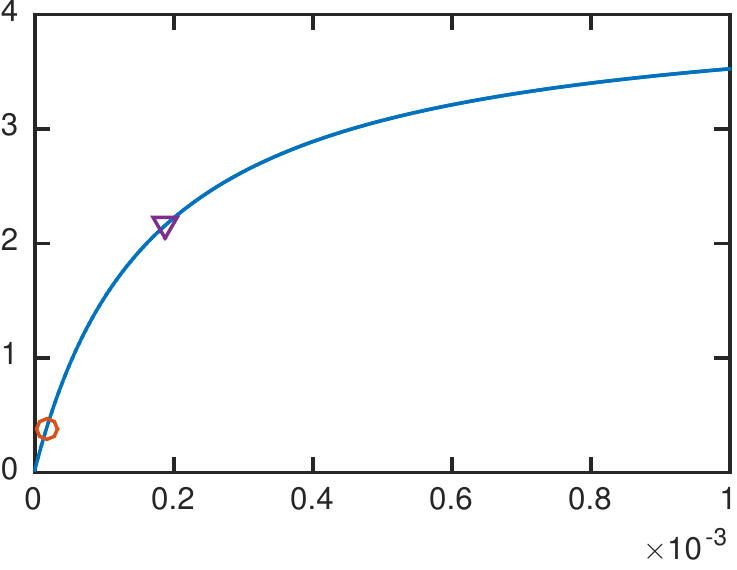}
      \put(-163,88){\rotatebox[origin=c]{90}{$\ftrans(G_{BF})$ }}
     \put(-87,2){$G_{BF}$}
     \end{center}
    \caption{}
    \label{fig:FunctionResponsesE}
  \end{subfigure}
   \caption{Visualisation of the granulopoiesis model's mechanisms as functions of their variables (solid blue lines) with their respective homeostatic and half-effect values (purple triangles), when relevant. Red circles: homeostasis values. }
  \label{fig:FunctionResponses}
\end{figure}

Figure~\ref{fig:CHOP} shows the result of the neutrophil response comparison of the model's prediction to the clinical data.  Unlike experimental settings where information on the HSCs, the marrow neutrophils, and the bound G-CSF concentrations are unavailable, the model's solutions for $Q(t)$, $N_R(t)$, and $G_2(t)$ are easily obtainable and provide insight into not only the mechanisms responsible for myelosuppresion during chemotherapy, but also ways in which this toxicity might be avoided. In Figure~\ref{fig:OtherPopsCHOP}, the HSCs, neutrophils in the marrow reservoir, and bound and unbound G-CSF are all seen to converge to periodic responses. However, while the reduction in HSC concentrations is minimal (Figure~\ref{fig:OtherPopsCHOPA}) the neutrophil marrow reservoir is seen in Figure~\ref{fig:OtherPopsCHOPB} to become severely depleted. This depletion is caused by the delayed effects of the administration of chemotherapy but also the rapid transit of cells from the reservoir into the blood caused by the introduction of exogenous G-CSF four days post-chemotherapy (see Figure~\ref{fig:FunctionResponsesE} below). This in turn prevents ANC recoveries from depressed values, despite the administration of G-CSF. As in \cite{Vainstein2005} and \cite{Craig2015}, it is likely that delaying the beginning of prophylactic G-CSF support during chemotherapy would help to combat myelosuppresion, but this will is a future avenue of investigation.

It can also be illuminating to study how each of the model's functions correspond to the estimated parameters to obtain further insight on the mechanisms of granulopoiesis. Figure~\ref{fig:FunctionResponses} shows the functions $\kappaN (G_1)$, $\etaNP (G_1)$, $\etaNP^{chemo}(G_1)$, $\VN(G_1)$, and $\ftrans (G_{BF})$ and identifies their respective homeostatic levels. We can see that $\ftrans (G_{BF}$) in Figure~\ref{fig:FunctionResponsesE}, has a homeostasis concentration $\ftrans(G_{BF}^*)$ very close to $\ftrans(0)$. This reflects the ability of the granulopoietic system to respond rapidly in the case of emergencies \cite{Rankin2010} but also supports the hypothesis that early prophylactic support with G-CSF during chemotherapy may hasten the emptying of the reservoir due to the responsiveness of $\ftrans (G_{BF}(t))$ in particular.

\section{Discussion}
\label{sec:Discussion}

Clinically relevant translational models in medicine must not only accurately depict different and independent treatment regimes \cite{Vainstein2005}, they must also be able to reconstruct homeostatic and pathological cases which may be intervention independent. The granulopoiesis model we have developed is physiologically-relevant and, perhaps most importantly, provides insight beyond that which is clinically measurable. The updated pharmacokinetic model of G-CSF, novel in that it explicitly accounts for unbound and bound concentrations, correctly accounts for G-CSF dynamics whereas previous one compartment models all resulted in renal dominated dynamics. The new pharmacokinetic model also further allows us to comment on the principle mechanisms driving the production of neutrophils. Although the relatively small number of neutrophil progenitors do not have a significant effect on G-CSF kinetics, our results suggest that differentiation, proliferation and maturation speed are driven primarily by signalling from G-CSF bound to neutrophil progenitors, and not from signalling of G-CSF bound to mature neutrophils. We can further characterise the principle processes governing myelosuppression during the concurrent administration of chemotherapy and prophylactic G-CSF, which we have determined lies in the simultaneous depletion of the marrow reservoir by high doses of exogenous G-CSF combined with fewer neutrophils reaching the reservoir due to the cytotoxicity of the anti-cancer drug.

The modelling reported here combines a number of original approaches to the conceptualisation of  physiological, pharmacokinetic, and pharmacodynamics models and to the estimation of parameters and model verification. For example, traditional least squares estimation was redefined using functions which ensured robustness and allowed for comparisons of predictions to data over richly sampled intervals instead of at fewer data points. Moreover, the model's physiological realism served as a means of evaluating the suitability of optimised parameter values so we were not relying solely on goodness-of-fit, which can obfuscate the biological relevance of results \cite{vanderGraaf2011}. The inclusion of the detailed characterisations of physiological mechanisms in our model therefore serves as a litmus test of suitability in addition to providing intuition about the processes driving granulopoiesis.

The broader implications of the approaches outlined in this work extend into various domains. The derivation of a delay differential equation model with variable aging rate from an age-structured PDE, as described in Section~\ref{sec:PDEDerivation}, is mathematically significant and its intricate nature has previously led to previous modelling errors. As mentioned, the fitting procedures outlined in Sections~\ref{sec:GCSFParameters}, \ref{sec:NeutFit}, and \ref{sec:ChemoEstimation} motivate the development of more refined least squares methods and parameter estimation techniques. Additionally, the novel pharmacokinetic model of G-CSF has ramifications with respect to the usual approaches used by PK/PD modellers. The mischaracterisation of the elimination dynamics, despite the inclusion of internalisation terms, has led to models which contradict what is known of the physiology. While they can  characterise certain clinical situations, like the single administration of exogenous G-CSF, they fail when applied to more complex scenarios. Without accounting for the entire process of neutrophil development or using physiological rationale for a model's parameters, one is unable to judge whether a model captures the complicated dynamics of granulopoiesis. In the model we have developed, we have ensured the accuracy of its predictions and the appropriateness of its parameters through careful construction. In turn, this rational approach has implications for the clinical practice where it can serve to optimise dosing regimens in oncological settings and also serve to pinpoint the origins of dynamical neutrophil disorders like cyclic neutropenia, ultimately contributing to the improvement of patient care and outcomes.






\section*{Acknowledgements}


ARH and MCM are grateful to the National Science and Engineering Research Council (NSERC), Canada for funding through the Discovery Grant program. MC wishes to thank NSERC for funding from the PGS-D program. We are grateful to Fahima Nekka, Jun Li, Jacques B\'elair, and David Dale for their insight and support.

\bibliographystyle{plain}
\bibliography{NeutrophilManuscript16arxiv}

\end{document}